\documentclass[useAMS,usenatbib]{mn2e}
\usepackage{graphicx}
\usepackage{rotating}
\usepackage{supertabular}
\usepackage{amssymb}
\usepackage{amsmath}

\begin{document}

\title[The X-ray properties of clusters in the ESO Distant Cluster Survey]
  {The X-ray properties of optically-selected $z>0.6$ clusters in the ESO Distant Cluster Survey}
  
\author[O.~Johnson et al.]  
  {O.~Johnson$^{1}$\thanks{E-mail:cocj@roe.ac.uk},  P.~Best$^1$, D.~Zaritsky$^2$, D.~Clowe$^2$, A.~Arag\'on-Salamanca$^3$, 
\newauthor C.~Halliday$^4$, P.~Jablonka$^5$, B.~Milvang-Jensen$^6$, R.~Pell{\'o}$^7$, B.~M.~Poggianti$^8$, 
\newauthor G.~Rudnick$^9$, R.~Saglia$^{10}$, L.~Simard$^{11}$, S.~White$^9$\\
$^1$SUPA\thanks{Scottish Universities Physics Alliance} Institute for Astronomy, Royal Observatory Edinburgh, Blackford Hill, Edinburgh EH9 3HJ, UK\\
$^2$Steward Observatory, University of Arizona, 933 North Cherry Avenue, Tucson, AZ 85721, USA\\
$^3$School of Physics and Astronomy, University of Nottingham, University Part, Nottingham NG7 2RD\\
$^4$Osservatorio Astrofisico di Arcetri, Largo E. Fermi 5, 50125 Firenze, Italy\\
$^5$GEPI CNRS-UMR8111, Observatoire de Paris, section de Meudon, 5 Place Jules Janssen, F-92195 Meudon Cedex, France\\
$^6$Dark Cosmology Centre, Niels Bohr Institute, University of Copenhagen, Juliane Maries Vej 30, DK-2100 Copenhagen, Denmark\\
$^7$Laboratoire de' Astrophysique, UMR 5572, Observatoire Midi-Pyrenees, 14 Avenue E. Belin, 31400 Toulouse, France\\
$^8$Osservatorio Astronomico, vicolo dell' Osservatorio 5, 35122 Padova, Italy\\
$^9$Max-Plank-Institute fur Astrophysik, Karl-Schwarschilde-Strasse 1, Postfach 1317, D-85741 Garching, Germany\\
$^{10}$Max-Plank-Institut fur extraterrestrische Physik, Giessenbachstrasses, D-85748 Garching, Germany\\
$^{11}$Herzberg Institute of Astrophysics, National Research Council of Canada, Victoria, BC V93 2E7, Canada}

\date{Accepted 2006 July 13. Received 2006 June 29; in original form 2006 March 06}

\maketitle

\begin{abstract}

We present XMM-Newton observations of three optically-selected $z>0.6$ clusters from the ESO Distant Cluster Survey (EDisCS), comprising the first results of a planned X-ray survey of the full EDisCS high-redshift sample.  The EDisCS clusters were identified in the Las Campanas Distant Cluster Survey as surface brightness fluctuations in the optical sky and their masses and galaxy populations are well described by extensive photometric and spectroscopic observations.  We detect two of the three clusters in the X-ray and place a firm upper limit on diffuse emission in the third cluster field.  We are able to constrain the X-ray luminosity and temperature of the detected clusters and estimate their masses.  We find the X-ray properties of the detected EDisCS clusters are similar to those of X-ray-selected clusters of comparable mass and --- unlike other high-redshift, optically-selected clusters --- are consistent with the $T-\sigma$ and $L_{X}-\sigma$ relations determined from X-ray selected clusters at low redshift.  The X-ray determined mass estimates are generally consistent with those derived from weak lensing and spectroscopic analyses.  These preliminary results suggest that the novel method of optical selection used to construct the EDisCS catalog may, like selection by X-ray luminosity, be well-suited for identification of relaxed, high-redshift clusters whose intracluster medium is in place and stable by $z\sim0.8$.  
 
\end{abstract}


\section{Introduction}

Surveys of distant galaxy clusters provide vital information about the development of the Universe on many scales.  The evolution of the cluster mass function with redshift is among the key testable predictions of hierarchical cosmologies, while the properties of the intracluster medium (ICM) reflect both the physics of baryonic structure formation and the effects of feedback from evolving galaxy populations.  The galaxies within dense environments comprise convenient and unique samples in which to study galaxy formation and evolution as well as the dependence of these processes on environment.  Additionally, the properties and demographics of active galactic nuclei (AGN) hosted by cluster members may provide insight into the triggering mechanisms of nuclear activity and the role of AGN feedback in the development of host galaxies and the ICM (e.g., Martini et al., 2006; Johnson, Best \& Almaini, 2003).  In recent years, therefore, considerable effort has been devoted to the construction of catalogues of high-redshift clusters which will put statistically significant constraints on the evolution of clusters and cluster galaxy populations from early epochs.  

Clusters are commonly identified either through optical/IR searches for galaxy overdensities or through X-ray detection of the intracluster gas.  Optical/IR studies have the advantage of efficiently surveying very large volumes of sky, but have traditionally been marred by a high rate of spurious identifications of line-of-sight superpositions.  While modern surveys utilising optimal colour filters (e.g., Postman et al., 1996; Gladders \& Yee, 2005) can substantially mitigate this shortcoming, it is increasingly problematic at high redshift where the colour dinstinction between member and non-member galaxies decreases.  In contrast, X-ray observations allow unambiguous identification of clusters over well-defined survey areas as well as direct measurement of the properties of the ICM (e.g., Gioia et al., 1990; Rosati et al., 1998).  Recent observations with the Chandra and XMM-Newton observatories have detected clusters to $z\sim1.4$ (e.g., Mullis et al., 2005), and surveys currently underway with these instruments promise systematic exploration of the $z>1$ cluster population (e.g., Romer et al., 2001; Pierre et al., 2003; Green et al., 2004).  However, selection based on detection of a luminous ICM may significantly bias these samples toward inclusion of only the most massive, gas-rich, virialised clusters and/or systems in which shocks elevate the X-ray temperature and flux.  

A critical issue in the interpretation of ongoing cluster studies, then, is the quantification of the biases inherent in the various cluster selection techniques and their dependence on both detection wavelength and redshift.  Several authors have recently addressed the question of whether X-ray and optical detection techniques are equal probes of the cluster population (e.g., Donahue et al., 2001, 2002), generally concluding that they are not.  At lower redshift, it appears that optical surveys may be able to detect less massive and/or less evolved systems.  For example, Plionis et al. (2005) and Basilakos et al. (2004) investigate the low X-ray luminosity of many optically-selected cluster candidates in Sloan Digital Sky Survey data and conclude the optical candidates are poor groups and/or structures which have not yet virialised.  Toward higher redshift, however,  there is evidence that  the properties of massive, optically-selected clusters may differ systematically from those of X-ray-selected clusters of comparable mass.  Several ROSAT observations of optically-selected samples of clusters at $z\sim0.5$ suggest they have consistently low X-ray luminosities for their apparent optical richness compared to X-ray-selected samples (e.g., Castander et al., 1994; Bower et al., 1994; Holden et al. 1997).  Lubin et al. (2004) have obtained Chandra observations of optically-selected clusters at $z=0.76$ and $z=0.90$ and report that both exhibit X-ray luminosity and temperature which are low for their measured velocity dispersion according to the low-redshift $L_{X}-\sigma$ and $T-\sigma$ scaling relations.  Similarly, a Chandra survey of clusters at $0.6 < z < 1.0$ selected by the Red-Sequence Cluster Survey (RCS; Gladders \& Yee, 2005) found them to be cooler, less luminous, and less centrally condensed than X-ray selected clusters of similar optical richness (Hicks et al. 2004).

The physical implications of the variation in the X-ray properties of optically- and X-ray-selected high-redshift cluster samples may be significant, but are currently poorly understood.  If the optically-selected sources are representative of the high-redshift cluster population in general, their deviation from the X-ray-optical scaling relations would indicate significant evolution of the ICM since $z\sim0.8$.  That this evolution is not apparent in X-ray-selected samples might suggest that X-ray luminosity is indeed a biased and limited indicator of the cluster population at high redshift.  On the other hand, as suggested by Bower et al. (1997) and Lubin et al. (2004), optical cluster surveys at high redshift may, like those at lower redshift, simply be able to select less fully evolved systems.  At redshifts where hierarchical structure formation models suggest clusters should still be in the process of collapsing along filaments, significant populations of massive proto-cluster systems may exist which have not yet developed a luminous ICM. The masses implied by the velocity dispersions in such collapsing systems may be inflated by non-equilibrium dynamics, while the properties of the ICM, particularly in lower mass structures, may be significantly affected by non-gravitational heating by stellar winds, cluster AGN, and/or substructure merging.  If the optically-selected cluster surveys at high redshift have preferentially selected these forming systems, their deviation from the $L_{x}-\sigma$ and $T-\sigma$ relations would not necessarily imply evolution in the ICM properties of the general cluster population or indicate a strong bias in X-ray-selected samples of virialised clusters.     

\begin{table*}
\begin{minipage}{\textwidth}
\caption{Properties of EDisCS clusters observed with XMM-Newton.  Listed for each cluster are: position of the brightest cluster galaxy, cluster redshift, number of spectroscopically confirmed members, spectroscopically-determined velocity dispersion, and estimated cluster mass.  Mass estimates are derived from the weak lensing analysis of Clowe et al. (2006) and quoted at a radius of 1 Mpc.  No significant peak was detecting in the weak lensing mass reconstruction map for  for Cl1040-1155; the mass estimate quoted for this cluster is a 2$\sigma$ upper limit.  All other values are from Halliday et al. (2004).  \label{tab:clust_prop}} 
\begin{center}
\begin{tabular}{ccccccc}
\hline
Cluster	&RA	&Dec.	&Redshift	 &$N$	&$\sigma$	&$M_{lens}$\\
	&(J2000)	&(J2000)	&	&	&(km s$^{-1}$)		&($10^{14} M_{\odot}$)\\
\hline 
\hline
Cl1216-1201 &12 16 45.1	&-12 01 18	&0.79	&66	&$1018^{+73}_{-77}$	&6.2\\
Cl1054-1145 &10 54 24.5	&-11 46 20	&0.70 	&48 	&$589^{+78}_{-70}$		&3.7\\
Cl1040-1155 &10 40 40.4	&-11 56 04	&0.70	&30	&$418^{+55}_{-46}$		&$<2.2$\\
\hline
\end{tabular}
\end{center}
\end{minipage}
\end{table*}

\begin{table*}
\begin{minipage}{\textwidth}
 \caption{Properties of XMM observations of three EDisCS clusters. Listed for each cluster are: observation date, observation ID, position of XMM aim-point, scheduled exposure time, 7--15 keV background rate above which flares were filtered for pn and MOS, and remaining usable exposure time for pn and MOS.  \label{tab:xray_dat}} 
 \begin{center}
\begin{tabular}{cccccccccc}
\hline
Cluster	   	&Date			&Obs. ID		&R.A.	&Dec.	&Exp. (ks) 	&\multicolumn{2}{c}{Bkg. rate (cts s$^{-1}$)}		&\multicolumn{2}{c}{\lq On\rq\, time (ks)}\\
			&				&			&(J2000)	&(J2000)	&		&MOS	&pn				&MOS		&pn\\
\hline 
\hline
Cl1216-1201 	&06 Jul 2003		&0143210801	 &12 16 45.0	&-12 01 17	&33.9		&0.8		&3.0				&27.7 		&25.5\\
Cl1054-1145 	&29-30 Dec 2004	&0201330201	&10 54 21.2	&-11 46 18	&34.6		&0.8		&2.4				&31.6 		&27.3\\
Cl1040-1155 	&19 May 2005		&0201330101	&10 40 41.6	&-11 55 51	&33.9		&0.6		&2.2				&31.6		&25.5\\
\hline
\end{tabular}
\end{center}
\end{minipage}
\end{table*}

Developing a physical understanding of the variation between X-ray- and optically-selected high-redshift cluster samples is clearly vital to knowing how well these survey techniques reflect the underlying mass distribution, and may also provide valuable information about the mechanisms and timescales of mass congregation and ICM heating in forming structures.  Unfortunately, however, there are currently very few high-redshift clusters which have been extensively studied at both optical and X-ray wavelengths.  In addition, the vast majority of detailed X-ray measurements of high-redshift clusters to date have been of massive X-ray-selected systems which will evolve to populate the highest mass tip of the present-day cluster population.  To address these issues, we are conducting an X-ray survey of clusters drawn from the high-redshift sample of the ESO Distant Cluster Survey (EDisCS; White et al., 2005), a uniquely well-studied set of optically-selected $z>0.6$ clusters.  As EDisCS clusters were first identified in the Las Campanas Distant Cluster Survey (LCDCS; Gonzales et al., 2001) on the basis of optical intracluster light within a compact spatial filter, the survey is tuned to detection of dynamically relaxed systems independent of their X-ray flux.  The mass and dynamical state of structures in the EDisCS fields are known from well-determined velocity dispersions (Halliday et al., 2004; Milvang-Jensen et al. in preparation) as well as weak-lensing mass reconstructions (Clowe et al., 2006) for each cluster.  Finally, the mass distribution of the EDisCS sample differs significantly from that of existing X-ray-selected samples, extending to much lower masses and, when evolved to $z=0$, matching the distribution of nearby clusters significantly better (Poggianti et al., 2006).  

In this paper, we present XMM-Newton observations of the first three EDisCS clusters.  We describe the EDisCS programme and the target clusters in \S\ref{ediscs} and the reduction and analysis of the X-ray data in \S\ref{xray}.  In \S\ref{results}, we report on the X-ray properties of each cluster in turn before discussing the ensemble results in \S\ref{scaling} and \S\ref{discussion} and summarising our findings in \S\ref{summary}.  Throughout the paper we assume a $\Lambda$CDM cosmology with $\Omega_m$=0.3, $\Omega_{\Lambda}=0.7$, and $H_0=70$ km s$^{-1}$ kpc$^{-2}$.  Unless specified otherwise, errors throughout are quoted at the 90\% confidence level.  

\section{The ESO Distant Cluster Survey}
\label{ediscs}

The principal goal of EDisCS is the quantitative analysis of the evolution of cluster galaxy populations since $z\sim0.9$.  Targets were drawn from the LCDCS (Gonzales et al., 2001), in which cluster candidates were detected as low surface brightness peaks in source-subtracted drift-scan observations of a $90^{\circ}$ by 1.5$^{\circ}$ strip of sky and assigned estimated redshifts from the apparent magnitude of the brightest cluster galaxy (BCG).   The EDisCS \lq mid-$z$\rq\, and  \lq hi-$z$\rq\, samples are comprised of ten bright LCDCS candidates at $z\sim0.5$ and $z\sim0.8$, respectively, for which VLT imaging could confirm the presence of an apparent cluster with a possible red sequence (Gonzalez et al., 2002).  For these clusters, we have obtained a high-quality, homogenous dataset of deep optical and near-IR photometry, deep optical spectroscopy, wide-field imaging, and high-resolution Hubble Space Telescope Advanced Camera for Surveys (HST/ACS) imaging (see White et al., 2005, and references therein for a full description of the EDisCS data).  Multi-object spectroscopic observations have identified between 10 and 70 members in each cluster, yielding robust measurements of cluster velocity dispersions as well as indications of cluster substructure (Halliday et al., 2004; Milvang-Jensen et al., in preparation).  In addition, photometric redshifts based on both optical and IR colours are available across the cluster fields (Pell{\'o} et al., in preparation) and a weak-shear lensing analysis of the photometric data has provided mass reconstructions over the $\sim 6.5\arcmin\times6.5\arcmin$ VLT/FORS2 FOV (Clowe et al., 2006).   

We plan a detailed X-ray survey of nine EDisCS cluster fields with a redshift range of $0.6\lesssim z \lesssim0.8$ (median $z = 0.7$).  Our X-ray sample is comprised of the eight of ten \lq hi-$z$\rq\ clusters and one \lq mid-$z$\rq\ cluster with spectroscopically-confirmed $z\gtrsim 0.6$.  The number of spectroscopically confirmed members, $N$, in the target clusters range from 14 to 66 (median $N=22$) and imply a range in velocity dispersion of $320$ km s$^{-1} \lesssim \sigma \lesssim1020$ km s$^{-1}$ (median $\sigma\simeq570$ km s$^{-1}$).  Four additional $z\gtrsim0.6$ structures will be serendipitously observed in these fields, extending the sample to higher redshift ($z_{max}\simeq0.96$) and lower velocity dispersion ($\sigma_{min}=242$ km s$^{-1}$).  In this paper we present XMM observations of the first three clusters: Cl1216-1201, Cl1054-1145, and Cl1040-1155.  These fields were chosen for this initial X-ray study from among the five presented in the first EDisCS spectroscopic paper by Halliday et al. (2004) and cover a broad range in velocity dispersion.  Their optical properties are discussed in turn in \S\ref{results} and summarised in Table \ref{tab:clust_prop}.  

\section{X-ray data}
\label{xray}

We obtained XMM-Newton observations of the Cl1216-1201, Cl1054-1145, and Cl1040-1155 fields with 
scheduled exposure times of $\sim 30$ ks each.  The three detectors comprising the European Photon Imaging Camera (EPIC) --- two Metal Oxide Semiconductor (MOS) CCDs and one pn camera --- were all functioning nominally throughout these exposures, which were taken in Full Window mode through the Medium filter.  Details of the observations and their reduction are summarized in Table \ref{tab:xray_dat}.  

\subsection{Data preparation}
\label{data_preparation}

We reduced the raw data using version 6.0.0 of the XMM Science Analysis System (SAS).
We identified periods of background flaring in 7--15 keV lightcurves and removed them;   
the remaining usable exposures are listed in Table \ref{tab:xray_dat}.  
We applied the standard recommended filters, excluding events with PATTERN $>12$ or PI energies outwith the
0.2--15 keV range.  To create event lists suitable for spectral analysis, we used more conservative filters, removing
pn events with PATTERN $>4$ and all events with FLAG $\ne 0$.  We also chose these conservative filters
in pn imaging analysis as they significantly reduce visible artefacts, particularly at soft energies.

We made images for each of the EPIC cameras in the full (0.5--8 keV), soft (0.5--2 keV), 
and hard (2--8 keV) bands by binning the filtered events into 2'' pixels.   We calculated energy conversion 
factors (ECFs) for these bands from appropriate calibration data as
described in XMM Survey Science Centre memo SSC-LUX\_TN-0059, Issue
3.0, and list them in Table \ref{tab:ecfs}.  This procedure adopts a powerlaw spectrum with a photon index $\gamma=1.7$ and
an absorbing column of $3\times10^{20}$ cm$^{-2}$.  We generated exposure maps for these
images with the {\tt eexpmap} task, which assumes monochromatic spectra with energies at the mid-point of each band.

We combined EPIC images by summing the individual instrument images after setting 
all pixels with zero exposure to zero counts.  We generated merged exposure maps by 
multiplying the individual exposure maps by the instrumental effective area before summing. 
While these maps are sufficient to \lq flat-field\rq\, the counts images for the purposes of source detection and
rough spatial analysis, they are not accurate enough to provide reliable fluxes and therefore are not used in 
the derivation of source properties.

\begin{table}
\caption[XMM energy conversion factors for the 0.5 -- 8, 0.5 -- 2, and 2 -- 8 ke
V bands]{XMM energy conversion factors (ECFs) for the full, soft, and hard bands,
in units of $10^{11}$ counts cm$^{2}$ erg$^{-1}$.  These values are calculated from 
appropriate calibration data as described in the text.  \label{tab:ecfs}}
\begin{center}
\begin{tabular}{cccc}
\hline
Band (keV)      &MOS 1 ECF      &MOS 2 ECF      &pn ECF \\
\hline\hline
0.5--8.0        &1.067  &1.071  &3.396\\
0.5--2.0        &1.980  &1.977  &6.596\\
2.0--8.0        &0.510  &0.518  &1.457\\
\hline
\end{tabular}
\end{center}
\end{table}  

\subsection{Imaging analysis}
\label{imaging_analysis}

We adaptively smoothed the full-band images using the {\tt csmooth} tool available from the Chandra X-ray Centre so that only features locally significant at the 99\% level remain visible.  We identified point sources in the unsmoothed full-band images as detailed in \S\ref{point_sources} and excised all events within the 3$\sigma$ detection ellipses.  

We made radially-averaged surface brightness distributions of the diffuse emission by summing exposure-corrected counts from the source-free images in annuli.  We summed the source counts in 4\arcsec\, annuli to radii of 80\arcsec\, (598 kpc at $z=0.79$ and 571 kpc at $z=0.70$),  measured the background level between 80 and 100\arcsec\,, and subtracted it.  We modelled the distributions with a standard $\beta$-function of the form 
\begin{equation}
S(r) = S_{o} [1 + (\frac{r}{r_c})^2 ]^{-3\beta+\frac{1}{2}}, 
\end{equation}
where $S(r)$ is the surface brightness at radius $r$, $S_{o}$ is the central surface brightness, $r_{c}$ is the core radius, and $\beta$ is the slope at large radii.  We minimised the fits in the CIAO Sherpa package using the Chi Gehrels statistic (Gehrels, 1986) and estimated confidence intervals through projection of the statistic surface using the {\tt proj} task.  We note that this approach does not consider the extent of the XMM PSF.

\subsection{Spectral analysis}
\label{spectral_analysis}

As discussed at length in Lumb (2002), Read \& Ponman (2003), and Nevalainen et al. (2005), the complex spectral and spatial variation of the XMM background presents a significant challenge in analysis of extended and/or faint sources.  The background flux is comprised of an unvignetted particle component exhibiting sharp spectral and spatial stucture as well as a vignetted cosmic background which is approximately constant across the field of view.  In the current analysis, we compared the result of  background subtraction using: i) single background regions ii) multiple background regions and iii) the Read \& Ponman (2003) background maps, following the prescription of Arnaud et al. (2002).  The best fitting parameters obtained using all three methods were consistent within the uncertainties, which were substantial for method iii) due to the poor statistical quality of the spectra.  The results presented are based on the spectra constructed using multiple background regions.  

We extracted source and background spectra from point-source free event lists within optimised extraction regions estimated from the spatial analysis described above.  We created spectra and response files for each of the three EPIC cameras using the standard SAS tools, and fit them simultaneously to an absorbed MEKAL model of hot, diffuse gas emission (Mewe, Gronenschild, \& van den Oord, 1984; Mewe, Lemen, \& van den Oord, 1985; Kaastra, 1992; Liedahl, Osterheld, \& Goldstein, 1995) within the XSPEC package.  In fitting the spectra we excluded the 1.45--1.55 keV band to avoid highly variable Al K florescence associated with the detectors and detector housings, as suggested in Nevalainen et al. (2005), and we restricted the fit to energies above 0.5 and below 8 keV.  Prior to fitting, we grouped the spectra such that there are a minimum of 25 counts in each bin to allow use of the $\chi$-squared statistic.  We again estimated confidence levels from the statistic surface.   

 \subsection{Detection and characterisation of point-sources}
\label{point_sources}

We detected X-ray sources in the 0.5--8 keV full-EPIC images using versions of the \\ {\tt wtransform} and {\tt wrecon} algorithms distributed by the Chandra X-ray Center (Freeman et al., 2002), using wavelet scales of $\sqrt{2}$, $2\sqrt{2}$, $4$, $4\sqrt{2}$, $8$, $8\sqrt{2}$, and $16$ pixels.  We set the significance threshold conservatively at $1\times 10^{-8}$, such that less than one false source is expected per field.   We further limited the final sourcelists to include only sources with $> 10$ counts and a source significance value of $\sigma_{src}>4$, where $\sigma_{src}$ is defined in terms of source counts, $C_{src}$, and background counts, $C_{bkg}$, as $\frac {C_{src}} {1+\sqrt{0.75+C_{bkg}}}$. 

We determined source counts, background counts and mean exposure values for each detector and each band within apertures with 8\arcsec\, radii.  We obtained background counts from the output {\tt wtransform} background image and converted net counts within the aperture to total net counts by dividing by the fractional encircled energy (EE) within the aperture.  The EE is a function of off-axis angle and was determined using the Extended Accuracy Model of the XMM PSF.  We derived instrumental fluxes by dividing the total counts by the mean exposure value and the appropriate ECF.   Fluxes quoted in the paper are an error-weighted average of the fluxes meaured by each instrument.  Hardness ratios are defined as $HR=\frac{H-S}{H+S}$ where H and S are the 0.5--2 and 2--8 keV error-weighted count-rates given by each instrument, after background subtraction and exposure correction.  The on-axis, full-band flux limit for $3\sigma$ detection of point sources was $\sim1\times10^{-15}$ erg s$^{-1}$ cm$^{-2}$ for all three observations.  Catalogues of the X-ray point sources in each field are given in Appendix A and in machine-readable form at the Centre de DonnŽes astronomiques de Strasbourg (CDS).  

\subsection{Astrometric calibration and optical counterparts}
\label{optical_counterparts}

We identified optical counterparts to the X-ray point sources on astrometrically-corrected 60 minute exposures obtained with the Wide-Field Imager (WFI) at the ESO/MPG 2.2m telescope which cover the full XMM-Newton field of view (Clowe et al., in preparation).  We used the {\tt astrom} algorithm to correct the astrometry of the X-ray data by performing 6-coefficient fits to the positions of bright X-ray sources with unambiguous optical counterparts in $I$-band WFI images.  The rms of these fits were $\sim 1.5$\arcsec\, for all fields, consistent with the published positional accuracy of XMM.  We then identified candidate optical counterparts for all X-ray sources through comparison of the X-ray and $I$-band catalogues using a likelihood ratio technique similar to those employed by Sutherland \& Saunders (1992) and Rutledge et al. (2000).  For each X-ray source, we consider the catalogue of $M$ optical counterparts and define the likelihood ratio, $LR$, of the $i$th candidate counterpart as
\[LR_{i}={\frac{\exp{-r_{i}^{2}/2\sigma_{i}^{2}}}{\sigma_{i}^{2} N(<I_{i})}}\mathrm{,}\]
where $r_{i}$ is the distance between the X-ray source and the candidate, $\sigma_{i}$ is the uncertainty in the source position, and $N(<I_{i})$ is the number of optical sources with $I$-band magnitude less than or equal to that of the candidate.  To estimate the reliability of source association as a function of $LR$, we randomly shift the optical catalogue and recalculate the likelihoods, repeating the procedure 5000 times to obtain a smooth histogram of $LR$ values for false associations.  The reliability, $R$, of the $i$th optical candidate is then given by 
\[R_{i} = 1-N_{bkg} (\ge LR_{i})\mathrm{,}\] 
where $N_{bkg} (\ge LR_{i})$ is the number of false associations with $LR$ greater than or equal to that of the candidate.  We then calculate both the probability that there is no unique counterpart to the X-ray source:
\[P_{none} = \frac{\prod_{j=1}^M(1-R_j)}{S}\]
and the probability that the $i$th candidate is the true counterpart to the X-ray source:
\[ P_{id, i} = \frac{\frac{R_{i}}{1-R_{i}} \prod_{j=1}^{M} (1-R_{j})}{S}\mathrm{.}\]
The normalization, $S$, is found by summing over the full set of possibilities:
 \[S =  \sum_{i=1}^{M}\frac{R_{i}}{1-R_{i}} \prod_{j=1}^{M} (1-R_{j}) + \prod_{j=1}^{M} (1-R_{j})\mathrm{.}\]

Catalogues of the optical counterparts determined from the WFI data are given in Appendix B and in electronic form at the CDS.  Multi-band optical properties of the candidate optical counterparts within the VLT FOV are given in Tables \ref{tab:vlt1216}, \ref{tab:vlt1054}, and \ref{tab:vlt1040}.

\subsection{Identifying possible cluster AGN}

In support of the EDisCS science goal of examining evolution in cluster galaxy populations, we plan a survey of AGN hosted by EDisCS cluster members.  In this initial paper, we use the multi-band photometric and spectroscopic data available for X-ray sources detected in the central regions of the cluster fields to identify possible cluster AGN.  Photometric redshifts for all galaxies in the FORS fields have been determined using two independent photometric redshift codes (Bolzonella et al. 2000; Rudnick et al. 2001; Rudnick et al. 2003) and are presented in Pell{\'o} et al. (in preparation).  Each code returns its own probability histogram, $P(z)$, which is normalised to unity over $0<z<2$.  The probability of cluster membership, $P_{mem}$, is then defined as the integrated probability that $z_{phot} = z_{cluster} \pm 0.1$.  If both codes return an integrated probability greater than a threshold value determined through analysis of the spectroscopic identifications in each field, galaxies are flagged as possible cluster members (see Pell{\'o} et al., in preparation, for details).  In addtion, the Bolzonella code also returns a best-fitting redshift using a quasar template, $z_{QSO}$.  We present X-ray sources with optical properties plausibly consistent with cluster membership in \S\ref{agn1216}, \S\ref{1054agn}, and \S\ref{agn1040} below, but caution that these candidates for further study have been selected liberally.   

\section{Analysis of individual EDisCS fields}
\label{results}
\subsection{Cl1216-1201}
\label{1216res}

\begin{figure*}
\begin{minipage}{\textwidth}
\caption{X-ray contours overlaid on an $I$-band image of the central region of the Cl1216-1201 field. 
Small exes indicate sources for which spectra have been obtained; squares indicate spectroscopically confirmed cluster members.  North is up, East is to the left, and the XMM aim point is indicated by the cross.
The diffuse sources to the North are the main Cl1216-1201 component (East) and contaminating point source (West); the fainter source to the SSW is Cl1216-South.  The edge of the field of view of the spectroscopic observations lies just to the North of the southern component. 
\label{fig:1216_contour}}
\begin{center}
\includegraphics[width=0.35\textwidth]{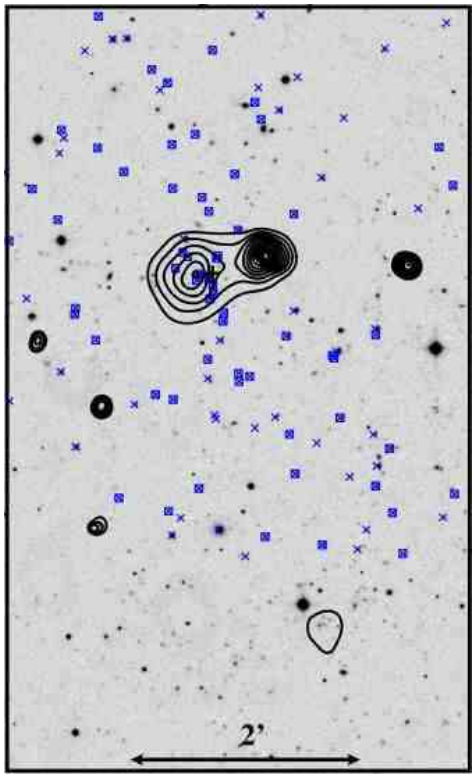}
\end{center}
\end{minipage}
\end{figure*}  

\begin{figure*}
\begin{minipage}{\textwidth}
\caption{Adaptively smoothed 0.5--8 keV X-ray image of the central region of the Cl1216-1201 field with the Clowe et al. (2006) weak lensing mass reconstruction of the field overlaid.  Contour intervals are $\sim 1\times10^8$M$_\odot$ kpc$^{-2}$ relative to the local mean, with negative contours shown in light grey.  Crosses indicate sources for which spectra have been obtained; squares indicate spectroscopically confirmed cluster members.  North is up and East is to the left.  The edge of the lensing map lies just to the North of the Southern component.
\label{fig:1216_mass}}
\begin{center}
\includegraphics[width=0.35\textwidth]{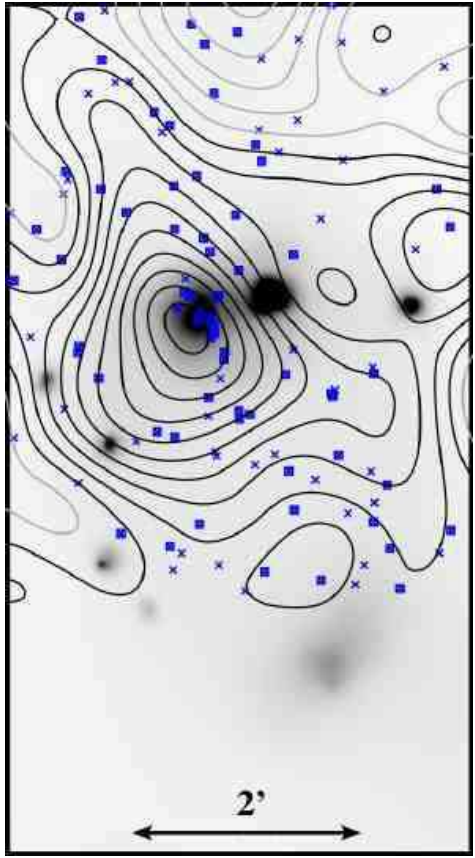}
\end{center}
\end{minipage}
\end{figure*}  

Cl1216-1201 at $z=0.79$ is among the richest and highest redshift EDisCS clusters, with 66 spectroscopically confirmed cluster members yielding a velocity dispersion of 1018$^{+73}_{-77}$ km s$^{-1}$ (Halliday et al., 2004).  The photometric data indicate a strong red sequence and a well-defined sequence of bluer, presumably star-forming, galaxies (White et al., 2005).  The mass reconstruction of the field indicates a mass peak coincident with both the BCG and the peak in galaxy density ((Clowe et al., 2006)), and there is also a strongly lensed arc associated with the cluster.  Examination of the three-dimensional distribution of the cluster galaxies reveals substructures, with one component centred on the BCG and a second extending to the NE (Halliday et al., 2004).  Corresponding filamentary structures are evident at moderate significance levels in the lensing map.  

\subsubsection{Cl1216-1201: Spatial analysis}

We show X-ray contours overlaid on an $I$-band image of the central region of the Cl1216-1201 field  in Figure \ref{fig:1216_contour}.  An extended source (Cl1216\_X46) is evident at the XMM aim point, coincident with a dense concentration of spectroscopically-confirmed cluster members which includes the BCG.  A second diffuse component (Cl1216\_X62), hereafter Cl1216-South, lies roughly 3.5\arcmin\, to the SSW of the XMM aimpoint and is significantly fainter and less centrally concentrated than the emission at the aim point (see \S\ref{1216south}).  The bright compact source (source Cl1216\_X9) embedded in the main cluster emission $\sim40$\arcsec\, to the West is coincident with a large, bright ($I$=18.2) foreground elliptical with a photometric redshift of $\sim 0.6$.  Where the Cl1216-1201 contours are unaffected by the point source flux, they appear regular and round, suggesting the main component of the cluster gas is reasonably relaxed despite the substructure detected in the galaxy distribution and mass map.  Figure \ref{fig:1216_mass} shows the smoothed X-ray flux in greyscale with contours from the weak lensing mass reconstruction of the field by Clowe et al. (2006) overlaid.  The weak lensing peak is offset slightly to the East of the X-ray peak and the BCG, and the contours extend along a NNE-SSW axis.  The field of view of the lensing observations does not extend far enough to the South to include Cl1216-South. 

\begin{figure*}
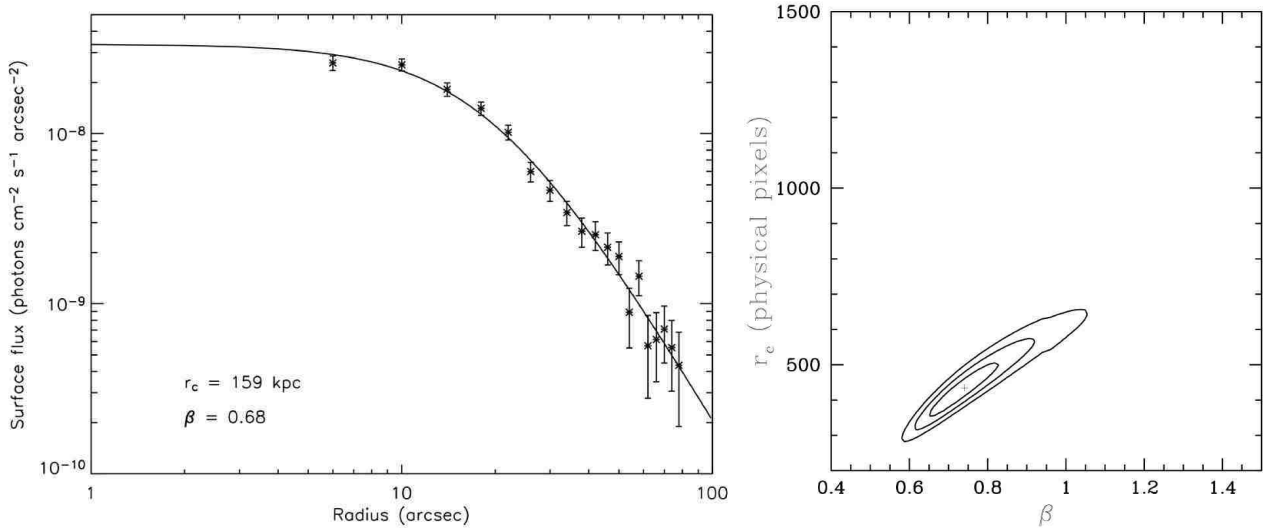

\begin{minipage}{\textwidth}
\caption{Left: Radial surface flux profile for Cl1216-1201 obtained by masking point sources and summing exposure-corrected, background-subtracted counts within annuli.  Imperfect subtraction of a contaminating point source may be evident beyond 40\arcsec.  \label{fig:1216_rprof}  Right: Corresponding constraints on $\beta$ and $r_c$.  The $r_c$ axis is in XMM physical pixels which are 0.05\arcsec\, on a side.  Contours mark 1, 2, and $3\sigma$ levels.  \label{fig:1216_regproj} }
\begin{center}
\hspace{-1.0cm}
\includegraphics[width=0.56\textwidth, angle=0]{cl1216_1d.epsf}
\includegraphics[width=0.43\textwidth, angle=0]{cl1216_regproj.epsf}
\end{center}
\end{minipage}
\end{figure*}

A radially averaged surface flux profile of the main cluster component centred at the peak of the X-ray emission ($\alpha=12^{\mathrm h}16^{\mathrm m}45^{\mathrm s}.4$, $\delta = -12^{\circ}$01\arcmin15\arcsec [J2000]),  is shown in Figure \ref{fig:1216_rprof} (left). The best-fitting $\beta$-model parameters were $r_c =159\pm^{20}_{18}$ kpc and $\beta = 0.68\pm^{0.06}_{0.05}$.  We note that these errors, obtained by projection of the $\chi^2$ surface along the parameter axis, underestimate the true uncertainty in $\beta$ and $r_c$ due to the correlation of these parameters, as shown in Figure \ref{fig:1216_regproj} (right).  Additionally, an upward kink in the data at $\sim 40$\arcsec\, suggests residual wings of the nearby point source may contaminate the cluster emission beyond this radius.  

As there were sufficient counts, we also modelled the surface brightness distribution in two dimensions.  Using the CIAO Sherpa package, we fit the full band image of the cluster emission with a model comprised of a two-dimensional $\beta$-model --- which includes eccentricity ($e$) and position angle ($\theta$) terms to allow for non-sphericity --- multiplied by the exposure map and convolved with a model of the XMM PSF.   We used the SAS {\tt calview} tool to generate a model of the PSF at 2 keV.  (The XMM PSF is not strongly energy-dependent at energies less than 5 keV and off-axis angles less than 4\arcmin).  We selected the Medium Accuracy model, which is the only two-dimensional calibration of the XMM PSF currently available, but note that it is known to provide a less accurate description of the radially-averaged PSF than the Extended Accuracy model (see XMM-Newton Calibration Access and Data Handbook, memo XMM-PS-GM\_20 Issue 2.1).  We rebinned the images, exposure maps, and model PSF into a common 2.2\arcsec\, pixel scale and normalised exposure maps to their maximum value to retain the statistics of the counts image as much as possible.  

The bright point source within the fit region was represented as a $\delta$-function and added to the model prior to the addition of instrumental effects.  We also included a constant background level, which we determined from a fit to a source-free region and fixed.  Neither the source-free background nor the exposure map, excluding chip gaps, varied significantly over the fitted region.   The low count-rate per bin required minimisation via the Cash statistic (Cash, 1979), and we computed confidence intervals from the statistic surface using the region projection task ({\tt rproj}) within Sherpa.  The best-fitting model was spherical ($e=\theta=0$) with $\beta=1.2\pm^{0.5}_{0.3}$ and $r_c=97\pm^{28}_{18}$ kpc.  We show the radially-averaged profiles of the raw counts image, the best-fitting model, and the model components in Figure \ref{fig:1216_2d}. 

The deviation of the model from the data near to the point source peak is likely to result from differences between the Medium Accuracy model XMM PSF and the observed profile of the point source.  As the source is nearly on-axis and the PSF should be approximately radially symmetric, we attempted to obtain a better fit using the radially-averaged High and Extended Accuracy models.  None of the model PSFs was found to sufficiently fit the peak of the point source, but choice of PSF model did significantly alter the best-fitting parameters of the cluster profile.  This uncertainty precludes rigourous constraint of the cluster properties from the two-dimensional fits; the best-fitting $\beta$ and $r_{c}$ parameters from these fits cannot be used to accurately model the cluster.  

The two-dimensional model is useful, however, in examining the extent of point source contamination in the one-dimensional fit.  Regardless of which PSF model is used, the point source wings remain above the background level to radii of nearly 25\arcsec.  By comparison, the region excluded by the $3\sigma$ ellipse produced by {\tt wtransform} for this source has a radius of $\sim30$\arcsec.  While the outmost bins beyond $\sim70\arcsec$ in Figure \ref{fig:1216_2d} indicate some fraction of emission in the point-source wings may be mistakenly ascribed to the cluster in a one-dimensional fit, it is clear that the majority of the point source emission is successfully removed within the 3$\sigma$ ellipse.  In addition, the two-dimensional fit demonstrates that the cluster emission is roughly spherical.  The one-dimensional fit should therefore provide a good estimate of the spatial distribution of the cluster emission. 

\subsubsection{Cl1216-1201: Spectral analysis}

\begin{figure*}
\begin{minipage}{\textwidth}
\caption{Radial surface brightness profile for Cl1216-1201 obtained through two-dimensional modelling of the uncorrected counts image.  The best-fitting model (solid curve) is comprised of cluster emission (dashed curve), a nearby point source (dot-dash curve), and a constant background level (dotted curve).  The grey curves indicate the best-fitting relation from the one-dimensional modelling (see Figure \ref{fig:1216_rprof}) and relations at the extreme of the 3$\sigma$ errors on $r_c$ and $\beta$ as indicated in Figure \ref{fig:1216_regproj}.  The vertical dotted lines mark the region around the point source which was excluded from the one-dimensional profile. \label{fig:1216_2d}}
\begin{center}
\hspace{-0.5cm}
\includegraphics[width=0.60\textwidth, angle=0]{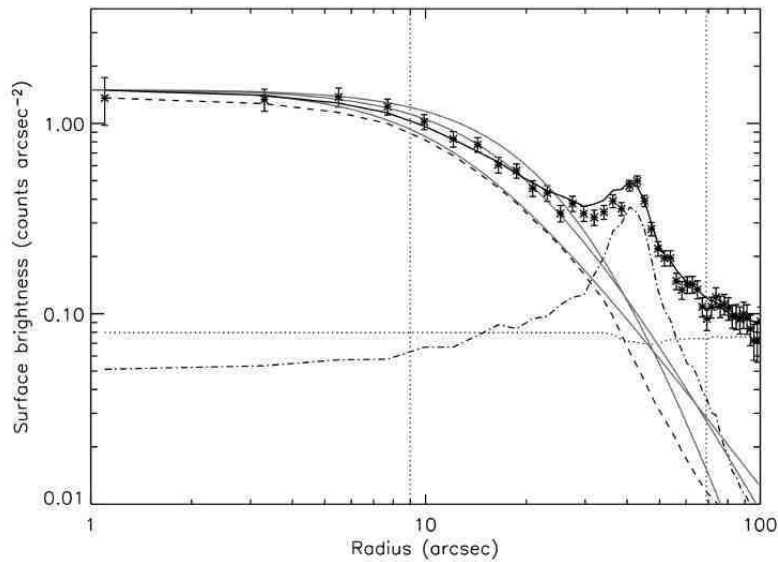}
\end{center}
\end{minipage}
\end{figure*}  

\begin{figure*}
\begin{minipage}{\textwidth}
\caption{X-ray spectra of Cl1216-1201 extracted from the MOS (lower points) and pn (upper points) cameras and best-fitting thermal emission model (histograms).  Residuals are shown in the lower panel. \label{fig:1216spec}}
\begin{center}
\hspace{-0.5cm}
\includegraphics[width=0.60\textwidth, angle=0]{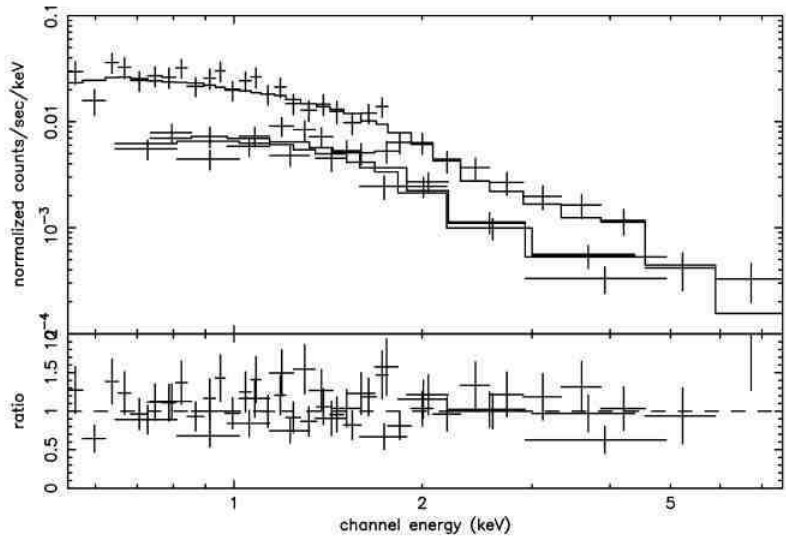}
\end{center}
\end{minipage}
\end{figure*}  

We extracted spectra of the cluster emission for the three EPIC cameras within circles centred at the XMM aimpoint with radii of 40\arcsec, as the imaging analysis indicates this region is clearly dominated by cluster emission.  Background counts were measured in five circular regions with radii of 115\arcsec\, centred within 7\arcmin\, of the aim-point and avoiding Cl1216-South.  The source aperture yielded 1503 counts where 194 are expected from the background.   Based on the properties of the bright point sources and the Medium Accuracy model of the XMM PSF, we estimate that $<5\%$ of the extracted counts could be contamination from the wings of the bright neighbouring source.  As we have extracted the spectrum only where the cluster emission appears clearly dominant (see Figure \ref{fig:1216_2d}), we expect the contamination to be negligible.  Use of a less conservative extraction region with a radius of 115\arcsec\, was also examined.  The spectrum from the larger extraction region yielded a consistent estimation of the cluster temperature, but exhibited more structure due to background emission and indicated an aperture-corrected luminosity which was $\sim20\%$ higher than that quoted below, likely due to the incomplete removal of the point source wings.

The spectra were fit to an absorbed MEKAL model with the absorption fixed at the Galactic value of $4.1\times10^{20}$ cm$^{-2}$ and the metallicity fixed at 0.3 solar.  The best-fitting model (reduced $\chi^2=1.0$) is plotted over the data in Figure \ref{fig:1216spec} and yields a temperature of $4.8\pm^{0.8}_{0.7}$ keV, a full-band luminosity of $L_{\mathrm{0.5-8 keV}}=2.5\pm{0.2}\times10^{44}$ erg s$^{-1}$, and a bolometric luminosity of $L_{bol}=3.0\pm{0.2}\times10^{44}$ erg s$^{-1}$ within the extraction radius.   If the emission follows the one-dimensional $\beta$ profile derived above, the spectral radius encloses 57\% of the total cluster luminosity, which extrapolates to $\sim5\times10^{44}$ erg s$^{-1}$.  

\subsubsection{Cl1216-1201: Mass estimate}

Under the assumptions that a cluster is spherically symmetric, in hydrostatic equilibrium, and is well described by an isothermal $\beta$ profile, and adopting a mean molecular mass of 0.59 times the proton mass, the total gravitating mass of the cluster at a given radius, $M(r)$, can be expressed in terms of the temperature of the gas, $T$, and the core radius, $r_{c}$, and slope, $\beta$ of the surface brightness profile:

\begin{equation}
 M(r)= 1.13\times 10^{14} \beta ~T ~r~ \frac{(r/r_c)^2}{1 + (r/r_c)^2},
 \end{equation}
 
\noindent where $T$ is in keV, $r$ and $r_c$ are in Mpc, and $M$ is in $M_\odot$ (Evrard et al., 1996).  The best-fit values derived from the spectral fit and the best-fitting one- and two-dimensional surface brightness models give masses within 1 Mpc of $\sim4\times10^{14}~M_{\odot}$ and $\sim7\times10^{14}~M_{\odot}$, respectively.  The range of these values illustrates the scale of the uncertainty in the X-ray derived mass estimate.  In comparison, assuming the mass distribution is reasonably approximated by a singular isothermal sphere --- i.e., $M(r)=2\sigma_{v}^2r/G$, where $\sigma_v$ is the  velocity dispersion in km s$^{-1}$ --- the weak lensing analysis of Clowe et al. (2006) implies a mass within 1 Mpc of $(5.4\pm1.2)\times10^{14}~M_{\odot}$, where the stated errors are 1$\sigma$.  The X-ray and weak lensing mass estimates are therefore broadly consistent.  A further comparison can be made with the system mass estimated from the velocity dispersion of cluster members, as in Finn et al. (2005):
\begin{equation}
M_{sys} = 1.2\times10^{15} (\frac{\sigma}{1000 \mathrm {km\, s^{-1}}} )^{3} \frac{1}{\sqrt{\Omega_\Lambda + \Omega_0 (1+z)^3}} \,	h^{-1} M_{\odot}
\end{equation}
In our assumed cosmology, the velocity dispersion of Cl1216-1201 implies a mass of $\sim (1.2\pm0.3)\times10^{15} M_{\odot}$ which is somewhat higher than the X-ray and weak lensing estimates.  We note, however, that the spectroscopic estimate is derived from spectra obtained across the FORS FOV and therefore not directly comparable to the estimates derived within fixed radii.

\subsubsection{Cl1216-1201: Cluster AGN}
\label{agn1216}

\begin{figure*}
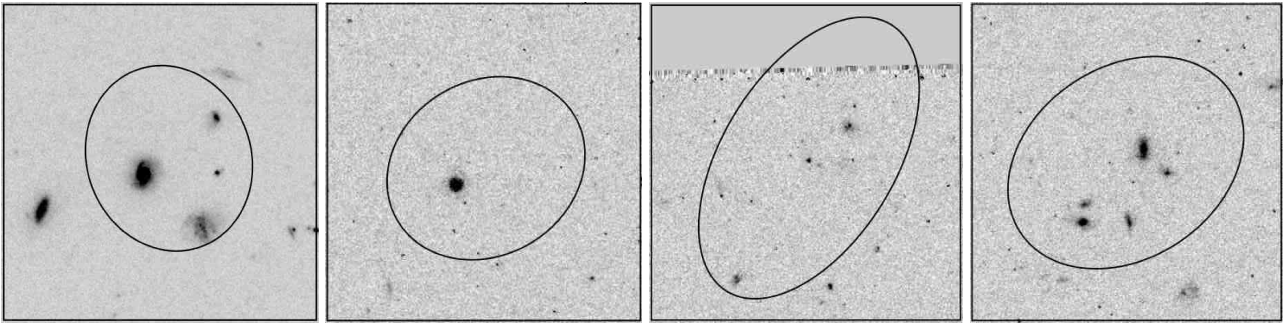

\begin{minipage}{\textwidth}
\caption{Candidate cluster AGN in Cl1216-1201.  From left: Cl1216\_X6, Cl1216\_X8, Cl1216\_X26, and Cl1216\_X59.  Boxes are 15\arcsec on a side, North is up, and East is to the left.  Ellipses are the 3$\sigma$ uncertainty of the XMM position.    \label{fig:1216agn}}
\hspace{-0.5cm}
\begin{center}

\includegraphics[width=0.24\textwidth, angle=90]{X6_bw.epsf}
\includegraphics[width=0.24\textwidth, angle=90]{X8_bw.epsf}
\includegraphics[width=0.24\textwidth, angle=90]{X26_bw.epsf}
\includegraphics[width=0.24\textwidth, angle=90]{X59_bw.epsf}
\end{center}
\end{minipage}
\end{figure*}

75 X-ray point sources are identified in the Cl1216-1201 field, of which 12 lie within the field of view of the deep, multi-band data.  The multi-band optical properties of these sources and the redshift estimates and probabilities of cluster membership returned by the photometric redshift codes are summarised in Table \ref{tab:vlt1216}.  None of the X-ray sources is a spectroscopically confirmed cluster member, but the four sources shown in Figure \ref{fig:1216agn} have optical properties which suggest they may be AGN hosted by cluster members.

\begin{itemize}

\item {Source Cl11216\_X6} is firmly associated ($P_{id}=92\%$) with a spiral galaxy $1\arcmin30\arcsec$ from the cluster core which contains a very bright central point source.  The source is not flagged as a possible cluster member based on photometric properties, but has a best-fitting redshift when fitted with a QSO template of $z_{QSO}=0.82$.  The X-ray source has a full-band flux of $S_{\mathrm{0.5-8 keV}}=1.6\times10^{-14}$ and $HR=-0.12$, consistent with a moderately obscured AGN.  

\item Similarly, {source Cl1216\_X8} is unambiguously associated with a galaxy found $3\arcmin30\arcsec$ from the cluster centre containing a very bright AGN which is not flagged as a possible cluster member but has $z_{QSO}=0.72$.   However, as the source lies outwith the FOV of the near-IR observations, the photometric redshift is based on only optical bands and is considered less reliable.  The X-ray source has a full-band flux of $S_{\mathrm{0.5-8 keV}}=2.1\times10^{-14}$ and $HR=-0.50$, consistent with an unobscured AGN.  

\item {Source Cl1216\_26} is ambiguously associated ($P_{id}=65\%$) with a galaxy lying $4\arcmin$ from the cluster centre which is flagged as a possible cluster member based on its photometric properties.  However, like Cl1216\_8, this source was not observed in the near-IR and its photometric redshift is based on only optical bands.  As both photometric redshift codes prefer $z\sim1.5$ solutions for this source and its optical counterpart is very faint, this source is most likely at a higher redshift than Cl1216.

\item {Source Cl1216\_X59}, lies $3\arcmin20\arcsec$ to the Northeast of the cluster centre and is ambiguously identified with a group of three sources based on the WFI imaging.  The HST view of this region reveals two fainter objects also within the X-ray 3$\sigma$ error ellipse.  The brightest source is flagged as possible cluster members based on its photometric properties.  The X-ray detection has a flux $S_{\mathrm{0.5-8 keV}}=4.7\times10^{-15}$ and $HR=0.36$, consistent with a heavily obscured AGN.  

\end{itemize}

\subsubsection{Cl1216-South}
\label{1216south}

\begin{figure*}
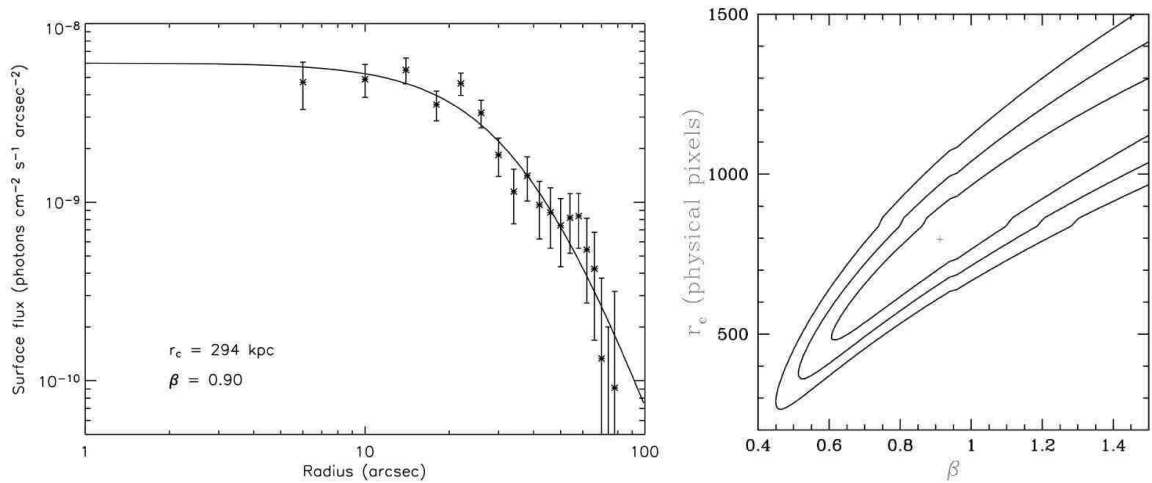

\begin{minipage}{\textwidth}
\caption{Left: Radial surface flux profile for Cl1216-South obtained by masking point sources and summing exposure-corrected, background-subtracted counts within annuli.  \label{fig:south_rprof}  Right: Corresponding constraints on $\beta$ and $r_c$.  The $r_c$ axis is in XMM physical pixels which are 0.05\arcsec\, on a side.  Contours mark 1, 2, and $3\sigma$ levels.  \label{fig:south_regproj} }
\hspace{-0.5cm}
\begin{center}
\includegraphics[width=0.50\textwidth, angle=0]{south_1d.epsf}
\includegraphics[width=0.39\textwidth, angle=0]{south_regproj.epsf}
\end{center}
\end{minipage}
\end{figure*}  

Unfortunately, the multiband imaging and FORS spectroscopy do not extend far enough to the South to confirm whether the collection of faint galaxies at the position of Cl1216-South are associated with the Cl1216-1201 structure at $z=0.79$.  The extension of the weak lensing mass-reconstruction along the NNE-SSW axis evident in Figure \ref{fig:1216_mass} is suggestive, however, as is the indication of a similar elongation in the distribution of spectroscopically selected cluster members.  While further spectroscopic observations will be necessary to determine the redshift of the Cl1216-South feature, we here include a discussion of its X-ray properties for completeness.

\noindent{\it 4.1.5.1 Cl1216-South: Spatial analysis}

A radially-averaged surface flux profile was constructed from annuli centred on the estimated centre of the Cl1216-South emission ($\alpha=12^{\mathrm h}16^{\mathrm m}39^{\mathrm s}.9$, $\delta = -12^{\circ}$04\arcmin21\arcsec [J2000]) as described in \S\ref{imaging_analysis} and is shown in Figure \ref{fig:south_rprof} (left).  The best-fitting parameters of a $\beta$-model fit are $\beta=0.90^{+0.12}_{-0.09}$ and $r_c=294^{+35}_{-35}$ kpc.  The degeneracy between $\beta$ and $r_c$ is evident in confidence contours, however, which are plotted in Figure \ref{fig:south_regproj} (right) and indicate the parameters are essentially unconstrained by this fit.

The Cl1216-South emission was modelled in two dimensions with a $\beta$-model plus a constant background level equal to that used in the fit to the main emission.  The best-fitting model is centred at $\alpha=12^{\mathrm h}16^{\mathrm m}40^{\mathrm s}.50$, $\delta = -12^{\circ}$04\arcmin25\arcsec.21 [J2000], is significantly elliptical ($e = 0.44$, $\theta = 0.83$), and yields $\beta=0.48^{+0.10}_{-0.06}$ and $r_c=79^{+81}_{-49}$ kpc.  Again, these errors do not reflect the true uncertainty in these parameters due to the degeneracy between $r_{c}$ and $\beta$.  The variation between these values and those implied by the one-dimensional fit results from the ellipticity of the Cl1216-South emission.  To test consistency of the two methods, we extracted a radially-averaged surface flux profile from an image of the best-fitting two-dimensional model in the same way we extracted the surface-flux profile from the data.  A fit to this profile yielded best-fitting parameters of $\beta=0.87$ and $r_c=271$ kpc which are entirely consistent with those found for the one-dimensional fit.   \\

\noindent {\it 4.1.5.2 Cl1216-South: Spectral analysis}

We extracted spectra of the cluster emission for the three EPIC cameras within circles centred at $\alpha=12^{\mathrm h}16^{\mathrm m}39^{\mathrm s}.90$, $\delta = -12^{\circ}$04\arcmin21\arcsec.12 [J2000] with radii of 50\arcsec.  This aperture yielded 1271 counts where 472 are expected from the background.   Use of a less conservative extraction region with a radius of 115\arcsec\, was also examined, and found to produce entirely consistent results.  

The spectra were fit to an absorbed MEKAL model with the absorption fixed at the Galactic value of $4.1\times10^{20}$ cm$^{-2}$ and the metallicity fixed at 0.3 solar.  Assuming Cl1216-South lies at $z=0.79$, the best-fitting model (reduced $\chi^2=0.9$) yields a temperature of $5.0\pm^{1.8}_{1.3}$ keV, a full-band luminosity of $L_{\mathrm{0.5-8 keV}}=1.7\pm{0.2}\times10^{44}$ erg s$^{-1}$ cm$^{-2}$, and a bolometric luminosity of $L_{bol}=2.0\pm{0.3}\times10^{44}$ erg s$^{-1}$ cm$^{-2}$ within the spectral extraction radius.   If the one-dimensional $\beta$ profile derived above correctly describes the radial shape at large radii, the spectral radius encloses 68\% of the emission and the total luminosity of the Southern component is $\sim3\times10^{44}$ erg s$^{-1}$ cm$^{-2}$.  

\begin{figure*}
\begin{minipage}{\textwidth}
\caption{X-ray spectra of Cl1216-South extracted from the MOS (lower points) and pn (upper points) cameras and best-fitting thermal emission model (histograms).  Residuals are shown in the lower panel.  \label{fig:1216_spec}}
\begin{center}
\hspace{-0.5cm}
\includegraphics[width=0.60\textwidth, angle=0]{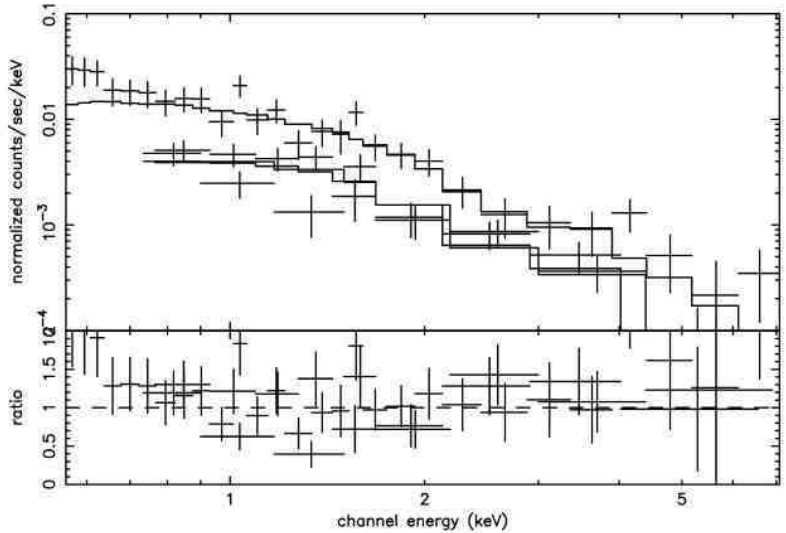}
\end{center}
\end{minipage}
\end{figure*}  

\subsection{Cl1054-1145}

\begin{figure*}
\begin{minipage}{180mm}
\caption{X-ray contours overlaid on an $I$-band image of the central region of the Cl1054-1145 field.  
Crosses indicate sources for which spectra have been obtained; squares indicate spectroscopically confirmed cluster members.  North is up and East is to the left. 
\label{fig:1054_contour}}
\begin{center}
\includegraphics[width=0.5\textwidth]{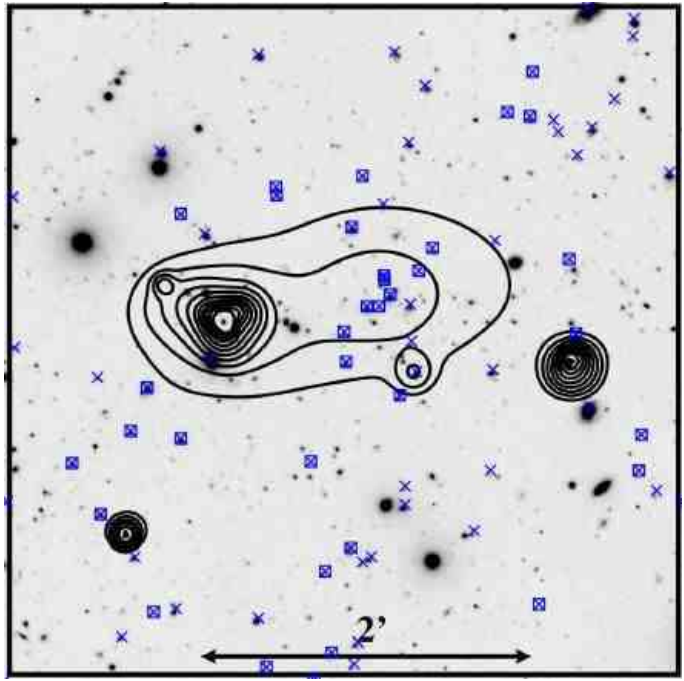}
\end{center}
\end{minipage}
\end{figure*}  

\begin{figure*}
\begin{minipage}{180mm}
\caption{Adaptively smoothed 0.5--8 keV X-ray image of the central region of the Cl1054-1145 field with the Clowe et al. (2006) weak lensing mass reconstruction of the field overlaid.  Contour intervals are $\sim 1\times10^8$M$_\odot$ kpc$^{-2}$ relative to the local mean, with negative contours shown in light grey.  Crosses indicate sources for which spectra have been obtained; squares indicate spectroscopically confirmed cluster members.  North is up and East is to the left.
\label{fig:1054_mass}}
\begin{center}
\includegraphics[width=0.5\textwidth]{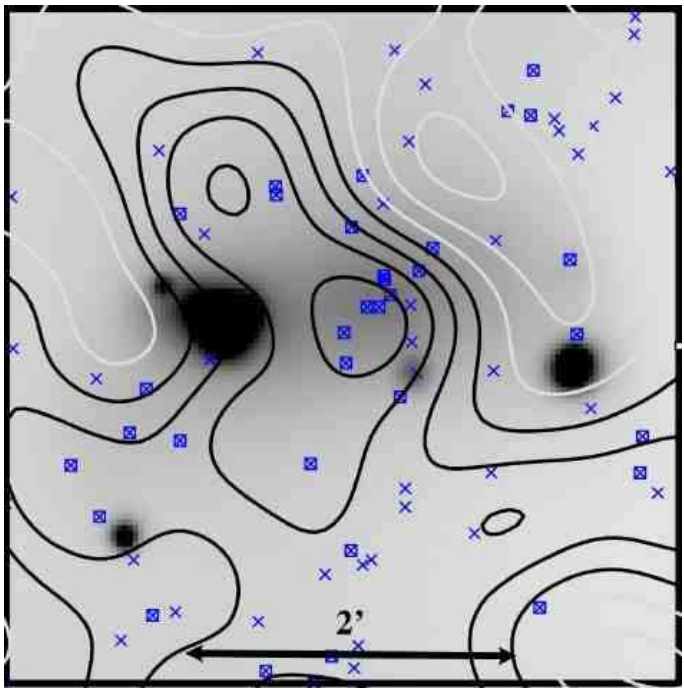}
\end{center}
\end{minipage}
\end{figure*}  

Cl1054-1145 at $z=0.70$ contains 49 spectroscopically confirmed members with a velocity dispersion of 589$^{+78}_{-70}$ km s$^{-1}$ (Halliday et al., 2004), a well-developed red sequence, and a population of bluer galaxies (White et al., 2005).  There is no evidence of substructure in the three-dimensional distribution of galaxies, but the lensing map of this cluster indicates that a large amount of mass may be currently infalling to the cluster (Clowe et al., 2006).   

\subsubsection{Cl1054-1145: Spatial analysis}
\begin{figure*}
\begin{minipage}{\textwidth}
\caption{Left: Radial surface flux profile for Cl1054-1145 obtained by masking point sources and summing exposure-corrected, background-subtracted counts within annuli.  \label{fig:1054_rprof}  Right: Corresponding constraints on $\beta$ and $r_c$.  The $r_c$ axis is in XMM physical pixels which are 0.05\arcsec\, on a side.  Contours mark 1, 2, and $3\sigma$ levels.  \label{fig:1054_regproj} }
\begin{center}
\includegraphics[width=0.50\textwidth, angle=0]{cl1054_1d.epsf}
\includegraphics[width=0.39\textwidth, angle=0]{1054_regproj.epsf}
\end{center}
\end{minipage}
\end{figure*}  
 
We overlay adaptively smoothed X-ray contours on an $I$-band image of the central region of the Cl1054-1145 field in Figure \ref{fig:1054_contour}.  Extremely faint difffuse emission is visible at the observation aimpoint, coincident with a concentration of spectroscopically-confirmed cluster galaxies including the BCG.  The diffuse emission is significantly contaminated by a bright point source $\sim1\arcmin$ to the E of the aimpoint, which appears to be an AGN associated with a faint galaxy known not to be a cluster member.  A second bright point-source lies $\sim1\arcmin$ to the WSW and may be an AGN associated with the cluster (see \S\ref{1054agn}).  A third faint peak is visible in the adaptively-smoothed image $\sim20\arcsec$ to the S of the peak of the diffuse emission which, if not an artefact of the smoothing process, may indicate very faint emission associated with a background galaxy at $z=0.76$.  The diffuse X-ray contours, where they are not contaminated by point-source emission, appear to be somewhat elongated along the E-W axis.  
Mass contours from the Clowe et al. (2006) weak-lensing mass reconstruction are plotted over an image of the smoothed X-ray flux in Figure \ref{fig:1054_mass} and indicate a significant mass peak coincident with the diffuse X-ray emission.

In Figure \ref{fig:1054_rprof}, we show the radially-averaged surface flux profile centred on the peak of the smoothed emission at $\alpha=10^{\mathrm h}54^{\mathrm m}24^{\mathrm s}.8$, $\delta = -11^{\circ}$46\arcmin21\arcsec [J2000], which we fit to a $\beta$-model as above.  The best-fitting parameters are $\beta=1.09^{+0.39}_{-0.31}$ and $r_c=370^{+116}_{-77}$ kpc though, again, the confidence contours indicate that the degeneracy between the parameters renders them essentially unconstrained.  Assuming a typical $\beta=0.7$, this fit implies $r_c\simeq256^{+71}_{-57}$ kpc and suggests the cluster may be less centrally concentrated than is typical.  Further study of the morphology of the diffuse emission, including two-dimensional modelling, was not possible due to the faintness of the source and the proximity of the bright point sources.  

\subsubsection{Cl1054-1145: Spectral analysis}

X-ray spectra were extracted within a 50\arcsec\, radius centred on the peak of the smoothed emission from point-source free event lists in which an additional region was excised around the faint peak visible to the south of the cluster emission in the smoothed image.   The aperture yielded 814 counts where 412 are expected from the background.  Based on the properties of the bright point sources and the Medium Accuracy model of the XMM PSF, we estimate that 10-15\% of the extracted counts may be contamination from the wings of the brighter sources.  

\begin{figure*}
\begin{minipage}{\textwidth}
\caption{X-ray spectra of Cl1054-1145 extracted from the MOS (lower points) and pn (upper points) cameras and best-fitting thermal emission model (histograms).  Residuals are shown in the lower panel. \label{fig:1054_spec}}
\begin{center}
\includegraphics[width=0.48\textwidth, angle=0]{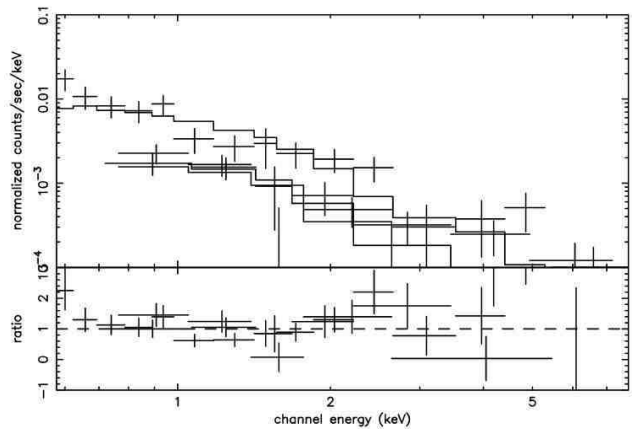}
\end{center}
\end{minipage}
\end{figure*}

The spectra were fit to an absorbed MEKAL model with the absorption fixed at the Galactic value of $3.5\times10^{20}$ cm$^{-2}$ and the metallicity fixed at 0.3 solar.  The best-fitting model (reduced $\chi^2=1.4$) is plotted over the data in Figure \ref{fig:1054_spec} and yields a temperature of $3.5^{+1.7}_{-1.6}$ keV, a full-band luminosity of $L_{\mathrm{0.5-8 keV}}=0.7\pm0.2\times10^{44}$ erg s$^{-1}$ cm$^{-2}$, and a bolometric luminosity of $L_{bol}=0.8\pm0.2\times10^{44}$ erg s$^{-1}$ cm$^{-2}$ within the spectral extraction radius.  Assuming the best-fitting values of $\beta$ and $r_c$ derived above, the spectral radius encloses 69\% of the cluster emission and the total luminosity of Cl1054-1145 is  $\sim1.2\times10^{44}$ erg s$^{-1}$ cm$^{-2}$.  Adopting the $\beta\equiv0.7$ profile, the fraction of enclosed flux falls to 48\% of the emission and the total luminosity rises to $\sim1.7\times10^{44}$ erg s$^{-1}$ cm$^{-2}$.     

\subsubsection{Cl1054-1145: Mass estimate}

Taking our derived values of $\beta$, $r_c$, and $T$ at face value and adopting Eq. 2, the estimated mass of Cl1054-1145 within 1 Mpc is $\sim3.8\times10^{14}$ M$_{\odot}$.   Assuming the $\beta\equiv0.7$ profile discussed above yields a somewhat lower mass of $\sim2.6\times10^{14}$ M$_{\odot}$.  These X-ray-derived estimates are both in  agreement with the weak-lensing estimate of $3.7^{+0.8}_{-0.9}\times 10^{14}$ M$_\odot$ though somewhat higher than the spectroscopically derived value of $\sim1.2\times10^{14} M_\odot$. However, we note that all of the derived X-ray properties are subject to large uncertainty due to the poor statistical quality of the data and the unquantified contamination by the nearby point sources.   It is also far from clear if the assumptions underpinning Eq. 2 --- spherical symmetry, hydrostatic equilibrium, and isothermality --- hold in Cl1054-1145.  

\subsubsection{Cl1054-1145: Cluster AGN}
\label{1054agn}

\begin{figure*}
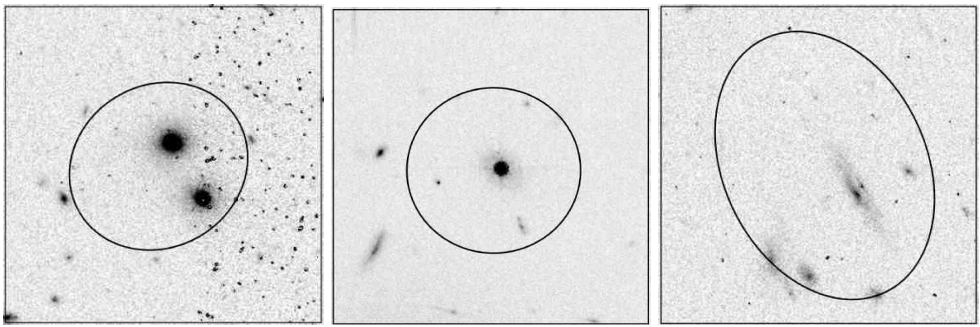

\begin{minipage}{\textwidth}
\caption{Candidate cluster AGN in Cl1054-1145.  From left: Cl1054\_X9, Cl1054\_X10, 
and Cl1054\_X26.  Boxes are 15\arcsec on a side, North is up, and East is to the left.  Ellipses are the 3$\sigma$ uncertainty of the XMM position.  \label{fig:1054_agn}}
\begin{center}
\includegraphics[width=0.24\textwidth, angle=0]{X9_bw.epsf}
\includegraphics[width=0.24\textwidth, angle=0]{X10_bw.epsf}
\includegraphics[width=0.24\textwidth, angle=0]{1054X26_bw.epsf}
\end{center}
\end{minipage}
\end{figure*}  

Of the 67 X-ray point sources detected in the Cl1054-1145 field, nine sources fall within the FORS FOV and three sources, shown in Figure \ref{fig:1054_agn}, have optical properties plausibly consistent with cluster membership. 

\begin{itemize}

\item {Source Cl1054\_X9} is ambiguously associated with two galaxies located $\sim4\arcmin$ SE of the BCG.  The probability of association $P_{id}$, is roughly 50\% for both galaxies.  The galaxy further to the South is given very low membership probabilities ($P_{mem}<2\%$), while the Northern galaxy is a spectroscopically confirmed cluster member with no obvious AGN signatures in its optical spectra.  The X-ray detection has a flux $S_{\mathrm{0.5-8 keV}}=8.5\times10^{-14}$ erg s$^{-1}$ cm$^{-2}$ and $HR=0.02$, consistent with a moderately obscured AGN. 

\item {Source Cl1054\_X10}, the compact source embedded in the Eastern side of the diffuse emission, is unambiguously identified with a galaxy with a bright AGN core which lies $\sim1\arcmin$ from the BCG and is flagged as a cluster candidate ($P_{mem} \sim75\%$).    The X-ray detection has a flux $S_{\mathrm{0.5-8 keV}}=5.4\times10^{-14}$ erg s$^{-1}$ cm$^{-2}$ and $HR-0.5$, consistent with an unobscured AGN.  However, as it is likely the optical colours are affected by the luminous AGN, the validity of the photometric redshift  based on a galaxy template is questionable. $z_{QSO}$ for this source is $\sim1.5$.

\item {Source Cl1054\_X26} is unambiguously associated with a galaxy $\sim3\arcmin45\arcsec$ from the cluster centre which is flagged as a possible cluster member, though with fairly low probability ($P_{mem} \sim30\%$).  The source has disturbed morphology and is faint in both the optical and X-ray bands, with an X-ray flux of $S_{\mathrm{0.5-8 keV}}=7.2\times10^{-15}$ erg s$^{-1}$ cm$^{-2}$ and $HR=0.36$, consistent with a heavily obscured AGN. 

\end{itemize}

\subsection{Cl1040-1155}

There are 30 spectroscopically-identified cluster galaxies in the Cl1040-1155 field at $z=0.70$ with the second lowest velocity dispersion in the EDisCS high redshift sample, $\sigma=418^{+55}_{-46}$ km s$^{-1}$.  Isopleths of galaxy density show a clear concentration around the BCG (White et al., 2005), but no significant mass peak is detected in the weak-lensing analysis which yields an upper limit on the mass which is consistent with the low velocity dispersion (Clowe et al., 2006).

\subsubsection{Cl1040-1155: Spatial analysis}
\begin{figure*}
\begin{minipage}{\textwidth}
\caption{X-ray contours overlaid on an $I$-band image of the central region of the Cl1040-1155 field.  Crosses indicate sources for which spectra have been obtained; squares indicate spectroscopically confirmed cluster members.  North is up and East is to the left. 
\label{fig:1040_contour}}
\begin{center}
\includegraphics[width=0.5\textwidth]{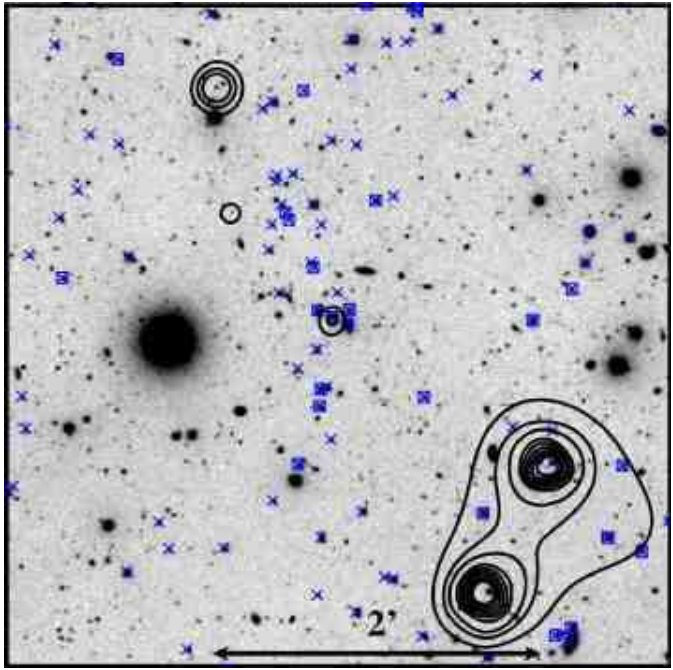}
\end{center}
\end{minipage}
\end{figure*}

\begin{figure*}
\begin{minipage}{\textwidth}
\caption{Radial surface flux profile for Cl1054-1145 obtained by masking point sources and summing exposure-corrected, background-subtracted counts within annuli. \label{fig:1040_rprof}}
\hspace{-0.5cm}
\begin{center}
\includegraphics[width=0.50\textwidth, angle=0]{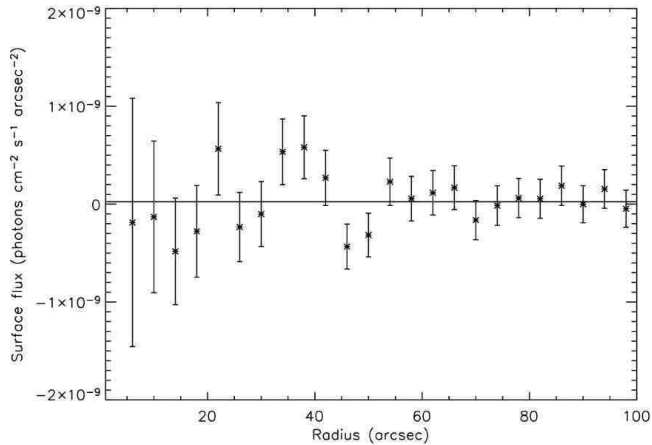}
\end{center}
\end{minipage}
\end{figure*}

In Figure \ref{fig:1040_contour} we show an $I$-band image of the central region of the Cl1040-1155 field overlaid with contours of adaptively smoothed X-ray flux.  There is no evidence for extended X-ray emission associated with the intracluster medium.  An extremely faint compact source is seen at the observation aimpoint which appears to be associated with one of the spectroscopically-confirmed cluster members near to the BCG.  The source was not detected by the wavelet detection algorithm; it may not meet the selected significance or total counts thresholds or it may be an artefact of the adaptive-smoothing.  A radially-averaged surface flux profile centred on the aimpoint of the XMM observation is shown in Figure \ref{fig:1040_rprof}.  It is well fit with a constant background level and confirms the lack of diffuse detection.  The data place a $3\sigma$ upper limit on diffuse emission on 0.5 Mpc scales of $0.2\times10^{44}$ erg s$^{-1}$.  

\begin{figure*}
\begin{minipage}{\textwidth}
\caption{Adaptively smoothed 0.5--8 keV X-ray image of the central region of the Cl1040-1155 field with the Clowe et al. (2006) weak lensing mass reconstruction of the field overlaid.  Contour intervals are $\sim 1\times10^8$M$_\odot$ kpc$^{-2}$ relative to the local mean, with negative contours shown in light grey.  Crosses indicate sources for which spectra have been obtained; squares indicate spectroscopically confirmed cluster members.  North is up and East is to the left.
\label{fig:1040_mass}}
\begin{center}
\includegraphics[width=0.5\textwidth]{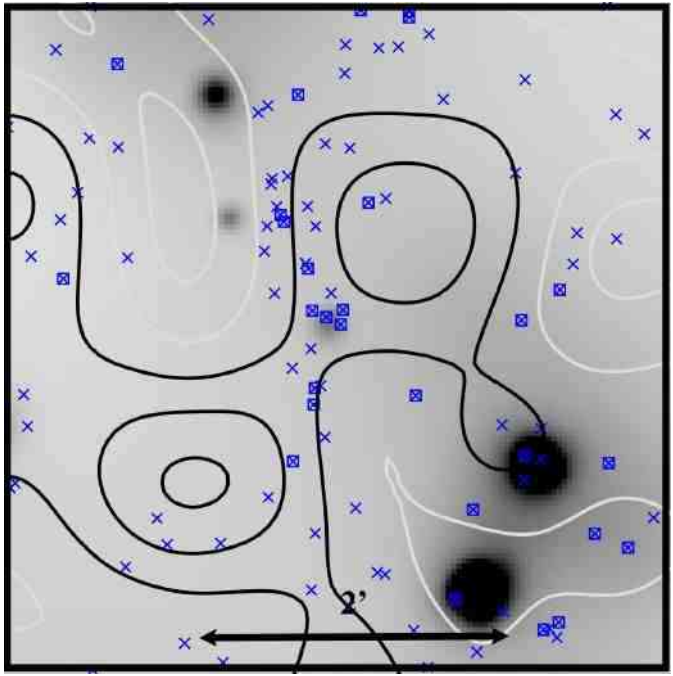}
\end{center}
\end{minipage}
\end{figure*}

Neither of the two marginal peaks detected in the weak lensing mass reconstruction is coincident with the position of the BCG, as shown in Figure \ref{fig:1040_mass}.  That Cl1040-1155 is not detected in the X-ray and only marginally detected in the weak lensing analysis suggests the overdensity of galaxies at $z=0.70$ in the field comprise a poor cluster or a group, (see also discussion in White et al., 2005).  

\subsubsection{Cl1040-1155: Cluster AGN}
\label{agn1040}

\begin{figure*}
\begin{minipage}{\textwidth}
\caption{Candidate cluster AGN in Cl1040-1155.  From left: Cl1040\_X12, Cl1040\_X23 
and Cl1040\_X67.  Boxes are 15\arcsec on a side, North is up, and East is to the left.  Ellipses are the 3$\sigma$ uncertainty of the XMM position. \label{fig:1040_agn}}
\hspace{-0.5cm}
\begin{center}
\includegraphics[width=0.24\textwidth, angle=0]{X12_bw.epsf}
\includegraphics[width=0.24\textwidth, angle=0]{X23_bw.epsf}
\includegraphics[width=0.24\textwidth, angle=0]{X67_bw.epsf}
\end{center}
\end{minipage}
\end{figure*}  

Eleven of the 67 X-ray sources detected in the Cl1040-1155 field fall within the FORS FOV.  Three of these, shown in Figure \ref{fig:1040_agn}, have optical properties plausibly consistent with cluster membership. 

\begin{itemize}

\item {Source Cl1040\_X12} is ambiguously associated with two objects $\sim1\arcmin30\arcsec$ from the cluster core.  The object to the West is a star, and is selected as the WFI counterpart to the X-ray source by the association algorithm.  However, a spectroscopically-confirmed cluster member visible further to the East in the HST image is also within the 3$\sigma$ error circle.  The X-ray detection has a flux $S_{\mathrm{0.5-8 keV}}=4.0\times10^{-14}$ erg s$^{-1}$ cm$^{-2}$ and $HR=0.03$, which is consistent with a moderately obscured AGN and unusually hard for stellar X-ray emission. 

\item {Source Cl1040\_X23} is unambiguously identified with a galaxy $\sim2\arcmin30\arcsec$ from the cluster centre which is flagged as a possible cluster member based on its photometric properties, though with low membership probabilities ($P_{mem} \sim25\%$).  The X-ray flux of the source is $S_{\mathrm{0.5-8 keV}}=1.1\times10^{-14}$erg s$^{-1}$ cm$^{-2}$ and $HR=0.07$, consistent with a moderately obscured AGN.

\item{Source Cl1040\_X67} is ambiguously associated with two sources $\sim4\arcmin15\arcsec$ from the cluster centre.  The source further to the North is extremely faint, while the Southern source is flagged as a possible cluster member.  The photometry for this source is based on only optical colours, however, and is therefore considered not reliable.   The X-ray flux of the source is $S_{\mathrm{0.5-8 keV}}=3.7\times10^{-14}$erg s$^{-1}$ cm$^{-2}$ and $HR=0.03$, consistent with a moderately obscured AGN.

\end{itemize}

\section{Cluster scaling relations}
\label{scaling}

\begin{figure*}
 \begin{minipage}{\textwidth}
 \caption{X-ray temperature versus bolometric X-ray luminosity for the Cl1216-1201 and Cl1054-1145 clusters (squares).  The solid line shows the best-fit relation of Xue \& Wu (2000), while the dashed line indicates this relation evolved to $z=0.8$ following the parameterization of Vikhlinin et al. (2002).   Points denoted by small diamonds are the optically-selected $z=0.76$ and $z=0.90$ clusters of Lubin et al. (2004), while other points are the X-ray selected $0.6 < z < 1.4$ clusters of Mullis et al. (2004), Gioia et al. (2004), Donahue et al. (1999), Gioia et al. (1999), Stanford et al. (2001), and Valtchanov et al. (2004).  
\label{fig:templum}}
\begin{center}
\includegraphics[width=.50\textwidth, angle=0]{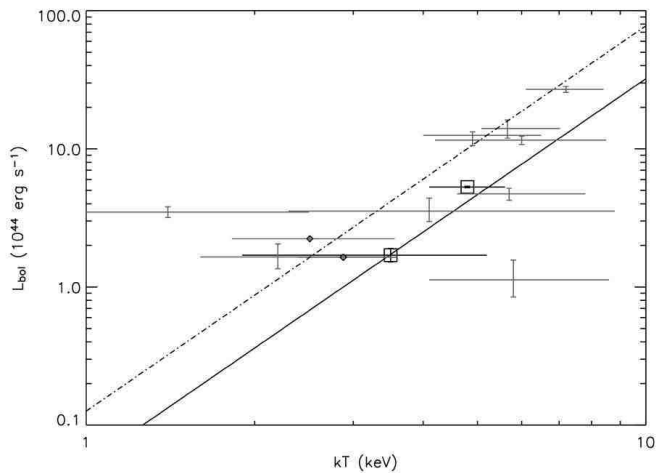}
\end{center}
\end{minipage}
\end{figure*}

\begin{figure*}
\begin{minipage}{\textwidth}
\caption{Bolometric X-ray luminosity versus optical velocity dispersion for the Cl1216-1201, Cl1054-1145, and Cl1040-1155 clusters.   Symbols are as described in Figure \ref{fig:templum}.  The Cl1040-1155 point is a $3\sigma$ detection limit based on the background level within a radius of 90\arcsec.  The solid line shows the best-fit relation of Xue \& Wu (2000) while the dotted line shows that of Mahdavi \& Geller (2001).  In addition to the optically-selected clusters of Lubin et al. (2004), RXJ1716.6+6708 (Gioia et al., 1999; $z=0.81$) also appears significantly under-luminous in X-rays for its velocity dispersion.
\label{fig:lumsig}}
\begin{center}
\includegraphics[width=.50\textwidth, angle=0]{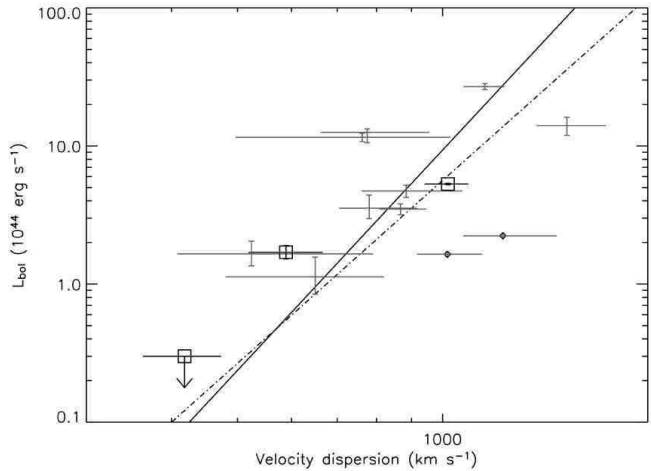}
\end{center}
\end{minipage}
\end{figure*}

\begin{figure*}
\begin{minipage}{\textwidth}
\caption{Optical velocity dispersion versus X-ray temperature for  in the Cl1216-1201, Cl1054-1145, and Cl1040-1155 clusters.  Symbols are as described in Figure \ref{fig:templum}.  As Cl1040-1155 was undetected in the XMM observations, its temperature is unknown as indicated by the dashed line.  The solid line shows the best-fit relation of Xue \& Wu (2000) and other data points are as in Figure \ref{fig:templum}.  In addition to the optically-selected clusters of Lubin et al. (2004), the X-ray selected clusters RXJ1716.6+6708 (Gioia et al., 1999; $z=0.81$) and XLSSC001 (Valtchanov et al., 2004; $z=0.61$) appear to have low X-ray temperatures for their velocity dispersions. 
\label{fig:tempsig}}
\begin{center}
\includegraphics[width=.50\textwidth, angle=0]{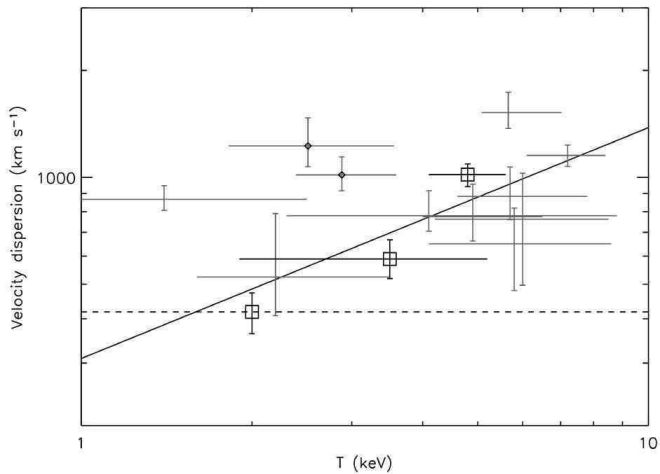}
\end{center}
\end{minipage}
\end{figure*}

The X-ray properties of the three EDisCS targets and the Southern component in the Cl1216-1201 field are summarised in Table \ref{tab:summary}.  In Figures \ref{fig:templum}, \ref{fig:tempsig}, and \ref{fig:lumsig}, we compare the X-ray and optical properties of these structures with the best-fitting scaling relations of Xue \& Wu (2000), derived from a standardised sample of 274 primarily low-redshift clusters taken from the literature.  The Xue \& Wu (2000) sample spans a range in redshift from $0.005 < z < 1$, but $\sim 65\%$ are at $z < 0.1$ and $< 2\%$ are at $z > 0.4$.  We have converted these relations from an Einstein-deSitter cosmology to the concordance $\Lambda$CDM cosmology assuming a mean redshift $\bar z=0.1$.  For comparison at high redshift, we have drawn from the literature a sample of clusters with $0.6 < z < 1.4$ for which X-ray temperature, X-ray luminosity and optical velocity dispersion are known (Lubin et al., 2004; Mullis et al., 2005; Gioia et al., 2004; Donahue et al., 1999; Gioia et al., 1999; Stanford et al., 2001; and Valtchanov et al., 2004).  Of these, only the two Lubin et al. (2004) clusters at $z=0.76$ and $z=0.90$ are optically selected.  We have standardised the plotted values where necessary by calculating bolometric luminosities from a Mekal model with the appropriate temperature and an abundance of 0.3 solar.  Where luminosities are given within a radius, we have accounted for flux outside the aperture using the best-fitting $\beta$-model, when given, or a typical profile ($\beta=0.7, r_{c}=140$ kpc).

\begin{table*}
\begin{minipage}{\textwidth}
\caption{Summary of the X-ray properties of EDisCS clusters and the Cl1216-South structure.   Columns are slope and core radius of the best-fitting surface brightness models, ellipticity, radius of the spectral aperture, source counts within the aperture, background counts within the aperture, X-ray temperature, 0.5--8 keV flux within the aperture, 0.5--8 keV luminosity within the aperture, and total bolometric luminosity.  $\beta$ and $r_{c}$ are from one-dimensional modelling except for Cl1216-South, where the two-dimensional model is preferred. Total luminosities are extrapolated from the spectral radius using the parameters of the best-fitting one-dimensional $\beta$-models.  Core radii are given in kpc, temperatures in keV, fluxes in $10^{-14}$ erg s$^{-1}$ cm$^{-2}$and luminosities in $10^{44}$ erg s$^{-1}$.  Quoted errors indicate 90\% confidence levels; errors on $\beta$ and $r_c$ underestimate the true uncertainty due to the strong degeneracy of these parameters (see Figures \ref{fig:1216_rprof}, \ref{fig:south_rprof}, and \ref{fig:1054_rprof}).  \label{tab:summary}}
\begin{center}
\begin{tabular}{lcccccccccc}
\hline
Source     			&$\beta$	      			&$r_c$			&$e$	&$r_{s}$		&$C_{src}$	&$C_{bkg}$	&kT      	&$S_{\mathrm{0.5-8}}$	&$L_{\mathrm{0.5-8}}$	&$L_{bol}$\\
\hline\hline
Cl1216-1201		&$0.68^{+0.06}_{-0.05}$	&$159^{+20}_{-18}$	 	&0.0		&40\arcsec	&1503	&194	&$4.8^{+0.8}_{-0.7}$	&$8.8^{+0.6}_{-0.6}$	&$2.5^{+0.2}_{-0.2}$	&$\sim5$\\
Cl1054-1145		&$0.7$	&$256^{+71}_{-57}$	&---		&50\arcsec  	&814	&412	&$3.5^{+1.7}_{-1.6}$	&$6.1^{+0.8}_{-0.6}$ 	&$0.7^{+0.2}_{-0.2}$	&$\sim2$\\
Cl1040-1155		& ---					& ---					&---		&---			&---	&---	&---	&$<0.8$				&$<0.2$			&$<0.3$\\
\hline
Cl1216-South    	&$0.48^{+0.10}_{-0.06}$	 &$79^{+81}_{-49}$ 		&0.4	 	&50\arcsec 	&1271	&472	&$5.0^{+1.8}_{-1.3}$	&$3.2^{+0.6}_{-0.6}$	&$1.7^{+0.2}_{-0.2}$ 	&$\sim3$\\
\hline
\end{tabular}
\end{center}
\end{minipage}
\end{table*}

In Figure \ref{fig:templum}, we plot bolometric X-ray luminosity versus X-ray temperature for Cl1216-1201 and Cl1054-1145 along with the comparison sample of high-redshift clusters.  We compare these points with both the Xue \& Wu (2000) $L_{X}-T$ relation and its evolution to $z=0.8$ following the parameterization of Vikhlinin et al. (2002): $L(T,z)=L(T)\times(1+z)^{1.5\pm0.3}$.  Both detected EDisCS sources are consistent at the 90\% confidence level with the relation of Xue \& Wu, and appear slightly underluminous compared with the evolved relation.  The scatter among the plotted high-redshift clusters is substantial, however, and the X-ray properties of the EDisCS clusters, like those of the Lubin et al. (2004) optically selected clusters, are fully consistent with the properties of the X-ray selected sample as a whole.  

We note that the values of $L_{X}$ and $T$ derived for the Cl1216-South structure assuming it is associated with Cl1216 at $z=0.79$ are also consistent with the Xue \& Wu (2000) relation.  While we stress that further observations are needed to determine the true redshift of this source and therefore do not include it in Figure \ref{fig:templum}, we note that fits to the Cl216-South spectra yield values which are consistent with this relation within the 90\% confidence limits only if the assumed redshift is $z> 0.7$, with assumed $0.1 < z < 0.6$ fits producing values which are severely under-luminous for the implied temperatures in comparison with the Xue \& Wu (2000) relation.

In Figure \ref{fig:lumsig}. we plot the bolometric X-ray luminosity versus optical velocity dispersion for the targetted EDisCS clusters over both the best-fitting Xue \& Wu (2000) $L_{X}-\sigma$ relation and that of Mahdavi \& Geller (2001).  The variation between these two relations may result from the inclusion of lower-mass groups in the Mahdavi \& Geller cluster sample which are fit with a separate, shallower relation in the Xue \& Wu paper.  The Cl1216-1201 and Cl1054-1145 points lie near to the low-redshift relations and within the scatter of the other high-redshift points.   Cl1216-1201 is under-luminous for its velocity dispersion with respect to the Xue \& Wu (2000) relation by roughly a factor of 2, but is entirely consistent with the Mahdavi \& Geller (2001) relation.  The non-detection of Cl1040-1155 is potentially consistent with either of the low-redshift relations, as it places an upper limit on the luminosity which is roughly twice that expected from them.  

In contrast, and despite their unremarkable position on the $L_{X}-T$ plot, the optically-selected clusters of Lubin et al. (2004) stand well apart from both the low-redshift $L_{X}-\sigma$ relations and the bulk of the high-redshift clusters in Figure \ref{fig:lumsig}, exhibiting significantly lower X-ray luminosities than expected for their velocity dispersions.  The X-ray selected cluster RXJ1716+6708, discovered in the ROSAT all-sky survey, appears to have similarly disparate X-ray/optical properties.   On the basis of its X-ray morphology and optical/X-ray properties, Gioia et al. (1999) argue this $z=0.81$ system may be in the process of collapsing along a filament and thus have a velocity dispersion which is not representative of the luminosity and temperature of the X-ray gas. 

In Figure \ref{fig:tempsig} we plot optical velocity dispersion against X-ray temperature for the three EDisCS target clusters.  Both detected clusters are entirely consistent with the best-fitting Xue \& Wu (2000) $T-\sigma$ relation within the observational uncertainties and fall well within the scatter of the other high-redshift points.  As Cl1040-1155 is undetected in our XMM observation, its temperature is unconstrained.  However, provided it is cool ($<3$ keV), it too lies close to the Xue \& Wu (2000) relation and is consistent with the other high-redshift clusters.  Though the measurement uncertainties are larger, the Lubin et al. (2004) clusters and RXJ1716+6708 again appear to be outliers, in the sense that they are significantly cooler than both the prediction of the low-redshift $T-\sigma$ relation and the bulk of the X-ray selected high-redshift clusters.  (The fourth cluster which appears to be unusually cool in Figure \ref{fig:tempsig} is XLSSC001, a $z=0.61$ cluster identified in the XMM Large Scale Survey reported by Valtchanov et al. (2004).  However, these authors specify the velocity dispersion determined for this cluster is still preliminary.)

\section{Discussion}
\label{discussion}

\subsection{Intracluster gas}

The primary result of this initial X-ray study of the EDisCS sample is the apparent consistency of the ICM properties of the EDisCS clusters with those of X-ray selected cluster populations at both high and low redshift.  While based on only two detected clusters, this preliminary finding demonstrates that it is possible to select clusters which appear to be comparable to the X-ray luminous population at high redshift in a way which is insensitive to X-ray flux.  If found to hold throughout the full EDisCS sample, this result would suggest that the X-ray selected samples are {\it not} biased directly by their selection in the X-ray band.  Instead, both the novel selection criteria of the EDisCS sample --- which rely on detection of compact intracluster light rather than excess galaxy counts --- and X-ray surveys may effectively identify the more fully virialised structures at high redshift.  The apparent lack of evolution in the cluster scaling relations for these structures suggests that these detection techniques are well-suited for selection of relaxed, high-redshift clusters whose ICM is in place and stable by $z\sim0.8$.  That the X-ray determined masses of Cl1216-1201 and Cl1054-11, derived under the assumption that they are relaxed structures in hydrostatic equilibrium, are in general agreement with those determined from the weak lensing analysis lends further support to this conclusion.  

These first results also suggest that the discrepancy which has been noted between other optically selected, high-redshift cluster samples and the X-ray luminous cluster population may not be due to a peculiarity of X-ray selection.  Rather, as has been suggested, optical matched-filter techniques may preferentially select a physically distinct population of massive structures at high redshift.  Both Lubin et al. (2004) and Bower et al. (1997) argue that the discrepancy between the high velocity dispersion and low X-ray luminosity of their optically selected clusters could be explained were they systems of lower virial temperature undergoing gravitational collapse.  Similarly, Hicks et al. (2004) note that the X-ray properties of their sample of RCS clusters  --- less centrally condensed morphologies and low X-ray temperatures and luminosities which nonetheless lie on the $L_{X}-T$ relationship of the general cluster population --- are consistent with their containing relatively low mass virialised cores imbedded within larger forming structures.  It is interesting to note in this context that RXJ1716.6+6708 --- the only X-ray selected cluster with ICM properties mimicking those of the optically selected clusters in Figures \ref{fig:tempsig} and \ref{fig:lumsig} --- is also believed to be in the process of collapsing.

\subsection{Cluster AGN}

In addition to the diffuse ICM emission, these data have allowed us to begin compiling a sample of candidate cluster AGN in the EDisCS sample.  Though no cluster AGN are yet spectroscopically confirmed, cross-correlation of the X-ray catalogues with the multi-band data available over the central $\sim 3$ Mpc by $\sim 3$ Mpc region of each cluster field has identified several promising candidates.  The extant data indicate a maximum of 3-4 cluster AGN per field.  The candidates are found $\sim 0.5 - 1.5$ Mpc from the cluster cores and, if associated with the clusters, comprise a sample of variously obscured Type I and II AGN with luminosities of $L_{\mathrm{0.5-8\, keV}}=10^{42-43}$ erg s$^{-1}$.  While spectroscopic confirmation of candidate cluster AGN and extension of this preliminary study to our larger sample will both be necessary to characterise the AGN content of the EDisCS clusters in a rigourous way, it is interesting that the preliminary data on these first clusters suggests that their AGN populations are typical of those found in the X-ray luminous cluster population in general.  In a study of over 100 clusters at $0.1<z<1$ selected from the Chandra archive, X-ray luminous clusters were found to host 2-3 AGN on average, at distances of 0.5 - 1 Mpc from the cluster centre (Dowsett, 2006).  Further study of the cluster AGN population within EDisCS will be conducted when planned X-ray and radio observations of the remainder of the sample are available. 

\section{Summary}
\label{summary}

In this paper we have presented XMM-Newton observations of three optically-selected, high-redshift clusters which comprise the first results of an X-ray survey of the full EDisCS $z > 0.6$ sample.  
We find Cl1216-1201 at $z=0.79$ to have a relaxed X-ray morphology, a temperature of $T=4.8^{+0.8}_{-0.7}$ keV, and a bolometric X-ray luminosity of $\sim 5\times 10^{44}$ erg s$^{-1}$ cm$^{-2}$.  We detect a second diffuse X-ray source $3.5 \arcmin$ to the SSE which, if also at $z=0.79$, has a similar temperature and a somewhat lower luminosity.  We marginally detect Cl1054-1145 at $z=0.70$ and derive a temperature of $T=3.5^{+1.7}_{-1.6}$ keV and a bolometric luminosity of  $\sim 2\times 10^{44}$ erg s$^{-1}$ cm$^{-2}$.  Cl1040-1155 at $z=0.70$ is not detected in our 30 ks XMM exposure; on a scale of 0.5 Mpc diameter, we place a $3\sigma$ upper limit on its luminosity of $L_{\mathrm{0.5-8}}<0.2\times 10^{44} $erg s$^{-1}$.  

In both X-ray-detected clusters the diffuse X-ray emission is coincident with galaxy overdensities containing the cluster BCG and peaks in the weak-lensing mass reconstructions of the field.  Furthermore, mass estimates based on the X-ray data for Cl1216-1201 and Cl1054-1145 are comparable to those derived from the lensing analysis.  Despite evidence of some level of extended substructure in mass maps, velocity dispersion histograms, and --- in the Cl1216-1201 field ---   the possible detection of a second X-ray component, the multiband evidence is suggestive of relatively relaxed systems in which the galaxies and the intracluster medium are in virial equilibrium within the same potential well.  We stress, however, that these conclusions are based on only two cluster detections, one of which is of poor statistical quality, and require confirmation in further, deeper observations of the remainder of the sample.

Comparison of the X-ray source-lists with the multi-band photometry and spectroscopy has identified several promising candidate cluster AGN, though none are yet spectroscopically confirmed.  We have identified 3-4 candidates per field; if associated with the target clusters, they comprise a population of Type I and Type II objects with typical X-ray luminosities of $L_{\mathrm{0.5-8\, keV}}=10^{42-43}$ erg s$^{-1}$ residing 0.5 - 1.5 Mpc from the cluster centres.  As reported in Dowsett et al. (2006), these findings are typical of X-ray luminous clusters. 

The principal result of this preliminary work is the consistency of our optically-selected clusters with both the low-redshift X-ray/optical scaling relations and the X-ray selected cluster population at high redshift.  This result stands in contrast to the disparate X-ray/optical properties reported for optically-selected high-redshift clusters in the literature, which appear systematically cooler and less luminous than expected for their velocity dispersions.  This early analysis suggests that the EDisCS, like surveys based on X-ray luminosity, may effectively select virialised high-redshift clusters in which the ICM is already well-formed and stable by $z\sim0.8$.  The discrepancy between the X-ray/optical properties of our sample and those of previously reported high-redshift clusters selected using optical matched-filter techniques may result from the latter being less fully virialised structures still undergoing collapse.  

Our results suggest X-ray surveys may provide an unbiased view of virialised massive clusters at high-redshift, but also stress the need for surveys in both optical and X-ray bands to fully sample the mass distribution at epochs where massive clusters are still forming.  The planned extension of this study to the full EDisCS high-redshift sample with deeper X-ray exposures will allow us to confirm these preliminary results and also to extend the high-redshift $L_{X}-\sigma$ and $T-\sigma$ relations to much lower velocity dispersions.  The EDisCS sample may offer a unique opportunity to target virialised structures at high-redshift which are intrinsically similar to the X-ray luminous population but with lower masses which are more representative of the precursors of common present-day clusters.      

\section{Acknowledgements}  
This work is based on observations collected at the European Southern Observatory, Chile,
as part of large programme 166.A--0162 (the ESO Distant Cluster Survey).
OJ acknowledges PPARC for support through a Postdoctoral Fellowship and thanks Michael Brown for helpful discussions.  PNB acknowledges the Royal Society for support through its University Research Fellowship Scheme.  The authors thank the anonymous referee for a careful and helpful review of the draft.

\appendix

\section{X-ray point source catalogues}

In Tables \ref{tab:xray1216}, \ref{tab:xray1054}, and \ref{tab:xray1040} we present the X-ray catalogues of point sources in the EDisCS fields.  Columns list: X-ray source ID; the X-ray R.A. and Dec. (J2000) and the positional error in arcseconds; the net X-ray counts in the full 0.5--8 keV band; the source significance, $\sigma_{src}$, as defined in \S\ref{point_sources};  full-, soft-, and hard-band fluxes in units of $10^{-15}$ erg s$^{-1}$ cm$^{-2}$; and X-ray hardness ratio.  These data are also available in machine-readable form at the CDS. 

\begin{table*}
\begin{minipage}{\textwidth}
\caption[Cl1216-1201 X-ray point source catalogue]{Catalogue of X-ray point sources in the Cl1216-1201 field.  Sources Cl1216\_46 and Cl1216\_62 have been omitted.  These detections correspond to the diffuse emission regions Cl1216-1201 and Cl1216-South which are discussed in \S\ref{1216res}.
\label{tab:xray1216}}
\begin{tabular}{lcccccccccc}
\hline
ID	&R.A. (J2000)		&Dec. (J2000)	&Error		&Counts	&$\sigma_{src}$	&$S_{\mathrm{0.5-8~keV}}$	&$S_{\mathrm{0.5-2~keV}}$	&$S_{\mathrm{2-8~keV}}$	&$HR$\\
\hline\hline
Cl1216\_1 &12:17:35.54 &-12:02:06.7 &1.6 &377.4 &30.1 &$82.0\pm3.1$ &$30.2\pm1.3$ &$48.5\pm3.8$ &$-0.43\pm0.04$\\
Cl1216\_2 &12:17:21.30 &-12:07:58.5 &1.8 &145.6 &17.3 &$62.2\pm3.9$ &$24.3\pm1.7$ &$30.6\pm4.1$ &$-0.52\pm0.06$\\
Cl1216\_3 &12:17:21.27 &-12:01:38.6 &1.6 &620.7 &41.4 &$91.8\pm2.7$ &$29.8\pm1.1$ &$74.7\pm3.7$ &$-0.23\pm0.03$\\
Cl1216\_4 &12:16:54.11 &-12:00:58.0 &1.6 &129.4 &12.2 &$24.6\pm1.7$ &$6.8\pm0.6$ &$22.9\pm2.3$ &$-0.08\pm0.07$\\
Cl1216\_5 &12:16:48.76 &-12:03:32.4 &1.6 &167.1 &13.9 &$18.3\pm1.1$ &$6.8\pm0.5$ &$11.1\pm1.4$ &$-0.42\pm0.06$\\
Cl1216\_6 &12:16:48.69 &-12:02:29.1 &1.6 &207.3 &13.5 &$16.0\pm0.9$ &$4.8\pm0.4$ &$15.0\pm1.3$ &$-0.12\pm0.06$\\
Cl1216\_7 &12:16:46.45 &-12:07:47.9 &1.6 &129.7 &10.5 &$15.3\pm1.1$ &$5.1\pm0.4$ &$11.4\pm1.5$ &$-0.29\pm0.07$\\
Cl1216\_8 &12:16:46.02 &-11:57:37.1 &1.6 &256.1 &19.0 &$21.4\pm1.0$ &$8.5\pm0.5$ &$11.3\pm1.2$ &$-0.50\pm0.04$\\
Cl1216\_9 &12:16:42.75 &-12:01:06.9 &1.5 &1406.9 &65.6 &$110.4\pm2.1$ &$30.8\pm0.8$ &$114.1\pm3.1$ &$-0.04\pm0.02$\\
Cl1216\_10 &12:16:37.39 &-12:01:11.9 &1.6 &370.2 &22.6 &$27.3\pm1.1$ &$9.9\pm0.5$ &$18.5\pm1.4$ &$-0.36\pm0.04$\\
Cl1216\_11 &12:16:30.69 &-12:11:17.8 &1.7 &245.1 &19.9 &$36.1\pm1.8$ &$11.5\pm0.7$ &$29.9\pm2.5$ &$-0.21\pm0.05$\\
Cl1216\_12 &12:16:30.61 &-11:57:03.8 &1.6 &229.4 &17.4 &$21.2\pm1.1$ &$8.1\pm0.4$ &$11.5\pm1.3$ &$-0.47\pm0.05$\\
Cl1216\_13 &12:16:29.24 &-12:04:24.1 &1.5 &1671.1 &80.1 &$139.9\pm2.5$ &$58.6\pm1.1$ &$61.8\pm2.5$ &$-0.58\pm0.01$\\
Cl1216\_14 &12:16:22.41 &-12:08:01.4 &1.6 &197.8 &14.7 &$24.0\pm1.3$ &$6.9\pm0.5$ &$23.4\pm1.9$ &$-0.08\pm0.05$\\
Cl1216\_15 &12:16:14.13 &-12:01:14.4 &1.5 &1670.3 &95.1 &$174.4\pm3.1$ &$88.3\pm1.6$ &$11.3\pm1.5$ &$-0.94\pm0.01$\\
Cl1216\_16 &12:16:06.67 &-11:57:42.0 &1.6 &276.4 &21.1 &$34.6\pm1.6$ &$13.2\pm0.7$ &$19.2\pm1.9$ &$-0.47\pm0.04$\\
Cl1216\_17 &12:17:43.79 &-11:53:46.8 &1.8 &141.6 &21.3 &$63.1\pm4.0$ &$28.4\pm1.8$ &$11.2\pm3.6$ &$-0.82\pm0.05$\\
Cl1216\_18 &12:17:35.01 &-11:59:23.6 &1.6 &122.3 &13.6 &$28.2\pm2.0$ &$10.4\pm0.8$ &$14.8\pm2.5$ &$-0.48\pm0.07$\\
Cl1216\_19 &12:17:29.75 &-11:57:39.5 &1.7 &101.9 &9.7 &$19.1\pm1.6$ &$6.4\pm0.6$ &$18.8\pm2.9$ &$-0.15\pm0.09$\\
Cl1216\_20 &12:17:14.88 &-11:55:08.8 &1.7 &133.8 &12.2 &$20.9\pm1.4$ &$7.1\pm0.6$ &$15.0\pm2.0$ &$-0.31\pm0.07$\\
Cl1216\_21 &12:17:07.02 &-11:56:05.1 &1.7 &109.6 &10.3 &$13.5\pm1.1$ &$4.7\pm0.4$ &$9.1\pm1.4$ &$-0.35\pm0.08$\\
Cl1216\_22 &12:17:01.45 &-12:02:30.4 &1.8 &82.3 &6.7 &$7.6\pm0.8$ &$3.2\pm0.3$ &$6.8\pm1.5$ &$-0.31\pm0.11$\\
Cl1216\_23 &12:16:53.67 &-11:53:06.0 &1.7 &106.1 &10.7 &$12.4\pm1.1$ &$5.5\pm0.5$ &$10.9\pm2.1$ &$-0.34\pm0.09$\\
Cl1216\_24 &12:16:51.12 &-11:51:19.6 &1.7 &87.3 &8.2 &$14.5\pm1.3$ &$4.8\pm0.5$ &$10.6\pm1.8$ &$-0.29\pm0.09$\\
Cl1216\_25 &12:16:51.05 &-12:01:53.3 &1.6 &168.1 &11.6 &$13.0\pm0.8$ &$5.1\pm0.4$ &$6.9\pm1.0$ &$-0.49\pm0.06$\\
Cl1216\_26 &12:16:50.51 &-11:57:32.8 &1.6 &61.4 &6.1 &$5.5\pm0.6$ &$2.6\pm0.3$ &$1.4\pm0.9$ &$-0.76\pm0.13$\\
Cl1216\_27 &12:16:50.10 &-11:56:17.4 &1.7 &88.8 &8.1 &$8.6\pm0.8$ &$3.3\pm0.3$ &$5.0\pm1.0$ &$-0.44\pm0.09$\\
Cl1216\_28 &12:16:47.16 &-11:55:28.1 &1.8 &104.0 &9.5 &$12.8\pm1.0$ &$3.8\pm0.4$ &$12.4\pm1.6$ &$-0.10\pm0.08$\\
Cl1216\_29 &12:16:44.69 &-12:10:21.0 &1.8 &85.7 &7.6 &$11.0\pm1.1$ &$3.7\pm0.4$ &$8.2\pm1.4$ &$-0.28\pm0.10$\\
Cl1216\_30 &12:16:39.33 &-11:58:00.1 &1.7 &53.9 &7.2 &$10.1\pm1.2$ &$5.7\pm0.6$ &$-\pm-$ &$1.00\pm0.00$\\
Cl1216\_31 &12:16:24.26 &-11:57:13.4 &1.6 &225.8 &15.0 &$21.7\pm1.1$ &$7.4\pm0.5$ &$16.3\pm1.5$ &$-0.29\pm0.05$\\
Cl1216\_32 &12:16:23.82 &-11:50:43.5 &1.6 &85.7 &8.4 &$13.3\pm1.3$ &$5.2\pm0.5$ &$6.7\pm1.6$ &$-0.51\pm0.10$\\
Cl1216\_33 &12:16:12.39 &-11:55:50.9 &1.7 &130.6 &11.9 &$16.8\pm1.2$ &$5.1\pm0.4$ &$15.0\pm1.7$ &$-0.15\pm0.07$\\
Cl1216\_34 &12:16:05.31 &-12:00:01.8 &1.8 &79.8 &7.2 &$8.3\pm0.8$ &$2.4\pm0.3$ &$7.8\pm1.2$ &$-0.11\pm0.10$\\
Cl1216\_35 &12:16:04.89 &-11:54:25.7 &2.1 &46.9 &6.7 &$17.4\pm2.1$ &$5.7\pm0.8$ &$14.0\pm2.9$ &$-0.24\pm0.12$\\
Cl1216\_36 &12:15:56.16 &-11:56:18.1 &2.0 &50.4 &7.2 &$20.0\pm2.3$ &$4.6\pm0.8$ &$23.5\pm3.7$ &$0.12\pm0.12$\\
Cl1216\_37 &12:17:34.91 &-12:08:08.1 &1.8 &66.4 &6.7 &$15.9\pm1.7$ &$5.4\pm0.7$ &$13.9\pm2.9$ &$-0.22\pm0.12$\\
Cl1216\_38 &12:17:28.86 &-11:58:17.8 &1.7 &52.2 &6.4 &$12.6\pm1.5$ &$3.8\pm0.5$ &$7.8\pm2.0$ &$-0.33\pm0.13$\\
Cl1216\_39 &12:17:16.05 &-12:05:13.5 &1.7 &72.7 &8.6 &$8.9\pm1.0$ &$3.2\pm0.4$ &$7.2\pm1.6$ &$-0.28\pm0.12$\\
Cl1216\_40 &12:17:13.82 &-11:57:40.1 &1.8 &55.6 &5.9 &$7.2\pm0.9$ &$2.5\pm0.4$ &$8.9\pm2.1$ &$-0.05\pm0.14$\\
Cl1216\_41 &12:17:13.23 &-12:13:00.6 &1.7 &82.2 &10.6 &$24.4\pm2.1$ &$9.8\pm0.9$ &$10.4\pm2.7$ &$-0.58\pm0.09$\\
Cl1216\_42 &12:17:06.74 &-12:08:15.8 &1.7 &95.5 &8.9 &$13.3\pm1.1$ &$4.6\pm0.4$ &$9.5\pm1.6$ &$-0.32\pm0.09$\\
Cl1216\_43 &12:16:57.68 &-11:48:25.1 &1.7 &59.2 &6.7 &$29.5\pm3.1$ &$11.6\pm1.3$ &$14.8\pm3.6$ &$-0.52\pm0.10$\\
Cl1216\_44 &12:16:56.54 &-12:11:02.3 &1.7 &89.2 &9.4 &$13.1\pm1.2$ &$3.8\pm0.5$ &$12.1\pm1.9$ &$-0.11\pm0.10$\\
Cl1216\_45 &12:16:47.22 &-12:03:59.8 &1.8 &93.5 &5.9 &$7.2\pm0.7$ &$2.3\pm0.3$ &$5.5\pm0.9$ &$-0.26\pm0.10$\\
Cl1216\_47 &12:16:41.60 &-11:53:35.6 &1.8 &80.9 &7.4 &$9.1\pm0.9$ &$4.5\pm0.4$ &$0.9\pm1.1$ &$-0.90\pm0.11$\\
Cl1216\_48 &12:16:40.91 &-11:54:17.6 &1.8 &80.5 &7.9 &$8.7\pm0.9$ &$2.9\pm0.3$ &$6.6\pm1.2$ &$-0.28\pm0.10$\\
Cl1216\_49 &12:16:31.45 &-12:07:31.2 &1.6 &148.6 &13.0 &$15.0\pm1.0$ &$6.5\pm0.5$ &$5.2\pm1.1$ &$-0.67\pm0.06$\\
Cl1216\_50 &12:16:28.96 &-12:10:19.9 &1.9 &71.7 &6.9 &$9.1\pm1.0$ &$2.8\pm0.4$ &$11.8\pm2.0$ &$0.02\pm0.11$\\
Cl1216\_51 &12:16:25.82 &-11:57:41.7 &1.6 &163.0 &16.6 &$13.9\pm0.9$ &$5.9\pm0.4$ &$5.5\pm1.1$ &$-0.62\pm0.06$\\
Cl1216\_52 &12:16:21.99 &-11:57:53.6 &1.9 &44.3 &4.8 &$7.4\pm1.0$ &$2.3\pm0.5$ &$11.8\pm2.1$ &$0.13\pm0.14$\\
Cl1216\_53 &12:15:59.63 &-12:06:57.7 &1.9 &46.8 &4.8 &$10.3\pm1.3$ &$3.2\pm0.5$ &$9.0\pm2.0$ &$-0.17\pm0.13$\\
Cl1216\_54 &12:15:56.78 &-11:55:40.8 &1.9 &33.5 &4.4 &$12.4\pm1.9$ &$4.1\pm0.7$ &$8.8\pm2.6$ &$-0.30\pm0.16$\\
Cl1216\_55 &12:17:30.43 &-11:55:58.6 &1.8 &68.5 &8.5 &$14.8\pm1.5$ &$5.7\pm0.6$ &$7.0\pm2.0$ &$-0.53\pm0.11$\\
Cl1216\_56 &12:17:30.54 &-11:53:23.0 &1.8 &54.5 &7.2 &$16.2\pm2.0$ &$5.6\pm0.8$ &$10.5\pm2.8$ &$-0.36\pm0.13$\\
Cl1216\_57 &12:17:05.34 &-12:02:40.5 &1.9 &49.7 &5.2 &$11.3\pm1.4$ &$3.9\pm0.6$ &$7.4\pm1.7$ &$-0.36\pm0.12$\\
Cl1216\_58 &12:17:04.05 &-11:58:38.0 &1.7 &59.6 &5.3 &$5.8\pm0.7$ &$2.4\pm0.3$ &$3.9\pm1.3$ &$-0.43\pm0.15$\\
Cl1216\_59 &12:16:53.25 &-11:58:38.7 &1.8 &54.0 &6.9 &$4.7\pm0.6$ &$0.8\pm0.2$ &$6.6\pm1.0$ &$0.36\pm0.12$\\
Cl1216\_60 &12:16:51.56 &-12:13:53.7 &1.7 &85.1 &11.3 &$16.1\pm1.6$ &$5.7\pm0.6$ &$9.8\pm2.1$ &$-0.40\pm0.10$\\
\end{tabular}
\end{minipage}
\end{table*}
\addtocounter{table}{-1}

\begin{table*}
\begin{minipage}{\textwidth}
\caption[Cl1216-1201 X-ray point source catalogue, continued.]{Catalogue of X-ray point sources in the Cl1216-1201 field, continued.}
\begin{tabular}{lcccccccccc}
\hline
ID	&R.A. (J2000)		&Dec. (J2000)	&Error		&Counts	&$\sigma_{src}$	&$S_{\mathrm{0.5-8~keV}}$	&$S_{\mathrm{0.5-2~keV}}$	&$S_{\mathrm{2-8~keV}}$	&$HR$\\
\hline\hline
Cl1216\_61 &12:16:49.73 &-12:09:33.7 &1.7 &55.3 &5.3 &$6.5\pm0.9$ &$2.4\pm0.4$ &$4.2\pm1.3$ &$-0.39\pm0.15$\\
Cl1216\_63 &12:16:39.24 &-11:52:43.9 &1.8 &57.9 &6.9 &$7.5\pm0.9$ &$2.3\pm0.3$ &$6.6\pm1.3$ &$-0.17\pm0.12$\\
Cl1216\_64 &12:16:35.19 &-12:06:43.9 &1.7 &90.0 &7.1 &$8.3\pm0.8$ &$3.5\pm0.4$ &$4.0\pm1.2$ &$-0.56\pm0.11$\\
Cl1216\_65 &12:16:24.05 &-12:09:07.5 &1.8 &60.9 &5.7 &$7.1\pm0.9$ &$1.5\pm0.4$ &$10.4\pm1.5$ &$0.26\pm0.13$\\
Cl1216\_66 &12:16:23.23 &-11:55:47.2 &1.8 &52.5 &7.0 &$13.1\pm1.5$ &$6.0\pm0.7$ &$3.4\pm1.6$ &$-0.75\pm0.10$\\
Cl1216\_67 &12:16:20.49 &-12:10:38.7 &1.8 &69.8 &7.5 &$9.4\pm1.1$ &$4.7\pm0.5$ &$4.2\pm2.1$ &$-0.63\pm0.15$\\
Cl1216\_68 &12:16:18.65 &-12:07:51.0 &1.7 &55.7 &6.3 &$7.1\pm0.8$ &$2.8\pm0.4$ &$4.2\pm1.3$ &$-0.45\pm0.13$\\
Cl1216\_69 &12:16:13.97 &-11:55:33.4 &1.8 &67.3 &8.0 &$8.5\pm0.9$ &$2.9\pm0.4$ &$6.3\pm1.3$ &$-0.29\pm0.11$\\
Cl1216\_70 &12:16:13.16 &-12:03:28.0 &1.7 &49.1 &5.1 &$5.3\pm0.7$ &$0.0\pm0.2$ &$11.1\pm1.4$ &$1.00\pm0.17$\\
Cl1216\_71 &12:17:00.40 &-12:12:44.9 &1.7 &43.1 &4.4 &$8.1\pm1.2$ &$2.6\pm0.5$ &$7.4\pm2.1$ &$-0.17\pm0.16$\\
Cl1216\_72 &12:16:59.95 &-12:11:34.5 &1.8 &45.5 &5.6 &$9.9\pm1.3$ &$1.3\pm0.4$ &$12.4\pm2.0$ &$0.41\pm0.14$\\
Cl1216\_73 &12:16:48.82 &-12:08:10.6 &1.9 &64.3 &7.1 &$6.8\pm0.8$ &$2.4\pm0.3$ &$5.9\pm1.4$ &$-0.25\pm0.13$\\
Cl1216\_74 &12:16:36.76 &-11:53:10.7 &1.9 &63.4 &5.6 &$9.8\pm1.1$ &$2.8\pm0.4$ &$10.0\pm1.7$ &$-0.05\pm0.11$\\
Cl1216\_75 &12:16:01.03 &-12:09:53.3 &2.2 &26.2 &5.9 &$23.3\pm3.7$ &$8.3\pm1.5$ &$15.6\pm4.6$ &$-0.36\pm0.15$\\
Cl1216\_76 &12:17:02.00 &-11:54:03.1 &1.7 &49.8 &5.9 &$6.0\pm0.9$ &$2.5\pm0.4$ &$6.2\pm1.8$ &$-0.24\pm0.15$\\
Cl1216\_77 &12:17:05.58 &-12:14:03.9 &1.7 &40.1 &4.8 &$9.5\pm1.4$ &$2.8\pm0.5$ &$8.1\pm2.0$ &$-0.16\pm0.15$\\
\end{tabular}
\end{minipage}
\end{table*}

\begin{table*}
\begin{minipage}{\textwidth}
\caption[Cl1054-1145 X-ray point source catalogue]{Catalogue of X-ray point sources in the Cl1054-1145 field.
\label{tab:xray1054}}
\begin{tabular}{lcccccccccc}
\hline
ID	&R.A. (J2000)		&Dec. (J2000)	&Error		&Counts	&$\sigma_{src}$	&$S_{\mathrm{0.5-8~keV}}$	&$S_{\mathrm{0.5-2~keV}}$	&$S_{\mathrm{2-8~keV}}$	&$HR$\\
\hline\hline
Cl1054-11\_1 &10:55:10.93 &-11:46:17.0 &1.6 &271.1 &22.8 &$39.2\pm1.8$ &$10.0\pm0.6$ &$43.6\pm2.9$ &$0.04\pm0.05$\\
Cl1054-11\_2 &10:55:01.63 &-11:41:52.7 &1.7 &125.2 &12.1 &$14.0\pm1.1$ &$4.7\pm0.4$ &$10.1\pm1.5$ &$-0.30\pm0.08$\\
Cl1054-11\_3 &10:54:54.62 &-11:37:18.8 &1.7 &177.8 &16.2 &$25.3\pm1.5$ &$10.8\pm0.7$ &$8.9\pm1.6$ &$-0.66\pm0.06$\\
Cl1054-11\_4 &10:54:45.21 &-11:50:29.2 &1.6 &257.6 &19.0 &$23.8\pm1.1$ &$9.2\pm0.5$ &$13.3\pm1.3$ &$-0.47\pm0.04$\\
Cl1054-11\_5 &10:54:44.94 &-11:44:10.3 &1.6 &276.1 &20.5 &$21.7\pm1.0$ &$11.4\pm0.5$ &$0.3\pm0.8$ &$-0.99\pm0.04$\\
Cl1054-11\_6 &10:54:44.65 &-11:50:55.6 &1.8 &94.4 &8.8 &$8.1\pm0.8$ &$3.0\pm0.3$ &$8.2\pm1.3$ &$-0.19\pm0.09$\\
Cl1054-11\_7 &10:54:41.31 &-11:45:28.2 &1.5 &1880.1 &97.4 &$141.8\pm2.3$ &$58.8\pm1.1$ &$62.7\pm2.3$ &$-0.58\pm0.01$\\
Cl1054-11\_8 &10:54:38.52 &-11:32:09.0 &1.6 &121.2 &19.3 &$108.2\pm7.3$ &$36.1\pm3.0$ &$81.4\pm9.5$ &$-0.28\pm0.07$\\
Cl1054-11\_9 &10:54:33.79 &-11:49:31.3 &1.5 &1111.2 &60.4 &$85.1\pm1.8$ &$31.8\pm0.8$ &$54.0\pm2.2$ &$-0.40\pm0.02$\\
Cl1054-11\_10 &10:54:28.58 &-11:46:28.9 &1.6 &793.4 &48.2 &$54.0\pm1.4$ &$21.5\pm0.6$ &$28.5\pm1.5$ &$-0.50\pm0.02$\\
Cl1054-11\_11 &10:54:28.07 &-11:42:59.2 &1.6 &199.9 &15.2 &$14.6\pm0.8$ &$6.2\pm0.4$ &$5.9\pm0.9$ &$-0.62\pm0.05$\\
Cl1054-11\_12 &10:54:27.82 &-11:56:24.5 &1.5 &415.3 &31.1 &$55.8\pm2.1$ &$19.3\pm0.9$ &$39.0\pm2.7$ &$-0.33\pm0.04$\\
Cl1054-11\_13 &10:54:27.41 &-11:57:01.7 &1.6 &255.8 &21.8 &$37.1\pm1.7$ &$14.5\pm0.8$ &$18.9\pm2.0$ &$-0.51\pm0.04$\\
Cl1054-11\_14 &10:54:17.60 &-11:52:02.0 &1.8 &135.5 &13.2 &$12.9\pm0.9$ &$5.2\pm0.4$ &$6.2\pm1.1$ &$-0.54\pm0.07$\\
Cl1054-11\_15 &10:55:07.47 &-11:46:03.1 &1.6 &224.1 &20.2 &$28.9\pm1.4$ &$9.4\pm0.6$ &$22.9\pm1.9$ &$-0.24\pm0.05$\\
Cl1054-11\_16 &10:55:05.44 &-11:39:45.7 &1.8 &104.2 &11.7 &$19.1\pm1.5$ &$7.0\pm0.6$ &$12.0\pm2.0$ &$-0.40\pm0.08$\\
Cl1054-11\_17 &10:54:59.53 &-11:45:51.9 &1.7 &138.0 &12.1 &$14.2\pm1.0$ &$6.1\pm0.4$ &$7.7\pm1.4$ &$-0.52\pm0.07$\\
Cl1054-11\_18 &10:54:54.14 &-11:45:39.9 &1.7 &148.4 &12.5 &$13.3\pm0.9$ &$5.0\pm0.4$ &$7.8\pm1.2$ &$-0.44\pm0.07$\\
Cl1054-11\_19 &10:54:51.74 &-11:37:02.6 &1.8 &57.0 &5.7 &$8.3\pm1.0$ &$3.1\pm0.4$ &$4.9\pm1.4$ &$-0.44\pm0.13$\\
Cl1054-11\_20 &10:54:48.57 &-11:47:17.1 &1.8 &104.0 &9.0 &$9.8\pm0.8$ &$1.7\pm0.2$ &$14.5\pm1.3$ &$0.37\pm0.08$\\
Cl1054-11\_21 &10:54:48.45 &-11:51:21.4 &1.6 &90.7 &8.3 &$9.2\pm0.8$ &$3.3\pm0.3$ &$5.9\pm1.3$ &$-0.38\pm0.10$\\
Cl1054-11\_22 &10:54:47.21 &-11:41:05.9 &1.8 &60.8 &5.7 &$6.4\pm0.7$ &$2.2\pm0.3$ &$4.4\pm1.0$ &$-0.32\pm0.12$\\
Cl1054-11\_23 &10:54:46.32 &-11:58:09.8 &1.7 &61.3 &6.5 &$14.3\pm1.5$ &$5.6\pm0.6$ &$6.5\pm2.0$ &$-0.55\pm0.11$\\
Cl1054-11\_24 &10:54:44.77 &-11:56:55.3 &1.7 &126.3 &11.6 &$19.7\pm1.4$ &$7.6\pm0.6$ &$9.4\pm1.8$ &$-0.53\pm0.07$\\
Cl1054-11\_25 &10:54:41.21 &-11:51:26.4 &1.8 &71.1 &5.8 &$6.5\pm0.7$ &$2.0\pm0.3$ &$5.9\pm1.0$ &$-0.14\pm0.11$\\
Cl1054-11\_26 &10:54:39.27 &-11:47:19.2 &1.8 &98.7 &8.8 &$7.2\pm0.6$ &$1.4\pm0.2$ &$11.9\pm1.1$ &$0.36\pm0.08$\\
Cl1054-11\_27 &10:54:33.68 &-11:58:56.6 &1.6 &81.5 &7.3 &$13.2\pm1.4$ &$5.6\pm0.6$ &$5.7\pm2.0$ &$-0.59\pm0.12$\\
Cl1054-11\_28 &10:54:31.07 &-11:47:45.6 &1.6 &113.4 &8.3 &$8.1\pm0.6$ &$2.6\pm0.3$ &$6.0\pm0.8$ &$-0.28\pm0.08$\\
Cl1054-11\_29 &10:54:23.72 &-11:58:26.5 &1.7 &65.6 &7.1 &$12.8\pm1.4$ &$4.1\pm0.5$ &$10.7\pm2.3$ &$-0.21\pm0.12$\\
Cl1054-11\_30 &10:54:22.86 &-11:59:25.2 &1.9 &64.2 &8.9 &$16.0\pm1.7$ &$6.0\pm0.7$ &$8.8\pm2.4$ &$-0.46\pm0.12$\\
Cl1054-11\_31 &10:54:22.61 &-11:42:25.4 &1.7 &136.8 &11.0 &$11.0\pm0.7$ &$5.0\pm0.3$ &$2.9\pm0.7$ &$-0.74\pm0.06$\\
Cl1054-11\_32 &10:54:20.24 &-11:52:51.1 &1.6 &165.1 &14.2 &$17.0\pm1.0$ &$5.5\pm0.4$ &$13.8\pm1.5$ &$-0.23\pm0.06$\\
Cl1054-11\_33 &10:54:19.95 &-11:46:43.8 &1.6 &287.5 &19.8 &$21.2\pm1.0$ &$8.3\pm0.4$ &$11.8\pm1.1$ &$-0.47\pm0.04$\\
Cl1054-11\_34 &10:54:04.98 &-11:53:00.7 &1.9 &75.1 &7.0 &$9.0\pm0.9$ &$4.2\pm0.4$ &$2.4\pm1.5$ &$-0.75\pm0.14$\\
Cl1054-11\_35 &10:53:54.48 &-11:45:26.8 &1.6 &51.7 &5.6 &$8.7\pm1.1$ &$2.7\pm0.4$ &$5.8\pm1.5$ &$-0.31\pm0.13$\\
Cl1054-11\_36 &10:53:50.43 &-11:53:35.6 &1.6 &129.6 &12.2 &$20.9\pm1.5$ &$8.1\pm0.6$ &$10.4\pm1.8$ &$-0.51\pm0.07$\\
Cl1054-11\_37 &10:53:33.38 &-11:50:32.2 &1.6 &101.4 &9.8 &$21.4\pm1.8$ &$8.7\pm0.7$ &$9.1\pm2.2$ &$-0.58\pm0.08$\\
Cl1054-11\_38 &10:55:09.06 &-11:49:42.8 &1.8 &54.5 &6.2 &$7.7\pm0.9$ &$2.1\pm0.3$ &$7.0\pm1.5$ &$-0.09\pm0.13$\\
Cl1054-11\_39 &10:55:06.68 &-11:36:48.1 &1.7 &59.0 &8.5 &$24.2\pm2.6$ &$9.4\pm1.1$ &$12.7\pm3.2$ &$-0.49\pm0.10$\\
Cl1054-11\_40 &10:55:04.37 &-11:51:28.9 &1.8 &51.3 &4.6 &$6.8\pm0.9$ &$3.1\pm0.4$ &$3.0\pm1.4$ &$-0.61\pm0.15$\\
\end{tabular}
\end{minipage}
\end{table*}
\addtocounter{table}{-1}

\begin{table*}
\begin{minipage}{\textwidth}
\caption[Cl1054-1145 X-ray point source catalogue, continued.]{Catalogue of point sources in the Cl1054-1145 field, continued.}
\begin{tabular}{lcccccccccc}
\hline
ID	&R.A. (J2000)		&Dec. (J2000)	&Error		&Counts	&$\sigma_{src}$	&$S_{\mathrm{0.5-8~keV}}$	&$S_{\mathrm{0.5-2~keV}}$	&$S_{\mathrm{2-8~keV}}$	&$HR$\\
\hline\hline
Cl1054-11\_41 &10:54:53.70 &-11:45:03.7 &1.8 &106.5 &10.1 &$10.2\pm0.9$ &$3.5\pm0.3$ &$6.8\pm1.1$ &$-0.35\pm0.09$\\
Cl1054-11\_42 &10:54:45.87 &-11:53:41.3 &1.7 &64.6 &6.2 &$7.5\pm0.8$ &$1.2\pm0.3$ &$13.3\pm1.6$ &$0.48\pm0.10$\\
Cl1054-11\_43 &10:54:30.53 &-11:51:00.6 &1.7 &44.7 &4.8 &$3.8\pm0.6$ &$1.6\pm0.2$ &$3.0\pm1.1$ &$-0.36\pm0.18$\\
Cl1054-11\_44 &10:54:29.13 &-11:40:08.4 &1.7 &60.4 &8.0 &$7.6\pm0.9$ &$2.9\pm0.4$ &$7.4\pm1.5$ &$-0.22\pm0.12$\\
Cl1054-11\_45 &10:54:26.32 &-11:51:20.4 &1.7 &73.6 &7.4 &$6.2\pm0.7$ &$2.2\pm0.3$ &$5.4\pm1.1$ &$-0.25\pm0.11$\\
Cl1054-11\_46 &10:54:25.75 &-11:51:40.7 &1.7 &93.0 &8.0 &$8.0\pm0.7$ &$3.1\pm0.3$ &$5.9\pm1.1$ &$-0.35\pm0.09$\\
Cl1054-11\_47 &10:54:21.84 &-11:44:06.9 &1.6 &43.5 &4.5 &$3.2\pm0.5$ &$0.9\pm0.2$ &$3.0\pm0.7$ &$-0.10\pm0.16$\\
Cl1054-11\_48 &10:54:20.18 &-11:57:12.7 &1.7 &111.6 &9.4 &$17.0\pm1.3$ &$4.6\pm0.5$ &$17.0\pm2.0$ &$-0.04\pm0.08$\\
Cl1054-11\_49 &10:54:15.76 &-11:44:52.5 &1.8 &99.3 &8.7 &$7.5\pm0.6$ &$2.7\pm0.3$ &$5.1\pm0.9$ &$-0.36\pm0.09$\\
Cl1054-11\_50 &10:54:11.69 &-11:44:40.0 &1.8 &63.7 &5.8 &$5.1\pm0.6$ &$2.2\pm0.3$ &$2.3\pm0.8$ &$-0.60\pm0.13$\\
Cl1054-11\_51 &10:54:06.81 &-11:50:51.0 &1.8 &68.8 &6.4 &$6.2\pm0.7$ &$2.9\pm0.4$ &$4.8\pm1.1$ &$-0.42\pm0.11$\\
Cl1054-11\_52 &10:53:38.76 &-11:38:46.7 &1.7 &56.3 &6.1 &$12.4\pm1.5$ &$2.7\pm0.5$ &$18.8\pm2.6$ &$0.26\pm0.11$\\
Cl1054-11\_53 &10:53:34.48 &-11:46:50.1 &1.7 &100.2 &10.3 &$19.0\pm1.6$ &$8.2\pm0.7$ &$7.6\pm2.3$ &$-0.62\pm0.10$\\
Cl1054-11\_54 &10:54:50.81 &-11:51:52.0 &1.8 &70.3 &7.6 &$7.5\pm0.8$ &$2.4\pm0.3$ &$6.0\pm1.2$ &$-0.22\pm0.11$\\
Cl1054-11\_55 &10:54:41.33 &-11:58:11.9 &1.9 &61.1 &6.6 &$10.5\pm1.2$ &$4.5\pm0.5$ &$3.5\pm1.5$ &$-0.67\pm0.12$\\
Cl1054-11\_56 &10:53:50.40 &-11:46:30.1 &1.9 &47.7 &6.5 &$6.0\pm0.9$ &$2.1\pm0.3$ &$4.4\pm1.3$ &$-0.31\pm0.15$\\
Cl1054-11\_57 &10:55:13.27 &-11:39:08.2 &2.0 &28.8 &4.8 &$21.1\pm3.2$ &$5.1\pm1.1$ &$23.9\pm5.0$ &$0.08\pm0.15$\\
Cl1054-11\_58 &10:55:08.22 &-11:43:52.0 &1.8 &53.2 &6.5 &$8.9\pm1.1$ &$2.0\pm0.3$ &$9.2\pm1.8$ &$0.08\pm0.13$\\
Cl1054-11\_59 &10:54:54.72 &-11:43:33.7 &1.7 &50.7 &5.5 &$4.5\pm0.7$ &$1.8\pm0.3$ &$4.0\pm1.4$ &$-0.28\pm0.17$\\
Cl1054-11\_60 &10:54:51.73 &-11:54:43.3 &1.8 &50.1 &4.6 &$6.3\pm0.9$ &$2.1\pm0.4$ &$7.4\pm1.4$ &$-0.06\pm0.13$\\
Cl1054-11\_61 &10:54:44.80 &-11:34:46.2 &2.2 &28.1 &5.7 &$10.1\pm1.8$ &$4.2\pm0.8$ &$5.3\pm3.4$ &$-0.52\pm0.25$\\
Cl1054-11\_62 &10:54:30.17 &-11:35:39.0 &1.8 &45.1 &4.9 &$8.3\pm1.1$ &$2.1\pm0.3$ &$8.6\pm1.8$ &$0.01\pm0.13$\\
Cl1054-11\_63 &10:53:51.11 &-11:43:39.1 &1.8 &38.7 &4.8 &$5.1\pm0.8$ &$2.1\pm0.3$ &$3.5\pm1.4$ &$-0.41\pm0.18$\\
Cl1054-11\_64 &10:54:49.50 &-11:42:21.0 &1.9 &29.0 &4.4 &$3.7\pm0.7$ &$1.4\pm0.3$ &$2.3\pm0.9$ &$-0.41\pm0.17$\\
Cl1054-11\_65 &10:54:36.90 &-11:43:28.6 &1.7 &52.0 &4.7 &$4.1\pm0.5$ &$1.4\pm0.2$ &$3.0\pm0.7$ &$-0.31\pm0.13$\\
Cl1054-11\_66 &10:53:32.80 &-11:47:48.1 &1.7 &40.1 &4.6 &$11.0\pm1.5$ &$2.9\pm0.5$ &$11.9\pm2.6$ &$0.02\pm0.14$\\
Cl1054-11\_67 &10:54:18.87 &-11:44:05.0 &1.7 &43.5 &4.6 &$3.1\pm0.5$ &$1.2\pm0.2$ &$3.0\pm0.9$ &$-0.23\pm0.16$\\
\end{tabular}
\end{minipage}
\end{table*}

\begin{table*}
\begin{minipage}{\textwidth}
\caption[Cl1040-1155 X-ray point source catalogue]{Catalogue of X-ray point sources in the Cl1040-1155 field. 
\label{tab:xray1040}}
\begin{tabular}{lcccccccccc}
\hline
ID	&R.A. (J2000)		&Dec. (J2000)	&Error		&Counts	&$\sigma_{src}$	&$S_{\mathrm{0.5-8~keV}}$	&$S_{\mathrm{0.5-2~keV}}$	&$S_{\mathrm{2-8~keV}}$	&$HR$\\
\hline\hline
Cl1040\_1 &10:41:33.34 &-11:55:11.2 &1.5 &909.1 &52.5 &$199.5\pm4.7$ &$76.8\pm2.0$ &$104.2\pm5.3$ &$-0.49\pm0.02$\\
Cl1040\_2 &10:41:08.70 &-12:03:30.6 &1.5 &1269.9 &66.2 &$186.8\pm3.8$ &$83.3\pm1.8$ &$50.3\pm3.1$ &$-0.74\pm0.01$\\
Cl1040\_3 &10:41:04.48 &-11:46:03.1 &1.6 &199.9 &15.3 &$34.8\pm1.9$ &$14.3\pm0.8$ &$14.4\pm2.2$ &$-0.60\pm0.05$\\
Cl1040\_4 &10:41:01.95 &-11:53:14.8 &1.5 &942.2 &48.8 &$93.8\pm2.2$ &$35.1\pm1.0$ &$57.8\pm2.7$ &$-0.42\pm0.02$\\
Cl1040\_5 &10:41:01.39 &-11:46:14.1 &1.5 &642.4 &41.6 &$102.2\pm3.0$ &$43.7\pm1.4$ &$35.3\pm2.9$ &$-0.66\pm0.02$\\
Cl1040\_6 &10:40:58.90 &-11:58:45.6 &1.6 &670.9 &31.7 &$61.4\pm1.7$ &$24.0\pm0.8$ &$33.9\pm2.0$ &$-0.48\pm0.03$\\
Cl1040\_7 &10:40:49.62 &-11:56:35.7 &1.6 &260.5 &16.8 &$23.1\pm1.1$ &$9.1\pm0.5$ &$12.7\pm1.3$ &$-0.48\pm0.04$\\
Cl1040\_8 &10:40:48.81 &-12:06:16.0 &1.6 &511.3 &30.8 &$97.1\pm3.1$ &$37.1\pm1.3$ &$53.4\pm3.6$ &$-0.47\pm0.03$\\
Cl1040\_9 &10:40:45.46 &-11:53:31.8 &1.6 &507.0 &27.1 &$39.5\pm1.3$ &$15.8\pm0.6$ &$20.2\pm1.4$ &$-0.51\pm0.03$\\
Cl1040\_10 &10:40:41.65 &-11:53:44.8 &1.6 &261.9 &15.1 &$20.0\pm1.0$ &$7.7\pm0.4$ &$11.3\pm1.2$ &$-0.46\pm0.05$\\
Cl1040\_11 &10:40:36.88 &-11:57:42.4 &1.5 &652.9 &30.5 &$48.8\pm1.4$ &$19.8\pm0.6$ &$24.0\pm1.5$ &$-0.54\pm0.03$\\
Cl1040\_12 &10:40:35.25 &-11:56:54.7 &1.6 &571.6 &25.8 &$40.1\pm1.3$ &$15.4\pm0.6$ &$23.9\pm1.5$ &$-0.44\pm0.03$\\
Cl1040\_13 &10:40:22.78 &-11:55:17.6 &1.6 &209.7 &14.0 &$17.1\pm0.9$ &$6.0\pm0.4$ &$11.9\pm1.2$ &$-0.34\pm0.05$\\
Cl1040\_14 &10:40:13.75 &-11:48:10.9 &1.7 &285.2 &19.1 &$33.3\pm1.7$ &$15.3\pm0.8$ &$3.4\pm1.4$ &$-0.89\pm0.04$\\
Cl1040\_15 &10:41:33.56 &-11:51:34.5 &2.1 &83.5 &9.7 &$41.7\pm3.6$ &$13.2\pm1.4$ &$34.8\pm5.3$ &$-0.21\pm0.09$\\
Cl1040\_16 &10:41:18.35 &-11:46:18.3 &1.8 &123.2 &13.9 &$35.0\pm2.4$ &$12.2\pm0.9$ &$24.3\pm3.3$ &$-0.34\pm0.07$\\
Cl1040\_17 &10:41:17.88 &-11:52:52.5 &1.8 &108.1 &10.0 &$15.1\pm1.3$ &$5.4\pm0.5$ &$10.0\pm1.8$ &$-0.37\pm0.09$\\
Cl1040\_18 &10:41:15.81 &-11:55:07.8 &1.7 &144.7 &10.7 &$16.6\pm1.3$ &$6.8\pm0.5$ &$12.5\pm2.0$ &$-0.37\pm0.08$\\
Cl1040\_19 &10:41:14.71 &-11:59:00.6 &1.8 &77.3 &5.9 &$9.6\pm1.0$ &$1.9\pm0.3$ &$12.9\pm1.8$ &$0.26\pm0.11$\\
Cl1040\_20 &10:41:03.79 &-12:01:52.8 &1.9 &77.7 &5.7 &$9.2\pm1.0$ &$3.9\pm0.4$ &$3.7\pm1.2$ &$-0.61\pm0.11$\\
Cl1040\_21 &10:41:01.47 &-11:59:33.3 &1.6 &221.6 &16.7 &$22.2\pm1.2$ &$9.3\pm0.5$ &$8.9\pm1.3$ &$-0.62\pm0.05$\\
Cl1040\_22 &10:40:48.86 &-11:59:38.0 &1.6 &196.3 &16.1 &$15.5\pm0.9$ &$4.1\pm0.3$ &$17.0\pm1.4$ &$0.02\pm0.06$\\
Cl1040\_23 &10:40:30.70 &-11:56:55.6 &1.7 &157.2 &9.3 &$11.1\pm0.8$ &$4.8\pm0.3$ &$4.2\pm0.9$ &$-0.64\pm0.07$\\
Cl1040\_24 &10:40:29.84 &-11:55:09.0 &1.6 &119.6 &8.5 &$8.9\pm0.7$ &$2.6\pm0.3$ &$8.7\pm1.0$ &$-0.09\pm0.08$\\
Cl1040\_25 &10:40:28.42 &-12:03:52.6 &1.6 &220.4 &17.1 &$29.5\pm1.6$ &$9.9\pm0.6$ &$23.2\pm2.2$ &$-0.26\pm0.05$\\
Cl1040\_26 &10:40:26.89 &-11:56:45.5 &1.7 &134.2 &8.5 &$10.1\pm0.7$ &$4.1\pm0.3$ &$5.0\pm0.9$ &$-0.53\pm0.07$\\
Cl1040\_27 &10:40:26.03 &-12:01:44.8 &1.6 &102.3 &7.7 &$12.2\pm1.0$ &$4.6\pm0.4$ &$7.2\pm1.4$ &$-0.44\pm0.09$\\
Cl1040\_28 &10:40:24.26 &-11:49:54.9 &1.7 &180.6 &12.0 &$17.9\pm1.1$ &$7.9\pm0.5$ &$6.0\pm1.2$ &$-0.68\pm0.06$\\
Cl1040\_29 &10:40:17.62 &-11:51:45.6 &1.7 &115.7 &8.5 &$11.7\pm1.0$ &$4.0\pm0.4$ &$8.6\pm1.3$ &$-0.30\pm0.08$\\
Cl1040\_30 &10:40:09.18 &-11:57:41.3 &1.7 &189.4 &15.4 &$28.8\pm1.7$ &$11.8\pm0.8$ &$21.0\pm2.5$ &$-0.39\pm0.06$\\
\end{tabular}
\end{minipage}
\end{table*}
\addtocounter{table}{-1}

\begin{table*}
\begin{minipage}{\textwidth}
\caption[Cl1040-1155 X-ray point source catalogue, continued.]{Catalogue of point sources in the Cl1040-1155 field, continued.}
\begin{tabular}{lcccccccccc}
\hline
ID	&R.A. (J2000)		&Dec. (J2000)	&Error		&Counts	&$\sigma_{src}$	&$S_{\mathrm{0.5-8~keV}}$	&$S_{\mathrm{0.5-2~keV}}$	&$S_{\mathrm{2-8~keV}}$	&$HR$\\
\hline\hline
Cl1040\_31 &10:39:58.40 &-11:49:03.2 &1.8 &94.2 &9.6 &$31.9\pm2.7$ &$15.5\pm1.3$ &$6.7\pm3.3$ &$-0.80\pm0.09$\\
Cl1040\_32 &10:39:58.00 &-11:58:55.9 &1.7 &88.7 &8.9 &$12.4\pm1.1$ &$6.0\pm0.5$ &$1.7\pm1.4$ &$-0.87\pm0.11$\\
Cl1040\_33 &10:39:53.42 &-11:56:21.4 &1.6 &147.9 &10.9 &$21.1\pm1.5$ &$10.6\pm0.7$ &$1.2\pm1.8$ &$-0.95\pm0.08$\\
Cl1040\_34 &10:41:20.43 &-11:50:23.5 &1.7 &62.0 &5.6 &$10.9\pm1.3$ &$5.1\pm0.5$ &$3.8\pm2.0$ &$-0.69\pm0.14$\\
Cl1040\_35 &10:41:15.24 &-12:06:24.0 &2.0 &58.8 &5.7 &$22.8\pm2.5$ &$11.5\pm1.1$ &$-\pm-$ &$1.00\pm0.00$\\
Cl1040\_36 &10:41:15.31 &-11:58:01.8 &1.7 &83.0 &6.2 &$10.3\pm1.1$ &$3.9\pm0.4$ &$5.9\pm1.4$ &$-0.45\pm0.11$\\
Cl1040\_37 &10:41:07.94 &-11:47:32.3 &1.7 &93.4 &8.0 &$13.6\pm1.3$ &$5.6\pm0.6$ &$8.8\pm2.2$ &$-0.44\pm0.11$\\
Cl1040\_38 &10:41:07.60 &-12:07:36.3 &1.7 &74.9 &7.8 &$15.7\pm1.6$ &$5.6\pm0.6$ &$9.6\pm2.2$ &$-0.40\pm0.11$\\
Cl1040\_39 &10:41:04.09 &-11:52:21.1 &1.7 &70.4 &5.5 &$7.8\pm0.9$ &$3.2\pm0.4$ &$3.7\pm1.2$ &$-0.55\pm0.12$\\
Cl1040\_40 &10:41:00.25 &-12:03:18.3 &1.8 &65.3 &5.4 &$11.4\pm1.3$ &$3.5\pm0.5$ &$9.9\pm1.9$ &$-0.17\pm0.12$\\
Cl1040\_41 &10:40:56.87 &-11:55:34.7 &1.6 &120.3 &10.8 &$10.6\pm0.8$ &$4.0\pm0.3$ &$6.3\pm1.1$ &$-0.44\pm0.08$\\
Cl1040\_42 &10:40:48.57 &-11:44:18.8 &1.7 &79.0 &9.6 &$13.6\pm1.4$ &$4.0\pm0.5$ &$11.8\pm2.0$ &$-0.16\pm0.10$\\
Cl1040\_43 &10:40:43.58 &-11:54:33.9 &1.8 &105.2 &7.0 &$7.8\pm0.7$ &$0.6\pm0.3$ &$16.7\pm1.3$ &$0.76\pm0.10$\\
Cl1040\_44 &10:40:41.48 &-11:49:42.9 &1.9 &75.7 &6.8 &$7.1\pm0.8$ &$2.4\pm0.3$ &$4.5\pm1.1$ &$-0.36\pm0.12$\\
Cl1040\_45 &10:40:38.56 &-11:46:11.4 &1.7 &79.7 &6.2 &$10.6\pm1.1$ &$4.1\pm0.5$ &$5.5\pm1.5$ &$-0.50\pm0.11$\\
Cl1040\_46 &10:40:16.86 &-11:55:58.9 &1.7 &92.0 &7.5 &$6.8\pm0.7$ &$2.6\pm0.3$ &$11.3\pm1.8$ &$0.05\pm0.10$\\
Cl1040\_47 &10:40:13.64 &-11:51:41.7 &1.7 &95.2 &8.9 &$10.2\pm0.9$ &$1.0\pm0.2$ &$18.1\pm1.7$ &$0.63\pm0.08$\\
Cl1040\_48 &10:40:10.56 &-12:03:45.7 &1.7 &129.7 &11.8 &$22.5\pm1.6$ &$8.6\pm0.7$ &$11.8\pm2.0$ &$-0.49\pm0.07$\\
Cl1040\_49 &10:39:46.78 &-11:51:44.4 &2.1 &60.2 &6.8 &$23.3\pm2.6$ &$10.5\pm1.2$ &$8.4\pm3.9$ &$-0.67\pm0.13$\\
Cl1040\_50 &10:41:25.55 &-12:01:53.4 &1.7 &63.6 &5.1 &$12.5\pm1.4$ &$4.6\pm0.6$ &$7.5\pm2.0$ &$-0.42\pm0.12$\\
Cl1040\_51 &10:41:19.99 &-11:49:38.5 &1.7 &67.2 &6.5 &$12.2\pm1.4$ &$5.8\pm0.6$ &$1.6\pm1.9$ &$-0.87\pm0.15$\\
Cl1040\_52 &10:41:05.76 &-11:51:18.7 &1.7 &60.5 &6.0 &$7.3\pm0.9$ &$3.2\pm0.4$ &$4.1\pm1.7$ &$-0.51\pm0.16$\\
Cl1040\_53 &10:40:56.99 &-12:08:00.9 &1.7 &59.5 &7.0 &$11.6\pm1.4$ &$5.5\pm0.6$ &$2.2\pm2.5$ &$-0.82\pm0.19$\\
Cl1040\_54 &10:40:36.53 &-11:52:56.9 &1.7 &65.7 &4.4 &$4.9\pm0.6$ &$1.5\pm0.2$ &$4.4\pm0.9$ &$-0.16\pm0.13$\\
Cl1040\_55 &10:40:15.79 &-11:51:33.0 &1.8 &70.0 &6.2 &$7.4\pm0.8$ &$3.1\pm0.4$ &$2.5\pm1.0$ &$-0.66\pm0.12$\\
Cl1040\_56 &10:40:14.22 &-11:58:07.4 &1.7 &67.7 &6.2 &$6.1\pm0.7$ &$2.4\pm0.3$ &$3.8\pm1.2$ &$-0.42\pm0.14$\\
Cl1040\_57 &10:40:13.41 &-12:05:36.9 &1.9 &52.4 &7.3 &$10.2\pm1.3$ &$2.2\pm0.4$ &$13.5\pm2.2$ &$0.22\pm0.12$\\
Cl1040\_58 &10:40:09.61 &-11:48:55.1 &1.8 &64.4 &6.1 &$8.2\pm1.1$ &$4.4\pm0.5$ &$-\pm-$ &$-1.01\pm0.19$\\
Cl1040\_59 &10:39:55.11 &-11:49:01.3 &1.9 &45.8 &6.6 &$16.8\pm2.3$ &$5.4\pm0.9$ &$13.5\pm3.1$ &$-0.23\pm0.13$\\
Cl1040\_60 &10:41:36.85 &-11:48:17.8 &1.8 &40.4 &5.9 &$18.5\pm2.6$ &$5.7\pm0.9$ &$16.7\pm4.3$ &$-0.15\pm0.15$\\
Cl1040\_61 &10:41:28.32 &-11:52:32.5 &1.8 &46.4 &5.3 &$12.0\pm1.6$ &$2.9\pm0.5$ &$10.5\pm2.6$ &$-0.06\pm0.15$\\
Cl1040\_62 &10:40:18.17 &-11:59:40.9 &1.7 &56.3 &5.1 &$5.4\pm0.7$ &$0.8\pm0.3$ &$9.2\pm1.2$ &$0.48\pm0.13$\\
Cl1040\_63 &10:40:15.37 &-12:03:51.4 &2.1 &34.6 &5.5 &$5.6\pm1.0$ &$2.0\pm0.4$ &$6.9\pm1.6$ &$-0.06\pm0.16$\\
Cl1040\_64 &10:40:14.86 &-12:00:51.4 &1.7 &49.6 &4.2 &$6.1\pm0.9$ &$0.5\pm0.2$ &$10.8\pm1.6$ &$0.68\pm0.14$\\
Cl1040\_65 &10:40:10.99 &-11:45:34.0 &1.9 &46.3 &4.7 &$8.3\pm1.2$ &$3.3\pm0.5$ &$4.5\pm2.0$ &$-0.50\pm0.17$\\
Cl1040\_66 &10:40:52.42 &-11:49:09.6 &2.0 &42.1 &5.0 &$6.0\pm0.9$ &$1.7\pm0.3$ &$5.1\pm1.4$ &$-0.14\pm0.17$\\
Cl1040\_67 &10:40:32.53 &-11:52:09.6 &1.8 &44.7 &4.3 &$3.7\pm0.6$ &$0.9\pm0.2$ &$3.9\pm0.9$ &$0.03\pm0.17$\\
\end{tabular}
\vspace{4cm}
\end{minipage}
\end{table*}

\renewcommand\topfraction{.1}

\section{Candidate optical counterparts}
In Tables \ref{tab:opt1216}, \ref{tab:opt1054}, and \ref{tab:opt1040} we present catalogues of the candidate optical counterparts determined from the WFI data.  For each X-ray source we list: the X-ray source ID; the R.A. and Dec. of the candidate optical counterpart (J2000); the probability of association, $P_{id}$; and the counterpart $V$, $R$, and $I$ magnitudes.  Where multiple candidate counterparts are found, their optical properties are listed in order of decreasing $P_{id}$.  These data are available in machine-readable form at the CDS.  

In Tables \ref{tab:vlt1216}, \ref{tab:vlt1054}, and \ref{tab:vlt1040}, we present additional optical information for candidate counterparts within the central region of each cluster covered by deep multi-band VLT observations.  For each X-ray source we list: the X-ray ID; the EDisCS catalogue ID (see White et al. 2005); the probability of association, $P_{id}$, determined from the WFI images; the $B$, $V$, $I,$ $J$,  and $K$ magnitudes; the redshift estimates, $z_{Bol}$ and $z_{Rud}$, and cluster membership probabilities, $P_{mem}$, returned by the Bolzonella et al. (2000) and Rudnick et al. (2001) photometric redshift codes; the redshift estimate derived from fits to a quasar template, $z_{QSO}$; and the cluster member flag derived from analysis of both redshift codes and the extant spectroscopic redshifts (Pell{\'o} et al., in preparation), with a "$\star$" denoting a flagged candidate and a "C" denoting a member which has already been spectroscopically confirmed. 

\begin{table*}
\begin{minipage}{\textwidth}
\caption[Candidate optical counterparts for X-ray sources in the Cl1216-1201 field.]{Candidate optical counterparts for X-ray sources in the Cl1216-1201 field. 
\label{tab:opt1216}}
\begin{center}
\begin{tabular}{lcccccc}
\hline
ID      &R.A. (J2000)  &Dec. (J2000)   &$P_{id}$  &$V$      &$R$      &$I$\\
\hline\hline
Cl1216\_1 &12:17:35.50 &-12:02:06.2 &0.99 &19.1 &18.6 &17.9\\
Cl1216\_2 &12:17:21.15 &-12:08:01.0 &0.91 &20.4 &19.5 &18.5\\
Cl1216\_3 &12:17:21.18 &-12:01:37.8 &0.90 &21.0 &20.1 &19.2\\
Cl1216\_4 &12:16:54.10 &-12:00:57.3 &0.97 &22.2 &21.3 &19.8\\
Cl1216\_5 &12:16:48.94 &-12:03:33.3 &0.97 &21.1 &20.4 &19.8\\
Cl1216\_6 &12:16:48.71 &-12:02:29.4 &0.92 &21.2 &20.7 &20.0\\
Cl1216\_7 &-- &-- &-- &-- &-- &-- \\
Cl1216\_8 &12:16:46.06 &-11:57:37.1 &0.98 &20.7 &20.4 &20.1\\
Cl1216\_9 &12:16:42.70 &-12:01:06.7 &0.85 &19.7 &18.9 &17.6\\
 &12:16:42.87 &-12:01:05.5 &0.14 &-- &20.5 &--\\
Cl1216\_10 &12:16:37.38 &-12:01:12.0 &0.98 &20.9 &20.3 &19.3\\
Cl1216\_11 &12:16:30.70 &-12:11:14.0 &0.59 &25.0 &23.2 &22.5\\
Cl1216\_12 &12:16:30.63 &-11:57:04.1 &0.94 &22.4 &21.0 &19.6\\
Cl1216\_13 &12:16:29.17 &-12:04:23.4 &0.99 &18.8 &18.4 &17.7\\
Cl1216\_14 &12:16:22.45 &-12:08:02.3 &0.87 &-- &23.3 &21.7\\
Cl1216\_15 &-- &-- &-- &-- &-- &-- \\
Cl1216\_16 &12:16: 6.70 &-11:57:42.1 &0.97 &22.0 &21.5 &21.0\\
Cl1216\_17 &12:17:43.75 &-11:53:46.3 &1.00 &15.3 &14.8 &13.2\\
Cl1216\_18 &12:17:34.97 &-11:59:26.0 &0.94 &22.7 &21.7 &20.9\\
Cl1216\_19 &12:17:29.64 &-11:57:37.1 &0.49 &22.3 &22.0 &20.9\\
 &12:17:29.85 &-11:57:40.1 &0.37 &25.1 &23.2 &21.5\\
Cl1216\_20 &12:17:14.84 &-11:55:08.1 &0.96 &21.9 &21.3 &21.5\\
Cl1216\_21 &12:17: 6.91 &-11:56:03.9 &0.85 &-- &23.2 &21.1\\
Cl1216\_22 &12:17: 1.46 &-12:02:30.6 &0.88 &24.1 &23.2 &22.3\\
Cl1216\_23 &12:16:53.67 &-11:53:07.1 &0.97 &21.9 &21.4 &20.5\\
Cl1216\_24 &12:16:51.22 &-11:51:22.1 &0.82 &22.3 &21.9 &21.2\\
Cl1216\_25 &12:16:51.08 &-12:01:52.2 &0.85 &24.3 &22.8 &22.7\\
Cl1216\_26 &12:16:50.44 &-11:57:32.6 &0.65 &-- &23.9 &23.5\\
 &12:16:50.31 &-11:57:31.2 &0.18 &24.4 &23.9 &23.5\\
Cl1216\_27 &12:16:50.09 &-11:56:16.3 &0.98 &21.1 &20.8 &20.2\\
Cl1216\_28 &12:16:47.19 &-11:55:31.9 &0.65 &23.2 &22.9 &21.9\\
Cl1216\_29 &12:16:44.74 &-12:10:18.9 &0.94 &22.1 &21.6 &20.9\\
Cl1216\_30 &12:16:39.38 &-11:58:01.0 &0.96 &22.3 &21.7 &20.8\\
Cl1216\_31 &12:16:24.34 &-11:57:14.0 &0.88 &22.8 &22.5 &21.4\\
Cl1216\_32 &12:16:23.75 &-11:50:42.7 &0.83 &23.1 &21.6 &20.2\\
 &12:16:23.86 &-11:50:44.4 &0.10 &-- &24.3 &--\\
Cl1216\_33 &12:16:12.30 &-11:55:46.8 &0.37 &24.2 &23.5 &22.5\\
Cl1216\_34 &12:16: 5.23 &-12:00:00.0 &0.48 &-- &23.9 &22.6\\
 &12:16: 5.28 &-12:00:04.0 &0.23 &-- &24.7 &--\\
Cl1216\_35 &12:16: 4.87 &-11:54:28.2 &0.99 &18.8 &17.9 &16.9\\
Cl1216\_36 &12:15:56.16 &-11:56:21.4 &0.54 &20.6 &20.2 &19.2\\
 &12:15:56.28 &-11:56:16.8 &0.22 &22.5 &22.2 &20.3\\
 &12:15:56.14 &-11:56:16.5 &0.22 &-- &22.5 &--\\
Cl1216\_37 &12:17:34.79 &-12:08:10.8 &0.49 &20.9 &19.4 &18.1\\
 &12:17:35.01 &-12:08:06.2 &0.41 &21.9 &20.2 &19.5\\
Cl1216\_38 &-- &-- &-- &-- &-- &-- \\
Cl1216\_39 &12:17:15.95 &-12:05:13.4 &0.95 &23.1 &21.8 &20.2\\
Cl1216\_40 &12:17:13.82 &-11:57:41.6 &0.93 &22.1 &21.5 &20.6\\
Cl1216\_41 &12:17:13.23 &-12:12:58.4 &0.96 &20.1 &19.6 &18.9\\
Cl1216\_42 &12:17: 6.73 &-12:08:13.5 &0.66 &22.8 &22.2 &21.3\\
 &12:17: 7.02 &-12:08:14.2 &0.29 &22.3 &21.4 &20.2\\
Cl1216\_43 &12:16:57.74 &-11:48:27.3 &0.97 &20.9 &20.8 &20.2\\
Cl1216\_44 &12:16:56.50 &-12:11:00.4 &1.00 &18.6 &17.4 &16.4\\
Cl1216\_45 &12:16:47.30 &-12:03:57.4 &0.73 &21.9 &21.4 &20.6\\
 &12:16:47.40 &-12:04:01.6 &0.15 &24.3 &23.2 &21.4\\
Cl1216\_46 &12:16:45.20 &-12:01:17.2 &0.64 &22.8 &20.9 &19.0\\
 &12:16:45.31 &-12:01:15.6 &0.27 &23.7 &21.9 &20.2\\
Cl1216\_47 &12:16:41.52 &-11:53:36.7 &1.00 &18.1 &17.1 &15.9\\
Cl1216\_48 &12:16:40.73 &-11:54:19.0 &0.68 &24.7 &23.5 &22.1\\
Cl1216\_49 &12:16:31.57 &-12:07:31.1 &0.95 &21.7 &21.0 &19.8\\
Cl1216\_50 &12:16:29.12 &-12:10:15.9 &0.19 &25.0 &24.0 &23.2\\
  \end{tabular}
\end{center}
\end{minipage}
\end{table*}

\addtocounter{table}{-1}

\begin{table*}
\begin{minipage}{\textwidth}
\caption[Candidate optical counterparts for X-ray sources in the Cl1216-1201 field.]{Candidate optical counterparts for X-ray sources in the Cl1216-1201 field, continued. 
\label{tab:opt1216}}
\begin{center}
\begin{tabular}{lcccccc}
\hline
ID      &R.A. (J2000)  &Dec. (J2000)   &$P_{id}$  &$V$      &$R$      &$I$\\
\hline\hline
Cl1216\_51 &12:16:25.94 &-11:57:41.8 &0.91 &21.3 &19.8 &18.2\\
Cl1216\_52 &12:16:22.27 &-11:57:52.5 &0.16 &-- &24.6 &--\\
Cl1216\_53 &12:15:59.47 &-12:06:59.8 &0.63 &22.1 &21.3 &20.4\\
 &12:15:59.86 &-12:06:58.7 &0.16 &23.5 &22.8 &21.8\\
  &12:15:59.68 &-12:06:59.4 &0.16 &-- &23.9 &--\\
Cl1216\_54 &12:15:56.89 &-11:55:42.8 &0.98 &20.3 &19.0 &18.0\\
Cl1216\_55 &12:17:30.41 &-11:55:58.1 &1.00 &17.3 &16.2 &14.3\\
Cl1216\_56 &12:17:30.33 &-11:53:21.2 &0.65 &21.6 &21.2 &20.1\\
 &12:17:30.55 &-11:53:23.8 &0.28 &24.7 &24.0 &--\\
Cl1216\_57 &12:17: 5.50 &-12:02:38.1 &0.55 &23.9 &23.2 &22.0\\
 &12:17: 5.36 &-12:02:38.3 &0.27 &-- &24.4 &--\\
Cl1216\_58 &12:17: 3.93 &-11:58:37.1 &0.86 &22.4 &22.2 &21.3\\
Cl1216\_59 &12:16:53.34 &-11:58:40.7 &0.39 &23.9 &23.1 &22.0\\
 &12:16:53.11 &-11:58:37.7 &0.39 &24.1 &23.0 &21.4\\
  &12:16:53.19 &-11:58:40.9 &0.15 &-- &24.1 &--\\
Cl1216\_60 &12:16:51.28 &-12:13:52.3 &0.92 &21.1 &20.4 &19.6\\
Cl1216\_61 &12:16:49.70 &-12:09:32.9 &0.93 &23.2 &21.3 &20.0\\
Cl1216\_62 &12:16:40.34 &-12:04:42.0 &0.92 &23.5 &22.2 &20.7\\
Cl1216\_63 &12:16:39.17 &-11:52:45.2 &0.97 &21.7 &21.1 &20.4\\
Cl1216\_64 &12:16:35.25 &-12:06:44.1 &0.93 &24.0 &22.6 &21.1\\
Cl1216\_65 &12:16:24.14 &-12:09:06.4 &0.84 &24.6 &23.5 &23.0\\
Cl1216\_66 &12:16:23.35 &-11:55:49.6 &0.49 &-- &24.3 &22.1\\
Cl1216\_67 &12:16:20.54 &-12:10:37.3 &1.00 &17.5 &16.6 &15.6\\
Cl1216\_68 &-- &-- &-- &-- &-- &-- \\
Cl1216\_69 &-- &-- &-- &-- &-- &-- \\
Cl1216\_70 &12:16:12.91 &-12:03:26.2 &0.26 &-- &24.7 &--\\
Cl1216\_71 &-- &-- &-- &-- &-- &-- \\
Cl1216\_72 &12:17: 0.25 &-12:11:35.4 &0.99 &17.1 &16.4 &15.5\\
Cl1216\_73 &12:16:48.53 &-12:08:11.7 &0.73 &21.1 &20.0 &18.8\\
 &12:16:49.03 &-12:08:08.9 &0.11 &24.9 &23.2 &22.7\\
  &12:16:48.97 &-12:08:12.1 &0.11 &-- &23.8 &22.6\\
Cl1216\_74 &12:16:36.94 &-11:53:08.3 &0.41 &24.7 &24.2 &22.6\\
Cl1216\_75 &12:16: 1.03 &-12:09:53.7 &0.98 &20.8 &20.3 &19.7\\
Cl1216\_76 &12:17: 2.01 &-11:53:58.7 &0.25 &22.9 &22.7 &22.3\\
 &12:17: 1.89 &-11:54:01.5 &0.25 &-- &24.4 &--\\
  &12:17: 1.86 &-11:53:59.0 &0.18 &23.3 &22.7 &22.0\\
Cl1216\_77 &12:17: 5.34 &-12:14:02.2 &0.99 &17.5 &16.6 &15.6\\
\end{tabular}
\end{center}
\end{minipage}
\end{table*}

\begin{table*}
\begin{minipage}{\textwidth}
\caption[Candidate optical counterparts for X-ray sources in the Cl1054-1145 field.]{Candidate optical counterparts for X-ray sources in the Cl1054-1145 field. 
\label{tab:opt1054}}
\begin{center}
\begin{tabular}{lcccccc}
\hline
ID      &R.A. (J2000)  &Dec. (J2000)   &$P_{id}$  &$V$      &$R$      &$I$\\
\hline\hline
Cl1054\_1 &10:55:11.03 &-11:46:16.4 &0.77 &24.2 &22.9 &21.8\\
 &10:55:10.86 &-11:46:19.8 &0.16 &24.5 &24.1 &--\\
Cl1054\_2 &10:55: 1.61 &-11:41:55.0 &0.83 &-- &21.6 &21.4\\
 &10:55: 1.61 &-11:41:48.6 &0.12 &23.8 &22.6 &22.3\\
Cl1054\_3 &10:54:54.70 &-11:37:19.3 &0.93 &-- &21.1 &20.9\\
Cl1054\_4 &10:54:45.30 &-11:50:31.6 &0.83 &24.2 &23.1 &22.3\\
Cl1054\_5 &10:54:44.95 &-11:44:12.8 &1.00 &-- &10.5 &10.0\\
Cl1054\_6 &10:54:44.80 &-11:50:56.0 &0.66 &23.8 &23.8 &--\\
 &10:54:44.58 &-11:50:59.5 &0.17 &-- &24.8 &--\\
Cl1054\_7 &10:54:41.32 &-11:45:28.3 &0.91 &-- &17.8 &17.6\\
Cl1054\_8 &-- &-- &-- &-- &-- &-- \\
Cl1054\_9 &10:54:33.76 &-11:49:32.5 &0.55 &21.9 &21.3 &20.2\\
 &10:54:33.84 &-11:49:29.9 &0.43 &23.2 &21.5 &20.0\\
Cl1054\_10 &10:54:28.66 &-11:46:28.4 &0.98 &21.0 &20.5 &20.2\\
Cl1054\_11 &10:54:28.29 &-11:43:00.1 &0.45 &23.3 &22.7 &22.2\\
 &10:54:28.12 &-11:42:58.1 &0.45 &24.6 &24.0 &--\\
Cl1054\_12 &10:54:27.82 &-11:56:25.0 &0.99 &19.6 &18.9 &18.1\\
Cl1054\_13 &10:54:27.43 &-11:57:01.1 &0.99 &20.5 &19.7 &18.9\\
Cl1054\_14 &-- &-- &-- &-- &-- &-- \\
\end{tabular}
\end{center}
\end{minipage}
\end{table*}

\addtocounter{table}{-1}

\begin{table*}
\begin{minipage}{\textwidth}
\caption[Candidate optical counterparts for X-ray sources in the Cl1054-1145 field.]{Candidate optical counterparts for X-ray sources in the Cl1054-1145 field, continued. 
}
\begin{center}
\begin{tabular}{lcccccc}
\hline
ID      &R.A. (J2000)  &Dec. (J2000)   &$P_{id}$  &$V$      &$R$      &$I$\\
\hline\hline

Cl1054\_15 &10:55: 7.63 &-11:46:04.0 &0.85 &20.2 &19.2 &18.1\\
 &10:55: 7.69 &-11:46:00.9 &0.12 &21.8 &20.7 &20.1\\
Cl1054\_16 &10:55: 5.46 &-11:39:45.1 &0.96 &20.7 &20.3 &19.5\\
Cl1054\_17 &-- &-- &-- &-- &-- &-- \\
Cl1054\_18 &10:54:54.30 &-11:45:41.8 &0.66 &21.5 &21.4 &21.5\\
 &10:54:53.91 &-11:45:40.3 &0.18 &23.5 &22.9 &--\\
 &10:54:53.98 &-11:45:42.4 &0.11 &24.6 &23.6 &22.4\\
Cl1054\_19 &10:54:51.70 &-11:37:02.6 &0.98 &20.6 &20.4 &19.8\\
Cl1054\_20 &10:54:48.61 &-11:47:18.8 &0.97 &22.3 &20.9 &19.4\\
Cl1054\_21 &10:54:48.47 &-11:51:21.8 &0.82 &23.5 &22.6 &22.0\\
 &10:54:48.54 &-11:51:18.1 &0.13 &24.4 &23.5 &22.6\\
Cl1054\_22 &10:54:47.08 &-11:41:05.6 &0.98 &20.6 &19.5 &18.7\\
Cl1054\_23 &10:54:46.32 &-11:58:10.9 &0.98 &21.9 &20.4 &19.6\\
Cl1054\_24 &10:54:44.78 &-11:56:55.9 &0.95 &22.8 &22.0 &20.9\\
Cl1054\_25 &10:54:41.29 &-11:51:27.6 &0.50 &24.9 &24.4 &--\\
 &10:54:40.94 &-11:51:28.8 &0.36 &22.9 &21.7 &20.8\\
Cl1054\_26 &10:54:39.29 &-11:47:19.3 &0.90 &24.9 &23.3 &22.2\\
Cl1054\_27 &10:54:33.75 &-11:58:57.3 &0.99 &20.7 &20.0 &18.9\\
Cl1054\_28 &10:54:31.11 &-11:47:45.6 &0.92 &24.5 &23.3 &--\\
Cl1054\_29 &10:54:23.91 &-11:58:27.2 &0.94 &20.6 &20.3 &19.9\\
Cl1054\_30 &10:54:22.55 &-11:59:25.6 &0.97 &19.1 &18.2 &17.4\\
Cl1054\_31 &10:54:22.68 &-11:42:25.2 &0.84 &-- &21.0 &20.8\\
 &10:54:22.44 &-11:42:23.1 &0.11 &23.8 &22.5 &21.1\\
Cl1054\_32 &10:54:20.50 &-11:52:50.4 &0.99 &16.2 &15.9 &15.7\\
Cl1054\_33 &10:54:20.00 &-11:46:43.4 &0.99 &19.7 &19.3 &18.9\\
Cl1054\_34 &10:54: 4.96 &-11:52:59.5 &0.96 &-- &21.8 &21.4\\
Cl1054\_35 &10:53:54.42 &-11:45:26.0 &0.94 &23.5 &22.7 &22.2\\
Cl1054\_36 &10:53:50.40 &-11:53:35.7 &0.97 &20.7 &20.6 &20.0\\
Cl1054\_37 &10:53:33.30 &-11:50:30.5 &0.98 &21.4 &20.6 &19.8\\
Cl1054\_38 &10:55: 8.98 &-11:49:38.9 &0.94 &-- &20.4 &20.0\\
Cl1054\_39 &10:55: 6.71 &-11:36:48.8 &0.92 &24.3 &22.9 &22.9\\
Cl1054\_40 &10:55: 4.43 &-11:51:30.8 &0.79 &24.3 &24.0 &--\\
Cl1054\_41 &10:54:53.66 &-11:45:01.7 &0.98 &19.4 &19.1 &18.6\\
Cl1054\_42 &10:54:45.86 &-11:53:41.8 &0.60 &-- &24.6 &22.7\\
 &10:54:45.94 &-11:53:45.0 &0.24 &25.4 &24.3 &--\\
Cl1054\_43 &-- &-- &-- &-- &-- &-- \\
Cl1054\_44 &10:54:29.22 &-11:40:10.0 &0.97 &22.3 &21.2 &20.3\\
Cl1054\_45 &10:54:26.27 &-11:51:18.1 &0.74 &-- &24.2 &--\\
Cl1054\_46 &10:54:25.71 &-11:51:39.9 &0.63 &-- &24.1 &--\\
 &10:54:25.54 &-11:51:38.7 &0.23 &24.1 &23.7 &--\\
Cl1054\_47 &10:54:21.85 &-11:44:09.9 &0.65 &-- &24.7 &--\\
Cl1054\_48 &10:54:20.09 &-11:57:12.3 &0.74 &23.2 &22.1 &20.2\\
 &10:54:20.10 &-11:57:14.4 &0.17 &-- &24.1 &--\\
Cl1054\_49 &10:54:15.51 &-11:44:50.8 &0.74 &23.2 &22.7 &22.7\\
Cl1054\_50 &10:54:11.38 &-11:44:38.5 &1.00 &13.7 &13.4 &12.6\\
Cl1054\_51 &10:54: 6.54 &-11:50:48.9 &0.64 &24.3 &22.9 &22.4\\
Cl1054\_52 &10:53:38.51 &-11:38:48.6 &0.60 &24.9 &23.6 &22.0\\
Cl1054\_53 &10:53:34.40 &-11:46:47.2 &0.98 &19.6 &19.1 &18.6\\
Cl1054\_54 &-- &-- &-- &-- &-- &-- \\
Cl1054\_55 &10:54:41.36 &-11:58:13.7 &0.93 &22.7 &22.2 &21.5\\
Cl1054\_56 &10:53:50.37 &-11:46:26.6 &0.63 &24.6 &23.4 &--\\
 &10:53:50.53 &-11:46:25.8 &0.15 &-- &24.7 &--\\
Cl1054\_57 &10:55:13.39 &-11:39:12.3 &0.97 &19.6 &18.6 &17.7\\
Cl1054\_58 &10:55: 8.19 &-11:43:51.4 &0.86 &23.0 &22.3 &22.1\\
Cl1054\_59 &10:54:55.03 &-11:43:34.3 &0.99 &15.6 &15.3 &14.7\\
Cl1054\_60 &10:54:51.95 &-11:54:42.9 &0.98 &19.6 &19.2 &18.6\\
Cl1054\_61 &10:54:44.59 &-11:34:46.4 &0.89 &22.7 &22.2 &21.2\\
Cl1054\_62 &-- &-- &-- &-- &-- &-- \\
Cl1054\_63 &-- &-- &-- &-- &-- &-- \\
Cl1054\_64 &-- &-- &-- &-- &-- &-- \\
Cl1054\_65 &-- &-- &-- &-- &-- &-- \\
Cl1054\_66 &-- &-- &-- &-- &-- &-- \\
Cl1054\_67 &10:54:18.82 &-11:44:01.8 &0.98 &20.5 &19.7 &19.1\\
\end{tabular}
\end{center}
\end{minipage}
\end{table*}

\begin{table*}
\begin{minipage}{\textwidth}
\caption[Candidate optical counterparts for X-ray sources in the Cl1040-1155 field.]{Candidate optical counterparts for X-ray sources in the Cl1040-1155 field.
\label{tab:opt1040}}
\begin{center}
\begin{tabular}{lcccccc}
\hline
ID      &R.A. (J2000)  &Dec. (J2000)   &$P_{id}$  &$V$      &$R$      &$I$\\
\hline\hline
Cl1040\_1 &10:41:33.34 &-11:55:10.3 &0.99 &19.2 &18.9 &17.9\\
Cl1040\_2 &10:41: 8.67 &-12:03:30.4 &1.00 &19.7 &18.5 &17.4\\
Cl1040\_3 &10:41: 4.48 &-11:46:06.4 &0.54 &22.5 &21.8 &20.9\\
 &10:41: 4.20 &-11:46:01.9 &0.41 &21.0 &20.7 &19.8\\
Cl1040\_4 &10:41: 2.02 &-11:53:14.0 &0.99 &20.1 &19.6 &19.3\\
Cl1040\_5 &10:41: 1.51 &-11:46:14.7 &0.97 &21.8 &21.0 &20.3\\
Cl1040\_6 &10:40:58.83 &-11:58:44.1 &0.65 &22.4 &22.0 &21.0\\
 &10:40:59.11 &-11:58:47.3 &0.25 &22.8 &22.3 &21.2\\
Cl1040\_7 &10:40:49.49 &-11:56:35.1 &0.81 &23.7 &23.1 &21.9\\
Cl1040\_8 &10:40:48.79 &-12:06:15.8 &0.99 &20.8 &19.9 &18.9\\
Cl1040\_9 &10:40:45.41 &-11:53:31.2 &0.99 &21.1 &20.4 &19.5\\
Cl1040\_10 &10:40:41.67 &-11:53:44.3 &0.98 &20.5 &20.2 &19.7\\
Cl1040\_11 &10:40:36.82 &-11:57:42.5 &0.98 &20.5 &20.3 &19.7\\
Cl1040\_12 &10:40:35.21 &-11:56:54.7 &0.99 &20.6 &20.0 &19.6\\
Cl1040\_13 &10:40:22.73 &-11:55:22.1 &0.32 &-- &24.1 &--\\
Cl1040\_14 &10:40:13.86 &-11:48:09.9 &1.00 &17.1 &16.0 &14.2\\
Cl1040\_15 &10:41:33.43 &-11:51:36.2 &0.90 &22.6 &22.0 &21.3\\
Cl1040\_16 &10:41:18.36 &-11:46:21.2 &0.77 &20.5 &20.3 &19.5\\
 &10:41:18.44 &-11:46:18.9 &0.19 &-- &23.1 &--\\
Cl1040\_17 &10:41:17.85 &-11:52:54.8 &0.75 &22.6 &22.5 &21.7\\
 &10:41:17.74 &-11:52:50.0 &0.11 &-- &24.6 &--\\
Cl1040\_18 &10:41:15.84 &-11:55:07.3 &0.99 &19.9 &19.6 &18.9\\
Cl1040\_19 &-- &-- &-- &-- &-- &-- \\
Cl1040\_20 &10:41: 3.59 &-12:01:53.5 &0.95 &21.1 &20.5 &20.3\\
Cl1040\_21 &10:41: 1.43 &-11:59:33.0 &0.99 &20.6 &20.0 &19.5\\
Cl1040\_22 &10:40:49.08 &-11:59:39.6 &0.97 &20.7 &19.5 &18.3\\
Cl1040\_23 &10:40:30.65 &-11:56:55.9 &0.94 &23.0 &22.4 &21.6\\
Cl1040\_24 &10:40:29.90 &-11:55:08.6 &0.76 &23.3 &23.1 &22.6\\
 &10:40:29.87 &-11:55:06.6 &0.14 &-- &24.3 &--\\
Cl1040\_25 &10:40:28.41 &-12:03:52.8 &0.81 &23.9 &22.2 &20.6\\
 &10:40:28.36 &-12:03:49.7 &0.16 &24.5 &23.0 &21.8\\
Cl1040\_26 &10:40:26.90 &-11:56:45.4 &0.76 &-- &24.7 &--\\
Cl1040\_27 &10:40:26.06 &-12:01:44.2 &0.80 &24.5 &23.2 &21.5\\
 &10:40:26.25 &-12:01:43.7 &0.10 &-- &23.8 &21.7\\
Cl1040\_28 &10:40:24.15 &-11:49:56.2 &0.95 &21.0 &20.6 &20.3\\
Cl1040\_29 &10:40:17.53 &-11:51:46.2 &0.81 &24.2 &23.4 &22.6\\
Cl1040\_30 &10:40: 9.04 &-11:57:40.2 &0.74 &20.7 &20.2 &19.9\\
 &10:40: 9.14 &-11:57:42.9 &0.21 &23.7 &22.3 &20.6\\
Cl1040\_31 &10:39:58.32 &-11:49:03.9 &1.00 &17.3 &16.1 &14.0\\
Cl1040\_32 &-- &-- &-- &-- &-- &-- \\
Cl1040\_33 &10:39:53.30 &-11:56:21.2 &1.00 &14.6 &14.6 &13.7\\
Cl1040\_34 &10:41:20.45 &-11:50:26.7 &0.86 &23.4 &22.5 &20.7\\
Cl1040\_35 &10:41:15.03 &-12:06:24.4 &1.00 &12.5 &11.9 &11.4\\
Cl1040\_36 &10:41:15.30 &-11:58:01.1 &0.95 &22.5 &22.0 &20.8\\
Cl1040\_37 &10:41: 8.03 &-11:47:33.4 &0.97 &20.9 &20.9 &19.9\\
Cl1040\_38 &10:41: 7.74 &-12:07:34.4 &0.70 &23.4 &22.7 &21.5\\
 &10:41: 7.46 &-12:07:34.7 &0.17 &-- &24.6 &22.6\\
Cl1040\_39 &10:41: 4.05 &-11:52:21.8 &0.98 &20.5 &20.2 &19.8\\
Cl1040\_40 &10:40:59.98 &-12:03:18.4 &0.49 &22.7 &21.9 &20.8\\
 &10:41: 0.36 &-12:03:17.9 &0.33 &-- &24.8 &--\\
Cl1040\_41 &10:40:56.87 &-11:55:33.6 &0.89 &21.9 &20.9 &19.9\\
Cl1040\_42 &10:40:48.54 &-11:44:17.5 &0.85 &21.2 &20.8 &19.7\\
 &10:40:48.59 &-11:44:20.4 &0.13 &-- &23.5 &21.2\\
Cl1040\_43 &10:40:43.50 &-11:54:34.2 &0.82 &-- &23.6 &21.7\\
Cl1040\_44 &10:40:41.42 &-11:49:42.6 &0.91 &23.5 &22.9 &22.2\\
Cl1040\_45 &10:40:38.60 &-11:46:12.0 &0.85 &23.6 &23.0 &21.5\\
Cl1040\_46 &10:40:16.81 &-11:55:59.9 &0.93 &24.0 &22.4 &20.8\\
Cl1040\_47 &10:40:13.62 &-11:51:38.9 &0.68 &22.4 &21.5 &19.9\\
 &10:40:13.62 &-11:51:42.8 &0.18 &-- &24.0 &--\\
Cl1040\_48 &10:40:10.44 &-12:03:47.8 &1.00 &16.1 &15.8 &15.1\\
Cl1040\_49 &10:39:46.66 &-11:51:49.0 &0.59 &22.2 &21.2 &19.3\\
 &10:39:46.60 &-11:51:44.2 &0.28 &-- &23.7 &22.7\\
Cl1040\_50 &-- &-- &-- &-- &-- &-- \\
\end{tabular}
\end{center}
\end{minipage}
\end{table*}

\addtocounter{table}{-1}

\begin{table*}
\begin{minipage}{\textwidth}
\caption[Candidate optical counterparts for X-ray sources in the Cl1040-1155 field.]{Candidate optical counterparts for X-ray sources in the Cl1040-1155 field, continued. 
}
\begin{center}
\begin{tabular}{lcccccc}
\hline
ID      &R.A. (J2000)  &Dec. (J2000)   &$P_{id}$  &$V$      &$R$      &$I$\\
\hline\hline
Cl1040\_51 &10:41:19.88 &-11:49:38.3 &0.97 &21.9 &21.5 &20.9\\
Cl1040\_52 &10:41: 5.92 &-11:51:20.2 &0.58 &-- &24.2 &--\\
Cl1040\_53 &10:40:57.07 &-12:08:00.3 &1.00 &18.5 &17.2 &15.0\\
Cl1040\_54 &10:40:36.65 &-11:52:55.6 &0.95 &21.8 &21.5 &21.1\\
Cl1040\_55 &10:40:15.64 &-11:51:31.7 &0.66 &25.3 &22.9 &21.0\\
 &10:40:16.02 &-11:51:33.0 &0.11 &-- &25.4 &--\\
Cl1040\_56 &10:40:14.23 &-11:58:10.1 &0.67 &-- &24.0 &22.8\\
Cl1040\_57 &-- &-- &-- &-- &-- &-- \\
Cl1040\_58 &10:40: 9.69 &-11:48:53.4 &0.83 &22.0 &21.4 &20.2\\
 &10:40: 9.55 &-11:48:57.6 &0.13 &24.0 &23.3 &21.7\\
Cl1040\_59 &10:39:55.13 &-11:48:59.8 &0.69 &23.2 &21.7 &20.5\\
 &10:39:55.12 &-11:49:06.1 &0.27 &21.7 &20.6 &19.7\\
Cl1040\_60 &10:41:36.97 &-11:48:16.4 &0.75 &24.9 &23.0 &22.4\\
 &10:41:36.89 &-11:48:20.9 &0.13 &-- &24.8 &--\\
Cl1040\_61 &10:41:28.15 &-11:52:35.5 &0.80 &21.5 &20.4 &19.4\\
 &10:41:28.44 &-11:52:34.1 &0.14 &-- &23.7 &--\\
Cl1040\_62 &10:40:18.16 &-11:59:36.8 &0.38 &-- &24.3 &22.4\\
Cl1040\_63 &10:40:15.39 &-12:03:55.3 &0.89 &21.9 &21.4 &21.1\\
Cl1040\_64 &10:40:14.70 &-12:00:52.3 &0.83 &-- &23.2 &21.1\\
Cl1040\_65 &10:40:10.98 &-11:45:35.7 &0.75 &25.2 &23.1 &21.7\\
 &10:40:11.21 &-11:45:31.1 &0.12 &24.0 &23.0 &21.4\\
Cl1040\_66 &10:40:52.47 &-11:49:12.4 &0.67 &21.6 &21.0 &20.0\\
 &10:40:52.19 &-11:49:08.8 &0.20 &22.4 &22.0 &21.1\\
Cl1040\_67 &10:40:32.25 &-11:52:09.5 &0.44 &23.2 &22.7 &21.8\\
 &10:40:32.77 &-11:52:09.1 &0.25 &-- &25.0 &--\\
\end{tabular}
\end{center}
\end{minipage}
\end{table*}

\begin{table*}
\begin{minipage}{\textwidth}
\caption{Additional optical properties of candidate counterparts to X-ray sources lying within the FOV of deep, multiband VLT observations in the Cl1216-1201 field.
\label{tab:vlt1216}}
\begin{tabular}{lccccccccccc}
\hline
ID	&Optical ID	&$P_{id}$  &$B$	&$V$	&$I$	&$J$	&$K$  &$z_{Bol}$ ($P_{mem}$) &$z_{Rud}$ ($P_{mem}$) & $z_{QSO}$ &Flag\\
\hline\hline
Cl1216\_4 &EDCSNJ1216541-1200570 &0.97 &22.0 &21.4 &20.3 &18.7 &16.9  &1.00 (27.9) &0.88 (34.6) &3.05 &\\
Cl1216\_5 &EDCSNJ1216489-1203334 &0.97 &20.9 &20.5 &20.1 &19.5 &18.4  &0.64 (25.6) &0.58 (5.33) &1.54 &\\
Cl1216\_6 &EDCSNJ1216487-1202295 &0.92 &21.0 &20.8 &20.3 &19.1 &17.5  &1.06 (4.47) &1.26 (0.66) &0.82 &\\
Cl1216\_8 &EDCSNJ1216461-1157369 &0.98 &20.7 &20.6 &20.3 &- &-  &0.88 (22.3) &1.94 (0.00) &0.72 &\\
Cl1216\_9 &EDCSNJ1216428-1201073 &0.85 &19.8 &19.0 &18.2 &16.9 &15.3  &0.63 (33.4) &0.58 (29.0) &3.53 &\\
 &EDCSNJ1216429-1201060 &0.14 &20.8 &20.3 &19.9 &19.1 &15.3  &0.30 (20.5) &0.58 (19.3) &1.55 &\\
Cl1216\_10 &EDCSNJ1216374-1201121 &0.98 &20.8 &20.5 &19.8 &18.7 &17.1  &0.97 (12.0) &1.04 (5.83) &0.58 &\\
Cl1216\_22 &EDCSNJ1217015-1202318 &0.88 &23.7 &23.1 &22.5 &- &-  &0.04 (16.3) &0.58 (9.00) &3.20 &\\
Cl1216\_25 &EDCSNJ1216511-1201521 &0.85 &23.6 &23.0 &22.6 &20.8 &19.0  &1.56 (0.01) &1.18 (0.00) &2.74 &\\
Cl1216\_26 &EDCSNJ1216505-1157327 &0.65 &24.0 &23.6 &22.8 &- &-  &1.66 (15.3) &1.46 (29.1) &1.87 &$\star$\\
 &EDCSNJ1216504-1157310 &0.18 &24.1 &23.8 &23.4 &- &-  &1.15 (14.8) &0.14 (2.00) &2.29 &\\
Cl1216\_30 &EDCSNJ1216392-1157594 &0.96 &22.3 &21.7 &21.3 &20.9 &19.2  &0.66 (34.6) &0.50 (3.33) &1.59 &\\
Cl1216\_45 &EDCSNJ1216473-1203575 &0.73 &21.7 &21.4 &21.1 &20.2 &19.2  &0.04 (0.07) &0.20 (1.66) &2.61 &\\
 &EDCSNJ1216474-1204016 &0.15 &24.6 &23.2 &22.1 &20.3 &19.2 &0.54 (35.4) &0.64 (16.6) &3.99 &\\
Cl1216\_46 &EDCSNJ1216453-1201176 &0.64 &22.1 &20.9 &19.5 &17.9 &16.2  &0.69 (82.3) &0.70 (79.0) &3.95 &C\\
 &EDCSNJ1216454-1201159 &0.27 &23.0 &21.8 &20.5 &18.9 &16.2  &0.66 (60.7) &0.68 (53.6) &3.99 &C\\
Cl1216\_59 &EDCSNJ1216533-1158406 &0.39 &23.7 &23.1 &22.6 &21.7 &20.3  &0.70 (27.1) &0.68 (45.3) &1.72 &\\
 &EDCSNJ1216531-1158378 &0.39 &23.8 &23.2 &21.9 &20.2 &20.3 &0.85 (43.3) &0.82 (50.8) &3.87 &$\star$\\
 &EDCSNJ1216532-1158409 &0.15 &24.5 &24.0 &23.4 &22.1 &20.3  &0.79 (21.1) &0.64 (35.8) &1.91 &\\

\hline
\end{tabular}
\end{minipage}
\end{table*}

\begin{table*}
\begin{minipage}{\textwidth}
\caption{Additional optical properties of candidate counterparts to X-ray sources lying within the FOV of deep, multiband VLT observations in the Cl1054-1145 field.
\label{tab:vlt1054}}
\begin{tabular}{cccccccccccc}
\hline
ID	&Optical ID	&$P_{id}$  &$B$	&$V$	&$I$	&$J$	 &$K$  &$z_{Bol}$ ($P_{mem}$) &$z_{Rud}$ ($P_{mem}$) & $z_{QSO}$ &Flag\\
\hline\hline
Cl1054\_7 &EDCSNJ1054412-1145286 &0.91 &18.1 &17.9 &17.6 &- &-  &0.30 (41.6) &1.30 (0.00) &0.99 &\\
Cl1054\_9 &EDCSNJ1054337-1149325 &0.55 &21.6 &21.1 &20.2 &18.9 &17.4  &1.03 (1.16) &1.06 (0.00) &3.01 &\\
 &EDCSNJ1054338-1149299 &0.43 &22.6 &21.4 &20.1 &18.6 &17.4  &0.69 (66.4) &0.64 (98.3) &3.92 &C\\
Cl1054\_10 &EDCSNJ1054286-1146286 &0.98 &21.1 &20.6 &20.2 &19.5 &18.1  &0.68 (86.7) &0.64 (64.1) &1.57 &$\star$\\
Cl1054\_26 &EDCSNJ1054392-1147198 &0.90 &24.1 &23.2 &21.9 &- &-  &0.94 (21.4) &0.84 (34.3) &3.78 &$\star$\\
Cl1054\_28 &EDCSNJ1054310-1147455 &0.92 &23.9 &23.0 &22.5 &20.2 &18.6  &1.30 (0.00) &1.18 (0.00) &3.29 &\\
Cl1054\_33 &EDCSNJ1054199-1146439 &0.99 &19.7 &19.3 &18.9 &18.3 &17.3  &0.67 (95.5) &0.28 (4.00) &1.60 &\\
Cl1054\_47 &EDCSNJ1054217-1144104 &0.65 &24.6 &24.0 &23.4 &22.7 &21.3  &0.67 (21.6) &0.58 (53.5) &1.56 &\\
Cl1054\_49 &EDCSNJ1054154-1144511 &0.74 &22.9 &22.5 &22.0 &20.5 &19.1  &0.20 (10.2) &1.28 (0.00) &2.64 &\\
Cl1054\_67 &EDCSNJ1054187-1144023 &0.98 &20.3 &19.7 &19.2 &18.3 &17.1  &0.35 (34.3) &0.48 (11.3) &3.49 &\\
\hline
\end{tabular}
\end{minipage}
\end{table*}

\begin{table*}
\begin{minipage}{\textwidth}
\caption{Additional optical properties of candidate counterparts to X-ray sources lying within the FOV of deep, multiband VLT observations in the Cl1040-1155 field.
\label{tab:vlt1040}}
\begin{tabular}{cccccccccccc}
\hline
ID	&Optical ID	&$P_{id}$  &$B$	&$V$	&$I$	&$J$	&$K$	 &$z_{Bol}$ ($P_{mem}$) &$z_{Rud}$ ($P_{mem}$) & $z_{QSO}$ &Flag\\
\hline\hline
Cl1040\_7 &EDCSNJ1040494-1156356 &0.81 &23.0 &22.6 &21.9 &20.2 &18.5  &1.20 (0.00) &1.30 (0.00) &2.64 &\\
Cl1040\_9 &EDCSNJ1040454-1153316 &0.99 &21.0 &20.4 &19.8 &18.4 &17.0  &0.24 (1.25) &0.96 (2.00) &3.00 &\\
Cl1040\_10 &EDCSNJ1040416-1153449 &0.98 &20.4 &20.2 &19.9 &18.8 &17.6  &0.22 (0.00) &1.98 (0.00) &2.58 &\\
Cl1040\_11 &EDCSNJ1040367-1157432 &0.98 &20.4 &20.2 &20.0 &18.9 &17.4  &1.31 (0.00) &1.78 (0.00) &0.85 &\\
Cl1040\_12 &EDCSNJ1040351-1156552 &0.99 &20.5 &20.2 &19.6 &19.1 &18.0  &0.81 (83.6) &0.20 (9.83) &1.85 &$\star$\\
Cl1040\_23 &EDCSNJ1040305-1156560 &0.94 &22.8 &22.4 &21.8 &20.9 &-  &0.78 (32.7) &0.86 (20.1) &2.05 &$\star$\\
Cl1040\_24 &EDCSNJ1040298-1155089 &0.76 &23.3 &23.1 &22.6 &- &-  &0.94 (9.91) &1.96 (0.00) &0.13 &\\
 &EDCSNJ1040298-1155066 &0.14 &24.6 &24.2 &23.5 &- &- &0.15 (14.8) &0.20 (9.33) &0.48 &\\
Cl1040\_26 &EDCSNJ1040268-1156449 &0.76 &24.7 &24.2 &23.7 &- &-  &1.50 (15.0) &1.02 (6.00) &1.87 &\\
Cl1040\_43 &EDCSNJ1040434-1154347 &0.82 &24.7 &23.6 &22.2 &20.1 &18.3  &0.94 (29.1) &0.92 (0.00) &4.62 &\\
Cl1040\_54 &EDCSNJ1040366-1152560 &0.95 &21.5 &21.4 &21.1 &20.2 &19.1  &0.25 (0.00) &0.34 (0.00) &0.84 &\\
Cl1040\_67 &EDCSNJ1040322-1152096 &0.44 &23.3 &22.8 &21.9 &- &-  &1.57 (16.6) &0.84 (26.8) &1.87 &$\star$\\
 &EDCSNJ1040327-1152092 &0.25 &24.8 &24.6 &24.3 &- &- &0.95 (15.0) &1.80 (0.33) &0.16 &\\

\hline
\end{tabular}
\end{minipage}
\end{table*}

\end{document}